\documentclass[useAMS,usenatbib,fleqn]{mn2e}
\usepackage{lastpage}
\usepackage{aas_macros}
\usepackage{times}
\usepackage{graphicx}
\usepackage{amsfonts}
\usepackage{amsmath,mathtools}
\usepackage{amssymb}
\usepackage{rotating} 
\usepackage{setspace}
\usepackage{pst-grad}
\usepackage{color}
\usepackage{enumitem}
\usepackage{verbatim}
\usepackage{soul}
\usepackage[normalem]{ulem}
\DeclareMathAlphabet{\mathbf}{OML}{cmm}{b}{it}
\DeclareMathAlphabet{\mathbfsf}{OT1}{cmss}{bx}{n}

\title[Efficient Photo-heating Algorithms] 
{Efficient Photo-heating Algorithms in Time-dependent Photo-ionization Simulations}
\author[Lee et al.]{Kai-Yan Lee\thanks{e--mail: klee@astro.su.se}, 
Garrelt Mellema,
and Peter Lundqvist \\
Department of Astronomy \& Oskar Klein Centre, AlbaNova, Stockholm University, SE-106 91
Stockholm, Sweden \\
}
\date{\today}
\pubyear{2015} \volume{000} \pagerange{0}
\begin{document}
\pagerange{\pageref{firstpage}--\pageref{LastPage}}
\maketitle
\label{firstpage}
\begin{abstract}
  We present an extension to the time-dependent photo-ionization code
  \textsc{C}$^2$\textsc{-Ray} to calculate photo-heating in an
  efficient and accurate way. In \textsc{C}$^2$\textsc{-Ray}, the
  thermal calculation demands relatively small time-steps for accurate
  results. We describe two novel methods to reduce the computational
  cost associated with small time-steps, namely, an adaptive time-step
  algorithm and an asynchronous evolution approach.  The adaptive
  time-step algorithm determines an optimal time-step for the next
  computational step. It uses a fast ray-tracing scheme to quickly
  locate the relevant cells for this determination and only use these
  cells for the calculation of the time-step.  Asynchronous evolution
  allows different cells to evolve with different time-steps. The
  asynchronized clocks of the cells are synchronized at the times
  where outputs are produced. By only evolving cells which may require
  short time-steps with these short time-steps instead of imposing
  them to the whole grid, the computational cost of the calculation 
  can be substantially reduced. We show that our methods work well for
  several cosmologically relevant test problems and validate our
  results by comparing to the results of another time-dependent
  photo-ionization code.

\end{abstract}


\section{Introduction}

Photo-ionization is a process of fundamental importance in
astrophysics as it is one of the most efficient ways to convert
radiation energy into thermal energy of a gas. In the photo-ionization
process the excess energy of the ionizing photons is transferred to
the liberated electrons which through collisions can then increase the
temperature of the gas. The main sources of ionizing photons in the
Universe are either young, massive stars or accretion disks around
compact objects. This transfer of energy thus often constitutes a
feedback process in which the formation of sources of ionizing photons
change their environment. For example in star formation regions the
photo-ionization of the parent molecular cloud can regulate further
star formation in that region. On cosmological scales, the
reionization of intergalactic medium (IGM) by radiation from stars in
the first generations of galaxies can likewise be considered to be a
radiative feedback process.

The temperature acquired by the photo-ionized gas depends on the one
hand on the energy distribution of ionizing photons from the source
and on the other hand on the effects of radiative cooling. The
ionization and the increase in temperature trigger radiative cooling
processes (often dominated by collisional excitation cooling) so that
within a cooling time, the temperature of the gas will achieve an
equilibrium value set by the balance of the photo-ionization heating rate (henceforth
photo-heating rate) and the cooling rate. In typical interstellar
medium (ISM) conditions, the cooling time is short compared to the
growth of an ionized region and the temperature inside the HII region
is close to this equilibrium value. For this reason photo-ionization
codes historically focused on solving for the equilibrium case, as for
example the well-known Cloudy code \citep{2013RMxAA..49..137F}.

However, for the low density and low, or even zero, metallicity
conditions found in the IGM during cosmic reionization, the cooling
time can greatly exceed the time in which the ionized region doubles
its size and the temperature structure will reflect the initial
photo-heating efficiency, thus carrying important information about
the source properties and gas density \citep{Theuns2002, Hui2003,
  Cen2009}. In this case, the calculation of the time-dependent
photo-heating should be performed carefully to achieve a correct
answer.

The impact of photo-heating onto the IGM temperature during cosmic
reionization can be probed by two methods, either by analysis of IGM
absorption lines in quasar spectra, at the highest redshifts dominated
by Lyman-$\alpha$ absorption, or by observations of the redshifted 21cm
signal from HI. 

Quasar spectra display absorption blueward of the Lyman-$\alpha$
(Ly$\alpha$) line center due to the presence of HI in the IGM. The
line width (or Doppler width) of these absorption features can be used
to measure the temperature of the IGM. Close to the end of
reionization the mean density of neutral hydrogen in the IGM is high
enough to scatter all of the Ly$\alpha$ radiation out of the
line-of-sight (the so-called Gunn-Peterson trough). However, through
their high luminosity quasars are capable of
creating more highly ionized regions of typical sizes $\sim
10$~proper Mpc around themselves, the so-called quasar near zone, in
which it is possible to identify Ly$\alpha$ absorption lines. From the
properties of these lines constraints on the IGM ionization state and
quasar properties can be derived~\citep{Fan, Bolton_z6, Willott,
  Mesinger07, Bolton_z7085, Schroeder13}, as well as constraints on
the IGM temperature in these regions \citep{Bolton_z6,
  Bolton_temperature, Bolton_temperature2}. The values of these
temperatures can then be used to put constraints on the redshift when
the reionization was complete~\citep{Raskutti_temperature}.

The 21cm line is a weak, low energy transition of neutral hydrogen,
which during the epoch of reionization is expected to produce an
observable signal below frequencies of $\sim 200$~MHz
\citep{2006PhR...433..181F}. The detection of signal is the major goal
of a range of new low frequency radio telescopes: 
GMRT\footnote{Giant Metrewave Radio Telescope, http://gmrt.ncra.tifr.res.in} \citep{2013MNRAS.433..639P},
LOFAR\footnote{Low Frequency Array, http://www.lofar.org} \citep{2013A&A...556A...2V},
MWA\footnote{Murchison Widefield Array, http://www.mwatelescope.org} \citep{2013PASA...30....7T},
PAPER\footnote{Donald
C.\ Backer Precision Array to Probe the Epoch of Reionization, http://eor.berkeley.edu/} \citep{2010AJ....139.1468P}
and 21CMA\footnote{ 21 Centimeter Array, http://21cma.bao.ac.cn/}.  This signal
maps out the distribution of neutral hydrogen on the largest scales in
the IGM and its detection will reveal the progress of reionization in
great detail. The intensity of the signal is measured against the
cosmic microwave background signal and described by the differential
brightness temperature which depends on the neutral hydrogen density
and the excitation or spin temperature, $T_\mathrm{s}$, of the neutral
gas \citep{2006PhR...433..181F}. For most of the epoch of reionization
the spin temperature is expected to be close to the kinetic
temperature of the gas. An early generation of x-ray sources can
mildly heat the gas through low level photo-ionization and thus leave
an imprint of their presence \citep[see e.g.][]{Pritchard2007}.

Therefore, understanding cosmic reionization not only requires
following the photo-ionization of the neutral hydrogen but also the
photo-heating of the gas. Modelling this process is best accomplished
with cosmological reionization simulations which follow both the
evolution of the dark matter and baryonic density distribution and the
transfer of ionizing photons emanating from the developing
(proto-)galaxies \citep[for a review
see][]{2011ASL.....4..228T}.

In such simulations, the radiative transfer part, in principle
requiring three spatial coordinates, two angles and a range of
frequencies, dominates the total computational cost. Therefore it is
important to develop and use efficient numerical methods for including
radiative transfer in cosmological simulations. Over the past 15 years
a range of time-dependent, three dimensional photo-ionization methods
for use in cosmological simulations have been developed. Many of these
have been benchmarked against each other in
\citet{2006MNRAS.371.1057I} and \citet{2009MNRAS.400.1283I}. As can be
seen in these comparisons, the focus of these codes has been on
accurately calculating the ionization fractions of hydrogen for which
there is good agreement among the different methods. Much less
attention has been paid to the accuracy of thermal evolution as also
illustrated by the much larger spread in the temperature results between the different codes.
The reason for this is partly that the ionized fraction is the more
important quantity in reionization studies but in addition accurate
photo-heating calculations require multi-frequency radiative transfer
and the inclusion of helium as a photon absorbing species.

Recently, several methods have been published in which the
photo-heating is treated more carefully, e.g.\ \textsc{CRASH}
\citep{CRASH3}, \textsc{TRAPHIC} \citep{TRAPHIC2} and
\textsc{RADAMESH} \citep{RADAMESH}. Although these codes all claim
accuracy, there has to date not been a comparison between their
results. In addition, the level of accuracy is expected to depend
on the choice of the computational time-step as shown in detail
in the study of \citet{2012A&A...539A.147M}. Computational efficiency
prefers longer time-steps but the impact of the choice of time-step
on the temperature results has not been investigated.

\textsc{C}$^2$-\textsc{Ray} \citep{C2-Ray} is a time-dependent
photo-ionization code which has been used extensively for reionization
simulations \citep[see for example][]{2006MNRAS.369.1625I,
  2011MNRAS.413.2093I, 2012MNRAS.424..762D, 2014MNRAS.439..725I}. Its
efficiency is based partly on the use of an implicit method which
allows the use of relatively large time-steps while retaining
accuracy. The updated version of \textsc{C}$^2$-\textsc{Ray}
\citep{C2-Ray2} includes an improved treatment of multi-frequency
radiative transfer as well as the inclusion of helium. However, as
shown in \citet{C2-Ray2}, accurate results for the photo-heating calculation
impose much stricter limits on the choice of the computational
time-step which impacts on the performance of the code.

In this paper we present a new, efficient method to obtain accurate
temperature results using the \textsc{C}$^2$-\textsc{Ray} code
although the general strategy can also be applied to other
time-dependent photo-ionization codes. The method relies on
an efficient method to determine the optimal time-step as well
as the use of different time-steps for different regions of the
computational grid. 

The lay out of the paper is as follows. In Section \ref{sec:C2-Ray},
we review the essential algorithm of the radiative transfer code
\textsc{C}$^2$\textsc{-Ray} and the impact of the choice of time-step
on the photo-heating rate calculation.  In Section \ref{sec:Adaptive
  time-step}, we introduce a new method for determining an optimal
time-step employing a fast ray-tracing method. In Section
\ref{sec:Asynchronized evolution}, we explain how we allocate
computational resources to different cells in the grid.  We proceed in Section
\ref{sec:Parallelization strategy} with the details of the
parallelization strategy.  Section \ref{sec:C2-Ray algorithm}
summarizes the overall algorithm of \textsc{C}$^2$\textsc{-Ray}.  We
examine our new method using several tests in Section \ref{sec:Tests}.
Lastly, we summarize our conclusions and suggest possible further
developments in Section \ref{sec:Conclusions}.


\section{\textsc{C}$^2$\textsc{-Ray}}\label{sec:C2-Ray}


The \textsc{C}$^2$\textsc{-Ray} algorithm consists of two major parts
- (1) radiative transfer of ionizing photons using a short
characteristics ray-tracing method and (2) solution of the coupled
photo-equations (photo-ionization and photo-heating equations) using a
combination of iterative and Runge-Kutta methods. The first version of
\textsc{C}$^2$\textsc{-Ray} \citep{C2-Ray} assumes hydrogen to be the
only photon-absorbing element and only implements soft photon
sources. The second version of \textsc{C}$^2$\textsc{-Ray}
\citep{C2-Ray2} was developed to handle a medium consisting of both
hydrogen and helium and implements both soft and hard photon
sources. In this section, we summarize the basis of the numerical
algorithm of this latter version and its limitations when calculating
photo-heating effects. Although \textsc{C}$^2$\textsc{-Ray} is a
general time-dependent photo-ionization code, we focus here on the
cosmological IGM case.


\subsection{Numerical radiative transfer}

\textsc{C}$^2$\textsc{-Ray} works on a three-dimensional structured
grid. Each cell in the grid stores the number density of gas particles
$n$, ionization fractions $x_{\mbox{\scriptsize{H{\sc I}}}},
x_{\mbox{\scriptsize{H{\sc II}}}}, x_{\mbox{\scriptsize{He{\sc I}}}},
x_{\mbox{\scriptsize{He{\sc II}}}}, x_{\mbox{\scriptsize{He{\sc
        III}}}}$ and the gas temperature $T$. The ionizing sources
are assumed to radiate isotropically.

Theoretically, the cosmological radiative transfer equation (in
comoving coordinates) considers the finite speed of light and
cosmological redshifting due to Hubble expansion
\citep{cosmological_radiative_transfer}.
\begin{equation}\label{eqn:cosmological_radiative_transfer_equation}
\frac{1}{c}\frac{\partial I_{\nu}}{\partial t} + \frac{\hat{n}\cdot\nabla I_{\nu}}{\bar{a}}
- \frac{H(t)}{c}\biggl(\nu\frac{dI_{\nu}}{d\nu}-3I_{\nu}\biggr)
= j_{\nu} - \kappa^{\mbox{\scriptsize{abs}}}_{\nu}I_{\nu}\,,
\end{equation}
where $c$ is the speed of light,
$I_{\nu}$ is the specific light intensity at frequency $\nu$,
$\hat{n}$ is the unit vector along the propagation direction,
$H(t)\equiv \dot{a}/a$ is the Hubble constant with $a$ the scale factor, 
$\bar{a}\equiv
\frac{1+z_{\mbox{\scriptsize{em}}}}{1+z}$ is the ratio of scale
factors between photon emission time 
(at redshift $z_{\mbox{\scriptsize{em}}}$)
and current time (at redshift $z$),
$j_{\nu}(\textbf{x},\hat{n})$ is the emission coefficient and
$\kappa^{\mbox{\scriptsize{abs}}}_{\nu}(\textbf{x},\hat{n})$ is the
absorption coefficient at position $\textbf{x}$. 

In order to deal efficiently with Equation \ref{eqn:cosmological_radiative_transfer_equation} 
it is simplified to the following form
\begin{equation}\label{eqn:simplified_radiative_transfer_equation}
  \hat{n}\cdot\nabla I_{\nu} = - \kappa^{\mbox{\scriptsize{abs}}}_{\nu}I_{\nu}\,.
\end{equation}
The solution of Equation \ref{eqn:simplified_radiative_transfer_equation}
along a propagation direction is
\begin{equation}
  I_{\nu}=I_{0,\nu}e^{-\int\kappa^{\mbox{\scriptsize{abs}}}_{\nu}(s)\;ds}=I_{0,\nu}e^{-\tau_{\nu}}\,,
\end{equation}
where $I_{0,\nu}$ is the initial specific intensity, $\tau_{\nu}$
is the IGM optical depth and $s$ is the distance variable along the
propagation direction.  The
numerical problem is reduced to calculating the optical depths of the
species H{\sc I}, He{\sc I} and He{\sc II} from the sources to the
position where we want to know the intensity.

The use of Equation~\ref{eqn:simplified_radiative_transfer_equation}
implies assuming that the mean free path of ionizing photons is
smaller than $c\Delta t$, the light travel distance within one time
step. This allows us to neglect light travel and cosmological
evolutionary effects.  Additionally Equation~\ref{eqn:simplified_radiative_transfer_equation} 
assumes that there are no photon
sources other than the source at the origin. The latter assumption
implies neglecting the effect of recombination radiation which can be
justified by either assuming these photons to escape or assuming them
to be absorbed locally and incorporating them through the on-the-spot
(OTS) approximation. For a discussion of the treatment of diffuse
radiation, we refer the interested reader to
\cite{recombination_radiation}.

To calculate the optical depths, \textsc{C}$^2$\textsc{-Ray} uses a
short characteristics ray-tracing method \citep{short_characteristics}.
The detailed implementation of this method is described in detail in
\cite{C2-Ray}.


\subsection{Photo-ionization and photo-heating rate calculation}

The absorption coefficient $\kappa^{\mbox{\scriptsize{abs}}}_{\nu}$ due to a species
of number density $n$ can also be written as $n\sigma_\nu$, where
$\sigma_\nu$ is the frequency-dependent photo-ionization cross-section of
the species. If we define $N$ to be the column density along a ray, we
can also write the optical depth as $\tau(\nu) = \sigma_\nu N$.

In a given computational cell the number of photons arriving depends
on the optical depth between the source and the cell. The number of
photons absorbed depends on the optical depth of the cell itself. The
photo-ionization rates $\Gamma$ and photo-heating rate $\mathcal{H}$ thus
depend on the values of these two optical depths. When we consider H{\sc I}, He{\sc I} and
He{\sc II} to be the photon-absorbing species, we need to calculate
the photo-ionization rates $\Gamma^{\mbox{\scriptsize{H{\sc I}}}}$,
$\Gamma^{\mbox{\scriptsize{He{\sc I}}}}$,
$\Gamma^{\mbox{\scriptsize{He{\sc II}}}}$ and the photo-heating rate $\mathcal{H}$.  
These thus depend on the individual and time- and frequency-dependent optical depths
$\tau^{\mbox{\scriptsize{H{\sc I}}}}(\nu)$,
$\tau^{\mbox{\scriptsize{H{e\sc I}}}}(\nu)$,
$\tau^{\mbox{\scriptsize{He{\sc II}}}}(\nu)$.
In the treatment of estimating photo-rates (henceforth
the collective term for the photo-ionization rates and photo-heating rate),
we partition the frequency domain into 47 frequency bins. Details of
this frequency partition are described in \cite{C2-Ray2}.

\subsubsection{Photo-ionization rates}
We next describe how $\Gamma^{\mbox{\scriptsize{H{\sc I}}}}$,
$\Gamma^{\mbox{\scriptsize{He{\sc I}}}}$ and
$\Gamma^{\mbox{\scriptsize{He{\sc II}}}}$ are obtained
for the case of a single source.
Since the photo-ionization rates behave additively, the
total photo-ionization rates contributed by all the sources to a
particular cell is simply the sum of the photo-ionization rates
contributed by each source. If we define the function
$\mathcal{G}^{\mbox{\scriptsize{total}}}_{j}$ as the total ionization event rate
per volume induced by the photons in frequency bin $j$, then
\begin{equation}
\mathcal{G}^{\mbox{\scriptsize{total}}}_{j} 
= \frac{G_j(\tau^{\mbox{\scriptsize{total}}}_{\mbox{\scriptsize{in}}})
-G_j(\tau^{\mbox{\scriptsize{total}}}_{\mbox{\scriptsize{out}}})}{V_{\mbox{\scriptsize{shell}}}}\,,
\end{equation}
where $V_{\mbox{\scriptsize{shell}}}$ is the volume of the spherical shell whose inner radius and outer radius 
correspond to the ray-traversed position of the cell.
The function $G_j(\tau^{\mbox{\scriptsize{total}}})$ is given by
\begin{equation}
G_j(\tau^{\mbox{\scriptsize{total}}})
=\int^{\nu_{j+1}}_{\nu_j}\frac{L_{\nu}}{h\nu}e^{-\tau^{\mbox{\scriptsize{total}}}(\nu)}\;d\nu\,,
\end{equation}
where $\tau^{\mbox{\scriptsize{total}}}(\nu)
=\tau^{\mbox{\scriptsize{H{\sc I}}}}(\nu)
+\tau^{\mbox{\scriptsize{He{\sc I}}}}(\nu)
+\tau^{\mbox{\scriptsize{He{\sc II}}}}(\nu)$ is the combined optical depth of all absorbing species at frequency $\nu$,
$L_{\nu}$ is the source luminosity and $h$ is the Planck constant.
We approximate this expression by
\begin{equation}
G_j(\tau^{\mbox{\scriptsize{total}}})
=\int^{\nu_{j+1}}_{\nu_j}\frac{L_{\nu}}{h\nu}e^{-\tau^{\mbox{\scriptsize{total}}}_j(\nu/\nu_j)^{-\eta_j}}\;d\nu\,,
\end{equation}
by imposing that
$\tau^{\mbox{\scriptsize{H{\sc I}}}}(\nu)$, $\tau^{\mbox{\scriptsize{H{\sc I}}}}(\nu)$ and $\tau^{\mbox{\scriptsize{H{\sc I}}}}(\nu)$ 
share the same power law index $\eta_j$ in frequency bin $j$~\citep{C2-Ray2}. This way the value of $G_j$
is controlled by just one parameter
$\tau^{\mbox{\scriptsize{total}}}_j \equiv \tau^{\mbox{\scriptsize{total}}}(\nu_j)$. 
This has the advantage that we can pre-calculate
$G_j$ for a large range of $\tau^{\mbox{\scriptsize{total}}}_j$'s
allowing the rates to be derived through interpolation in this table.
This way the repeated evaluation of the full photo-ionization
integrals can be avoided.

In each frequency bin $j$, the photo-ionization rate $\Gamma^{i}_j$ of
species $i$ can be calculated from $\mathcal{G}^{\mbox{\scriptsize{total}}}_j$ through
\begin{equation}\label{eqn:Gammaij}
\Gamma^{i}_j
=\frac{\tau^i_j}{\tau^{\mbox{\scriptsize{total}}}_j}\frac{\mathcal{G}^{\mbox{\scriptsize{total}}}_j}{n_i}\,.
\end{equation}
The number density $n_i$ is present here to change the dimension of photo-ionization rates from per volume to per particle. The ratio $\tau^i_j/
\tau^{\mbox{\scriptsize{total}}}_j$ correctly distributes the photo-ionization rates over the different species \citep[see appendix D in][]{C2-Ray2}.

Finally, the photo-ionization rate $\Gamma^i$ of species $i$ is the sum
over the relevant frequency bins,
\begin{equation}
\Gamma^{i}
=\sum_j\Gamma^{i}_j\,.
\end{equation}

\subsubsection{Photo-heating rate}
The photo-heating rate $\mathcal{H}$ for a single
source case is calculated in a similar
fashion. Also the photo-heating rate behaves additively
when one considers multiple sources. The function
$\mathcal{H}^{i}_{j}$ is defined as the rate of energy
injected from photons $j$ to species 
$i$, per unit volume
\begin{equation}
\mathcal{H}^{i}_{j} = 
\frac{H^{i}_{j}(\tau^{\mbox{\scriptsize{total}}}_{\mbox{\scriptsize{in}}})
-H^{i}_{j}(\tau^{\mbox{\scriptsize{total}}}_{\mbox{\scriptsize{out}}})}{V_{\mbox{\scriptsize{shell}}}}.
\end{equation}
The function $H^{i}_{j}(\tau^{\mbox{\scriptsize{total}}})$ is given by
\begin{align}
 H^{i}_{j}(\tau^{\mbox{\scriptsize{total}}}) &=
  \begin{dcases}
   \int^{\nu_{j+1}}_{\nu_j}h(\nu-\nu_i)\frac{L_{\nu}}{h\nu}e^{-\tau^{\mbox{\scriptsize{total}}}(\nu)}\;d\nu\, & \text{if } \nu_j \geq \nu_i \\
   0 & \text{if } \nu_j < \nu_i\,,
  \end{dcases}
\end{align}
where $\nu_i$ is the ionization threshold frequency of species $i$.
Note that because of the appearance of the $h(\nu-\nu_i)$ term, we
cannot work with one function as in the photo-ionization rate case.
Similar to the photo-ionization rates, we pre-calculate
$\mathcal{H}^{i}_{j}$ tables for each species $i$ for a large range of
$\tau^{\mbox{\scriptsize{total}}}_j$'s.  The photo-heating rate
$\mathcal{H}_{j}$ in $j$th bin is the
weighted sum of $\mathcal{H}^{i}_{j}$'s,
\begin{equation}
\mathcal{H}_{j}
=\sum_{i=1}^3\frac{\tau^i_j}{\tau^{\mbox{\scriptsize{total}}}_j}\mathcal{H}^{i}_{j}\,.
\end{equation}
Here we use the same weighting factors as in Equation~\ref{eqn:Gammaij} to find the contribution of each of the three photon-absorbing species.
The total photo-heating rate $\mathcal{H}$ is then given by
\begin{equation}
\mathcal{H}
=\sum_j\mathcal{H}_{j}.
\end{equation}

As in \citet{C2-Ray2} we employ the widely used approximation
from \citet{secondary_ionization} to correct the heating rates for the effect
of secondary ionizations and excitations associated with high energy
electrons.

\subsection{Effect of the time-step}
\label{sec:effect_of_time_step}

As described above, the photo-ionization rates
$\Gamma^{\mbox{\scriptsize{i}}}$'s and photo-heating rate
$\mathcal{H}$ through the optical depths depend on $x_{\mbox{\scriptsize{HI}}}$,
$x_{\mbox{\scriptsize{HeI}}}$ and $x_{\mbox{\scriptsize{HeII}}}$ which
are both position- and time-dependent. As a result the photo-rates
also are time-dependent.  This means that the accuracy of the results
for the ionization and thermal evolution depend on the choice for the
time-step. In principle one would have to choose the shortest timescale 
in the photo-ionization process to determine the time-step. 
However, as first shown in \cite{C2-Ray} for the hydrogen-only
case and in \cite{C2-Ray2} for the case of H+He, it is possible to use
a much longer time-step if one approximates the time-averaged
photo-rates $\bigl\langle\Gamma^{\mbox{\scriptsize{HI}}}
(x_{\mbox{\scriptsize{HI}}},x_{\mbox{\scriptsize{HeI}}},
x_{\mbox{\scriptsize{HeII}}})\bigr\rangle$,
$\bigl\langle\Gamma^{\mbox{\scriptsize{HeI}}}
(x_{\mbox{\scriptsize{HI}}},x_{\mbox{\scriptsize{HeI}}},
x_{\mbox{\scriptsize{HeII}}})\bigr\rangle$,
$\bigl\langle\Gamma^{\mbox{\scriptsize{HeII}}}
(x_{\mbox{\scriptsize{HI}}},x_{\mbox{\scriptsize{HeI}}},
x_{\mbox{\scriptsize{HeII}}})\bigr\rangle$ and
$\bigl\langle\mathcal{H}
(x_{\mbox{\scriptsize{HI}}},x_{\mbox{\scriptsize{HeI}}},
x_{\mbox{\scriptsize{HeII}}})\bigr\rangle$ by
$\Gamma^{\mbox{\scriptsize{HI}}}\bigl(\langle
x_{\mbox{\scriptsize{HI}}}\rangle, \langle
x_{\mbox{\scriptsize{HeI}}}\rangle,\langle
x_{\mbox{\scriptsize{HeII}}}\rangle\bigr)$,
$\Gamma^{\mbox{\scriptsize{HeI}}}\bigl(\langle
x_{\mbox{\scriptsize{HI}}}\rangle, \langle
x_{\mbox{\scriptsize{HeI}}}\rangle,\langle
x_{\mbox{\scriptsize{HeII}}}\rangle\bigr)$,
$\Gamma^{\mbox{\scriptsize{HeII}}}\bigl(\langle
x_{\mbox{\scriptsize{HI}}}\rangle, \langle
x_{\mbox{\scriptsize{HeI}}}\rangle,\langle
x_{\mbox{\scriptsize{HeII}}}\rangle\bigr)$ and
$\mathcal{H}\bigl(\langle x_{\mbox{\scriptsize{HI}}}\rangle, \langle
x_{\mbox{\scriptsize{HeI}}}\rangle,\langle
x_{\mbox{\scriptsize{HeII}}}\rangle\bigr)$.
In this $\langle
x_{\mbox{\scriptsize{HI}}}\rangle$, $\langle
x_{\mbox{\scriptsize{HeI}}}\rangle$ and $\langle
x_{\mbox{\scriptsize{HeII}}}\rangle$ are time-averaged values, evaluated
from an ansatz for how $x_{\mbox{\scriptsize{HI}}}$,
$x_{\mbox{\scriptsize{HeI}}}$ and $x_{\mbox{\scriptsize{HeII}}}$
evolve over time.

The tests in \cite{C2-Ray} and \cite{C2-Ray2} show that this approach
works as long as the time-step is shorter than the recombination time,
$\mathbin{t_{\mbox{\scriptsize{rec}}}\approx 1/(n_e\alpha_{\mbox{\scriptsize{H}}})}$, 
where $\mathbin{n_e}$ is the electron number density
and $\mathbin{\alpha_{\mbox{\scriptsize{H}}}}$ is the hydrogen recombination coefficient.
In our forthcoming work (Lee, in preparation)
we show analytically that if the time-step is less than
the recombination time and the medium contains
a sufficient amount of photon-absorbers, the following
relations hold
\begin{equation}
\bigl\langle\Gamma^{\mbox{\scriptsize{HI}}}(x_{\mbox{\scriptsize{HI}}},
x_{\mbox{\scriptsize{HeI}}},x_{\mbox{\scriptsize{HeII}}})\bigr\rangle
\approx \Gamma^{\mbox{\scriptsize{HI}}}\bigl(\langle x_{\mbox{\scriptsize{HI}}}\rangle,\langle x_{\mbox{\scriptsize{HeI}}}\rangle,\langle x_{\mbox{\scriptsize{HeII}}}\rangle\bigr),
\end{equation}
\begin{equation}
\bigl\langle\Gamma^{\mbox{\scriptsize{HeI}}}(x_{\mbox{\scriptsize{HI}}},
x_{\mbox{\scriptsize{HeI}}},x_{\mbox{\scriptsize{HeII}}})\bigr\rangle
\approx \Gamma^{\mbox{\scriptsize{HeI}}}\bigl(\langle x_{\mbox{\scriptsize{HI}}}\rangle,\langle x_{\mbox{\scriptsize{HeI}}}\rangle,\langle x_{\mbox{\scriptsize{HeII}}}\rangle\bigr),
\end{equation}
\begin{equation}
\bigl\langle\Gamma^{\mbox{\scriptsize{HeII}}}(x_{\mbox{\scriptsize{HI}}},
x_{\mbox{\scriptsize{HeI}}},x_{\mbox{\scriptsize{HeII}}})\bigr\rangle
\approx \Gamma^{\mbox{\scriptsize{HeII}}}\bigl(\langle x_{\mbox{\scriptsize{HI}}}\rangle,\langle x_{\mbox{\scriptsize{HeI}}}\rangle,\langle x_{\mbox{\scriptsize{HeII}}}\rangle\bigr)\,.
\end{equation}
However, the same analysis also shows that under the same conditions
\begin{equation}
\bigl\langle\mathcal{H}(x_{\mbox{\scriptsize{HI}}},
x_{\mbox{\scriptsize{HeI}}},x_{\mbox{\scriptsize{HeII}}})\bigr\rangle
\gg \mathcal{H}\bigl(\langle x_{\mbox{\scriptsize{HI}}}\rangle,\langle x_{\mbox{\scriptsize{HeI}}}\rangle,\langle x_{\mbox{\scriptsize{HeII}}}\rangle\bigr)\,.
\end{equation}
As a result, the use of time-averaged ionization fractions allows us
to calculate the ionization evolution correctly but can lead to
serious errors in the thermal evolution.

This effect can be understood intuitively in the following way. The
\textsc{C}$^2$\textsc{-Ray} formalism uses the time-averaged optical
depths of H{\sc I}, He{\sc I} and He{\sc II}. While the
photo-ionization rates depend only weakly on the optical depth values,
the photo-heating rate depends quite sensitively on them. Consider the
two extremes -- the optically thick limit and the optically thin limit.
In the optically thick limit, all photons are absorbed with equal
probability. Therefore the average energy added per photo-ionization
is large since it is dominated by the higher energy photons. In the
optically thin limit, due to the rather steep frequency dependence of
the ionization cross-section, low energy photons
have a higher probability of being absorbed and thus the average
energy added per photo-ionization is lower. 

A computational cell which becomes photo-ionized initially has a high
optical depth towards the source and therefore is initially heated
with a large value of the energy per photo-ionization. As it evolves
to a lower optical depth state, the average energy transferred per
photo-ionization drops. As a consequence most of the heating occurs in
the initial, optically thick phase, see for example Figure~5 in 
\citet{2006MNRAS.371.1057I}

If one chooses a long time-step compared to the ionization time of the
cell, e.g.\ 100 times the duration of the optically thick phase, the
time-averaged optical depth over the time-step will be
close to value of the optically thin phase. As a
result the \textsc{C}$^2$\textsc{-Ray} formalism will apply the
optically thin heating rate for the duration of the whole time-step. 
As this rate has a lower energy per photo-ionization, the total
heating during the time-step will be underestimated.

\begin{figure*}
  \centering
  \includegraphics[width=0.4\textwidth]{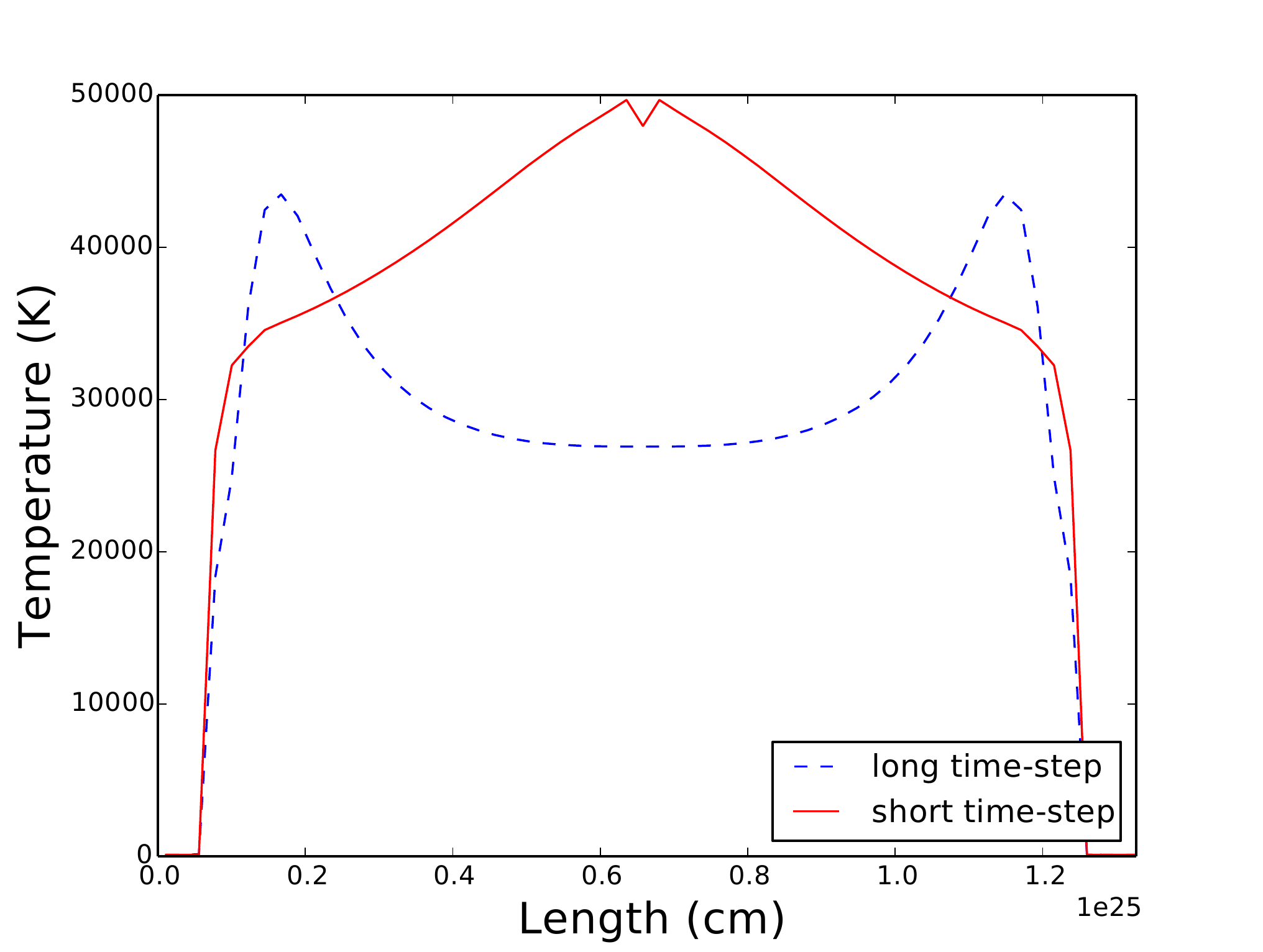}
  \includegraphics[width=0.4\textwidth]{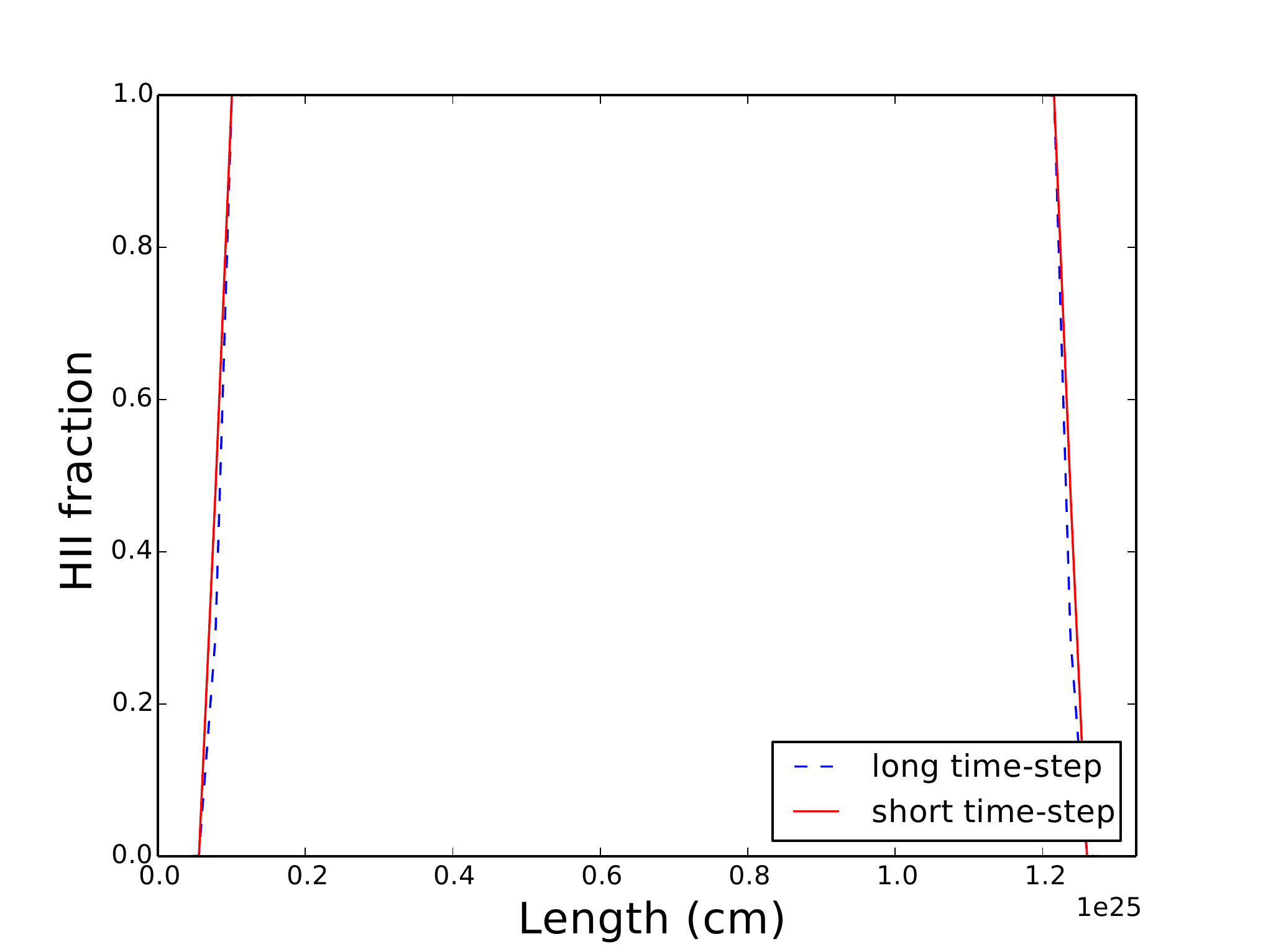}
  \includegraphics[width=0.4\textwidth]{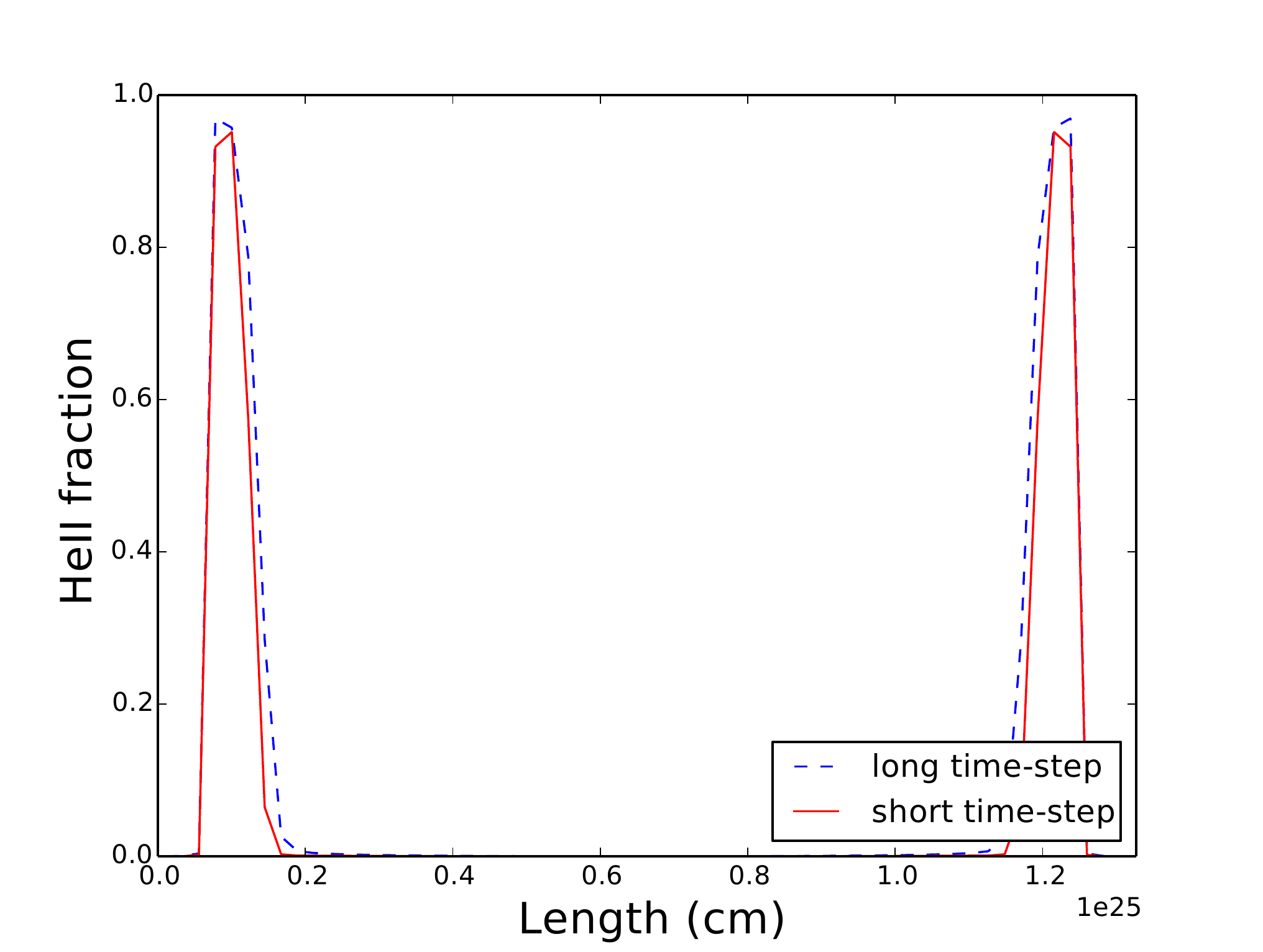}
  \includegraphics[width=0.4\textwidth]{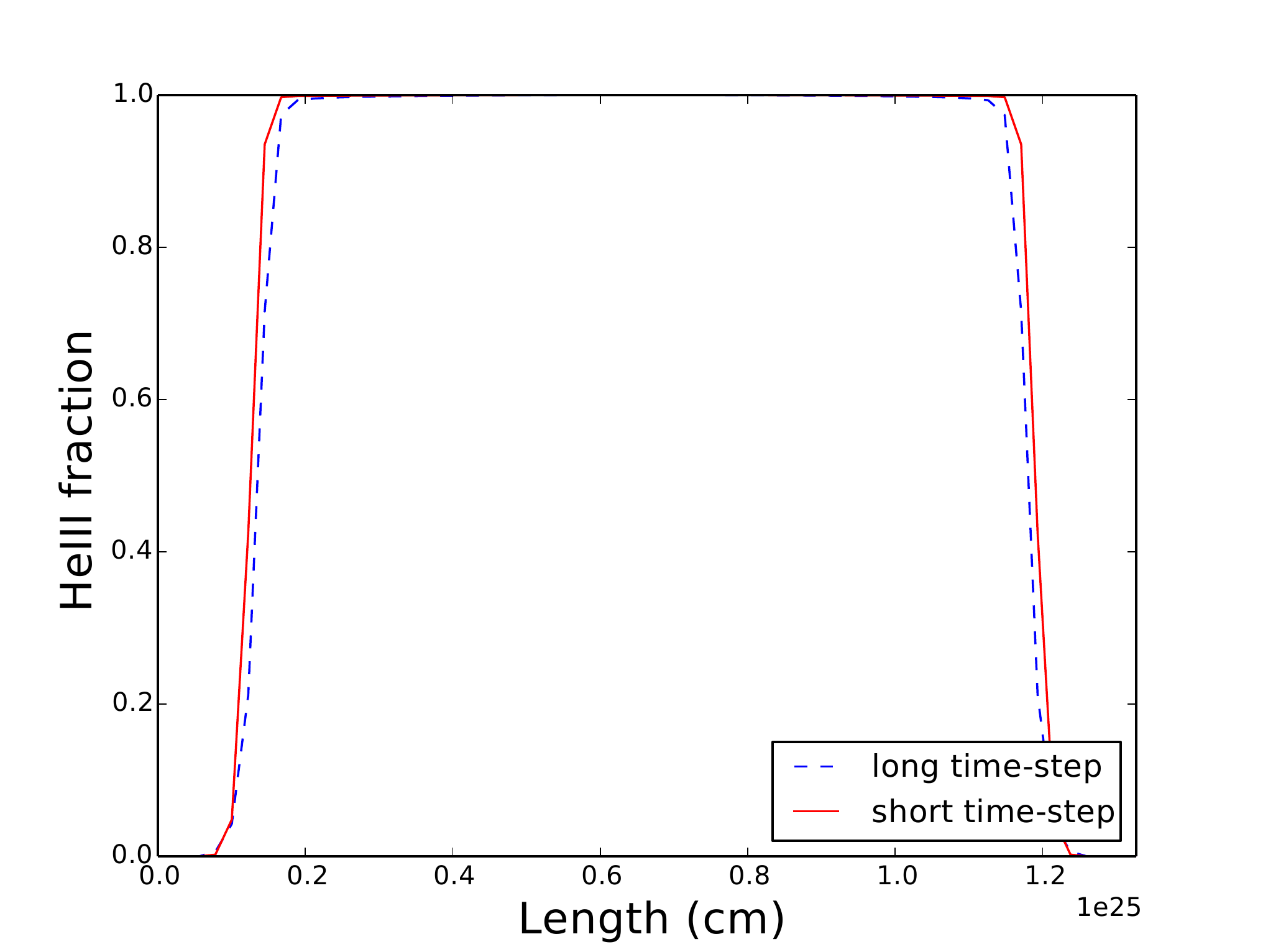}
    \caption{Temperature and ionization fraction results at $10^7$
      years for two simulations using different (constant)
      time-steps. Blue dashed lines: long time-step ($10^7$ years);
      red solid lines: short time-step ($10^{-6}$ years). Top left
      panel: Temperature profile. Top right panel: H{\sc II} fraction
      profile. Bottom left panel: He{\sc II} fraction profile. Bottom
      right panel: He{\sc III} fraction profile. The cubic volume is 4.2 Mpc
      across and is divided into $60^3$ cells. A black body source with an ionizing
      photon rate $5\times 10^{56}$~s$^{-1}$ and temperature 100,000~K
      is located at the centre of the volume. The initial temperature
      is 100~K and both hydrogen and helium are fully neutral. The plots show the values along a line containing the source location. The
      simulation using a long time-step severely underestimates the
      temperature in the centre of the H{\sc II} region and
      overestimates the temperature at the location of the ionization
      front. The small dip near the source in the temperature result
      for the short time-step case is caused by the same effect.}
  \label{fig:section2} 
\end{figure*}

This can be further illustrated through a numerical
experiment. Figure~\ref{fig:section2} shows the thermal and ionization
state at $\mathbin{10^7}$~years for two 3D hydrostatic
photo-ionization simulations with the same initial conditions.
The volume is 4.2~Mpc in size with $60^3$ cells. This size corresponds to 
30 $h^{-1}$ comoving Mpc at $z=9$ for a Hubble parameter of $h = H_0/100$~km~s$^{-1}$~Mpc$^{-1}= 0.7$.
A black body source
with an ionizing photon rate $5\times 10^{56}$~s$^{-1}$ and effective
temperature $T_\mathrm{eff}=10^5$~K is located at the centre of the volume. The
density field is homogeneous and constant in time at a value of $1.94\times~10^{-4}$~cm$^{-3}$.
Initially the temperature is 100~K and both hydrogen and helium are fully
neutral.

The blue dashed curve in Fig.~\ref{fig:section2} shows the result for a single long
time-step ($10^7$ years, a non-trivial fraction of recombination
timescale) and the red solid curve shows the result when using
$10^{13}$ short time-steps ($10^{-6}$ years, a non-trivial fraction of
heating timescale). The effect described in the previous paragraph is
clearly present. When using a long time-step, the inner parts do not
get heated sufficiently. The comparison also shows that the heating
near the ionization front is overestimated for the long time-step.
Very close to the source the hardest photons are not fully
absorbed by the gas and therefore the short time-step result shows a
little valley in the temperature profile near the source position.
Note that the two simulations do agree on the ionization fractions
(H{\sc II} in the top right panel, He{\sc II} in the bottom left panel
and He{\sc III} in the bottom right panel). The results of this test
are thus consistent with what we described above: the choice of the
time-step in \textsc{C}$^2$\textsc{-Ray} affects the thermal evolution
but hardly impacts the ionization evolution.

We conclude therefore that a reliable temperature result requires a
more careful choice of the numerical time-step. Such an
optimal time-step should ensure that 
\begin{align}
&\bigl\langle\mathcal{H}(x_{\mbox{\scriptsize{HI}}},x_{\mbox{\scriptsize{HeI}}},x_{\mbox{\scriptsize{HeII}}})\bigr\rangle
\approx \mathcal{H}\bigl(\langle x_{\mbox{\scriptsize{HI}}}\rangle,
\langle x_{\mbox{\scriptsize{HeI}}}\rangle,\langle x_{\mbox{\scriptsize{HeII}}}\rangle\bigr)\,,
\end{align}
without impacting the performance of the code unnecessarily.
The rest of this paper outlines our method to achieve this.

\section{Adaptive time-step algorithm}
\label{sec:Adaptive time-step}


\subsection{Time-step criteria}\label{sec:Time-step criteria}

The optimal size of the time-step will depend on the state of
simulation at the start of the simulation step. It can thus not be set at
the start of the simulation but needs to be calculated on the
fly. This actually requires performing part of the calculation to
obtain the rates of change of the quantities at hand (in our case
ionization fraction and temperature) and use these to establish the
optimal time-step for each cell. The smallest among these values
should then be selected to be the time-step of the next simulation
step.

Since the cause of the inaccurate temperature results lies in the fact
that cells change their ionization state too much over the time-step,
the obvious solution is to force ionization fronts not to proceed more
than one cell within a time-step. This way all cells which become
ionized in this time-step will experience the optically thick
heating phase. 

To implement this we introduce a free parameter $f$ so
that the change in $x_{\mbox{\scriptsize{HII}}}$, $\Delta
x_{\mbox{\scriptsize{HII}}}$, is at most $1/f$ in each time-step. Note
that this is an absolute condition, not a relative one. A relative
condition would limit the quantity $\Delta
x_{\mbox{\scriptsize{HII}}}/x_{\mbox{\scriptsize{HII}}}$.
\citet{2012A&A...539A.147M} investigated the merits of absolute and
relative constraints on the change in ionized fractions when setting
the time-step and concluded that absolute constraints are more
efficient. 

We find the required time-step $\Delta t$ by solving the ionization
equation (see equation (12) in \cite{C2-Ray})
\begin{equation}
  \Delta t = \frac{1}{\Gamma^{\mbox{\scriptsize{HI}}}}\ln
  \frac{x_{\mbox{\scriptsize{HII}}}^{\mbox{\scriptsize{begin}}}-1}
  {x_{\mbox{\scriptsize{HII}}}^{\mbox{\scriptsize{final}}}-1}\,,
\end{equation}
imposing the result
\begin{equation}
  x^{\mbox{\scriptsize{final}}}_{\mbox{\scriptsize{H{\sc II}}}} =
  \min\{x^{\mbox{\scriptsize{begin}}}_{\mbox{\scriptsize{H{\sc II}}}}
  + 1/f, 1-\epsilon\}\,,
\end{equation}
where $\mathbin{x^{\mbox{\scriptsize{begin}}}_{\mbox{\scriptsize{H{\sc II}}}}}$ and
$\mathbin{x^{\mbox{\scriptsize{final}}}_{\mbox{\scriptsize{H{\sc II}}}}}$ are the ionized hydrogen 
fractions at the beginning and the end of the simulation step.
In this $\epsilon$ is an arbitrarily small number introduced to stop the
time-step from diverging when
$x^{\mbox{\scriptsize{final}}}_{\mbox{\scriptsize{H{\sc II}}}}
\rightarrow 1$.  
We then adopt the minimum of all the $\Delta t$'s as the
time-step for the current simulation step.  

The complication we face is that the $\Gamma$ and $\mathcal{H}$ values
depend on the choice for the time-step. In order to determine the
optimal time-step we thus need to first choose a time-step, determine
the $\Gamma$ and $\mathcal{H}$ values and then find the optimal
time-step for these values. In principle one would need to iterate
this process. However, as we saw above, although the value for
$\mathcal{H}$ depends sensitively on the time-step, the values for the
$\Gamma$ do not. We therefore do not use the $\mathcal{H}$ values for
estimating the optimal time-step, but only the $\Gamma$ values. The
helium photo-ionization rates $\Gamma^{\mbox{\scriptsize{HeI}}}$ and
$\Gamma^{\mbox{\scriptsize{HeII}}}$ are not considered since we found
that time-steps estimated from $\Gamma^{\mbox{\scriptsize{HI}}}$ alone
suffice to get an accurate temperature evolution. We
experimented with different time-steps and found that a reasonable
optically thick $\langle\mathcal{H}\rangle$ is estimated if 
$x_{\mbox{\scriptsize{H{\sc I}}}}$ does not change substantially during
the time-step. The appropriate value of $f$ depends somewhat 
on the problem and the required accuracy. However, for typical IGM gas densities values we recommend values in the range $2 \leq f \leq 5$.
Reasonable values for the optically
thin $\langle\mathcal{H}\rangle$ can be obtained regardless of the
time-step.

Since only ionization front cells require a more strict time-step in
order to obtain accurate heating rates, we do not need to evaluate the
photo-ionization rate $\Gamma^{\mbox{\scriptsize{HI}}}$ for all
cells. It suffices to calculate them for the cells which are currently
being photo-ionized or in other words the ionization front cells
(ICs).  Furthermore, since a simulation volume can contain several
sources of ionizing photons, when evaluating the effect of a single
source we should not consider all ICs but only the ones which are
nearest to this source. These we will designate as nearest ionization
front cells (NICs).

Finding the values of $\Gamma^{\mbox{\scriptsize{HI}}}$ at the
locations of the NICs in principle requires ray-tracing from the source
to those locations in order to find the optical depths. However, we do
not need such accuracy for the optical depths if we only want to derive
a time-step. Instead, we find the locations of the NICs and assume that
the optical depths at the NIC are dominated by the NICs themselves, i.e.\ if
$\tau_\mathrm{NIC}$ is the optical depth from the source to the NIC
and $\Delta \tau_\mathrm{NIC}$ is the optical depth of the NIC itself,
we assume that $\Delta \tau_\mathrm{NIC} \gg \tau_\mathrm{NIC}$. In
this case, the only thing required is to find the location of the
NICs. Note that our time-step estimate also works
well in high resolution simulations when $\Delta
\tau_\mathrm{NIC} \sim \tau_\mathrm{NIC}\ll 1$. This
is because the $\Gamma^{\mbox{\scriptsize{HI}}}$
estimate always works well in the optically thin
regime $(\tau_\mathrm{NIC}\ll 1)$ and for optically
thin cells $(\Delta \tau_\mathrm{NIC}\ll 1)$. The next sections
explain how we find the locations of the NICs for the case of a single
source.


\subsection{Pyramidal ray tracing framework}\label{sec:Fast ray-tracing method}

Identifying the NICs for a given source is not trivial. The
obvious choice for this task is a ray-tracing method. However, the
short characteristics ray tracing method used by
\textsc{C}$^2$\textsc{-Ray} is not useful for this purpose as it
efficiently integrates quantities in all directions but does not
record `events' along a particular direction. For such a task, a
long characteristics method is the more natural choice. 

Consider a volume of $\mathbin{P^3}$ cells. In its
simplest form long characteristics ray tracing would require tracing
$\mathbin{6P^2}$ rays to cover all cells in the
volume. Along each ray one would need to identify the ICs, calculate
their distances to the source and compare these in order to identify
the NIC for that ray. Since the density of rays close to the source
will be high, the same NICs could be traversed by several rays,
leading to redundant work. Such a brute force approach would be
highly inefficient. 

Here we introduce a new long characteristics ray-tracing method for
identifying the NICs for a source. Our method relies on two key ideas.
First, we trace the lowest possible number of rays from a source to
identify all NICs. Second, we avoid calculation and comparison of
distances by storing the ICs in a sorted list such that for a given
ray the IC with the smallest index is the NIC. The combination of
these two ideas leads to a very efficient method. Below we describe
the algorithm in detail for the case of one source. The repeated use
of the method works well for the case of multiple sources.

First, we identify all the ICs in the simulation box. 
We define neutral as $x_{\mbox{\scriptsize{HI}}}\geq 1/(1+f)$
and ionized otherwise. Note that $f$ is the same parameter as defined in 
Section~\ref{sec:Time-step criteria}. 
We then define the set of all the ionized cells to be
$\mathbb{I}$.  We define another set of cells $\mathbb{IB}$ which
contains the union of all the $3\times 3\times 3$ cubes of cells in
which the centres of the cubes are ionized cells. The $\mathbb{IB}$
cells which do not belong to $\mathbb{I}$ are the ICs. That is to
say, the set $\mathbb{B}\equiv\mathbb{IB}\backslash \mathbb{I}$ is the
set of ICs.  

Next we need to establish a coordinate centre for the ray tracing.  We
define the source cell to be the origin of this coordinate system. If
periodic boundary conditions are assumed, we translate the coordinate
system so that the source is located in the
centre of the simulation volume. Since we do not trace radiation
which leaves the volume, these are not periodic boundary conditions
in a formal sense. However, since this procedure makes the horizon for 
the radiation different for each source, it does create a pseudo-periodic
character for the simulation. For transmissive boundary conditions
we do not change the relative position of sources within the
simulation volume.

In this coordinate system, the coordinates $(i,j,k)$ of the simulation
box for both cases thus follow the constraints $-M_x^-\leq i\leq
M_x^+$, $-M_y^-\leq j\leq M_y^+$ and $-M_z^-\leq k\leq M_z^+$, where
for example $M_x^-$ is the distance in terms of cell width from the 
source to edge of the simulation volume along the negative $x$-axis. 
Note that our current ray-tracing scheme works only on a fixed, regularly
spaced grid. Further improvements are required to use it with an
adaptive mesh refinement grid.

\begin{figure}
\centering
\includegraphics[width=0.5\textwidth]{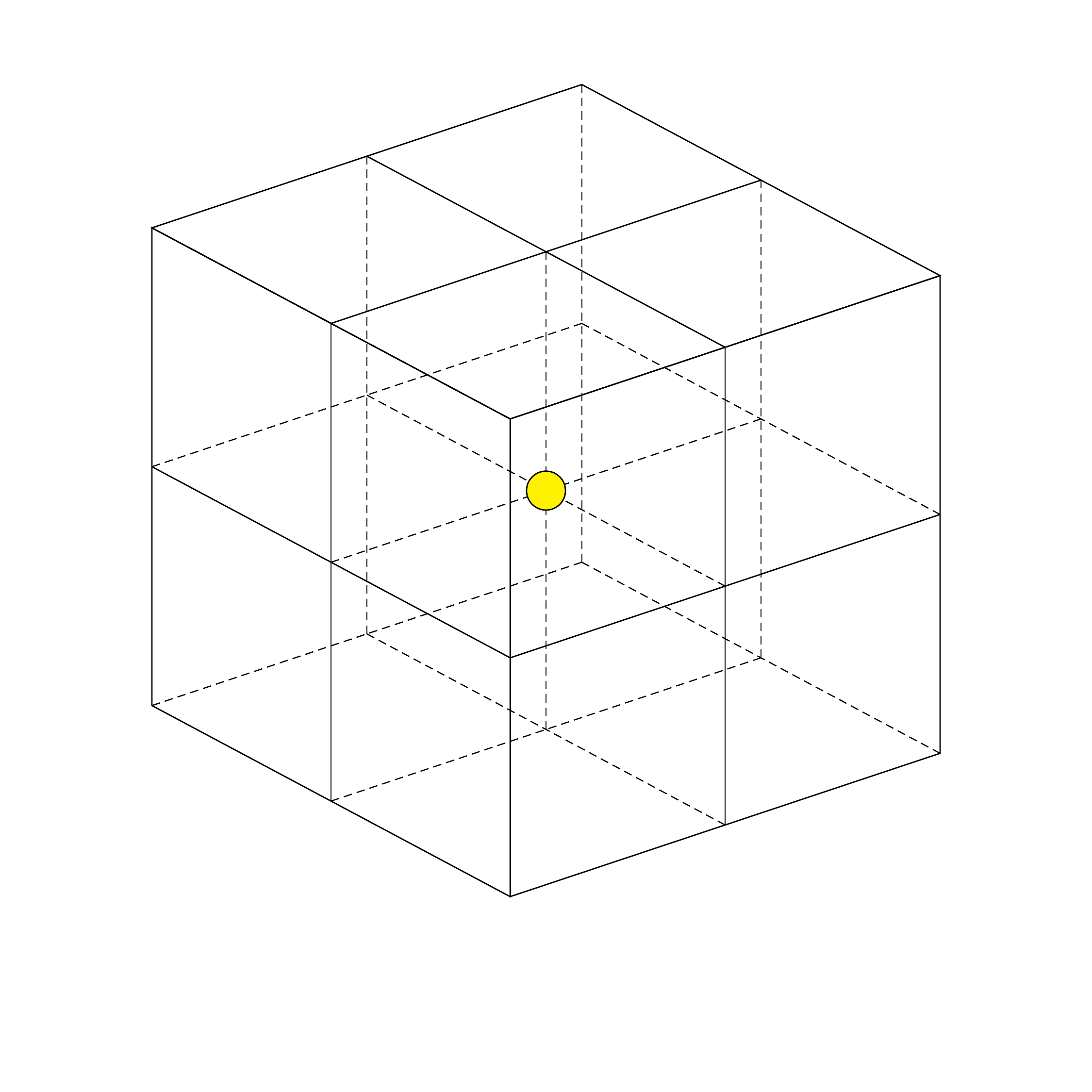}
\caption{The simulation volume is decomposed into 8 octants. The yellow
  spot indicates the position of the source. Note that for simplicity
  this graphical illustration only shows the case of the source being
  located in the centre of the volume. The domain decomposition for the case
  when the source does not lie in the centre of the volume follows the
  same principle.}
\label{fig:decomposition_8_intact}
\end{figure}

\begin{figure}
\centering
\includegraphics[width=0.5\textwidth]{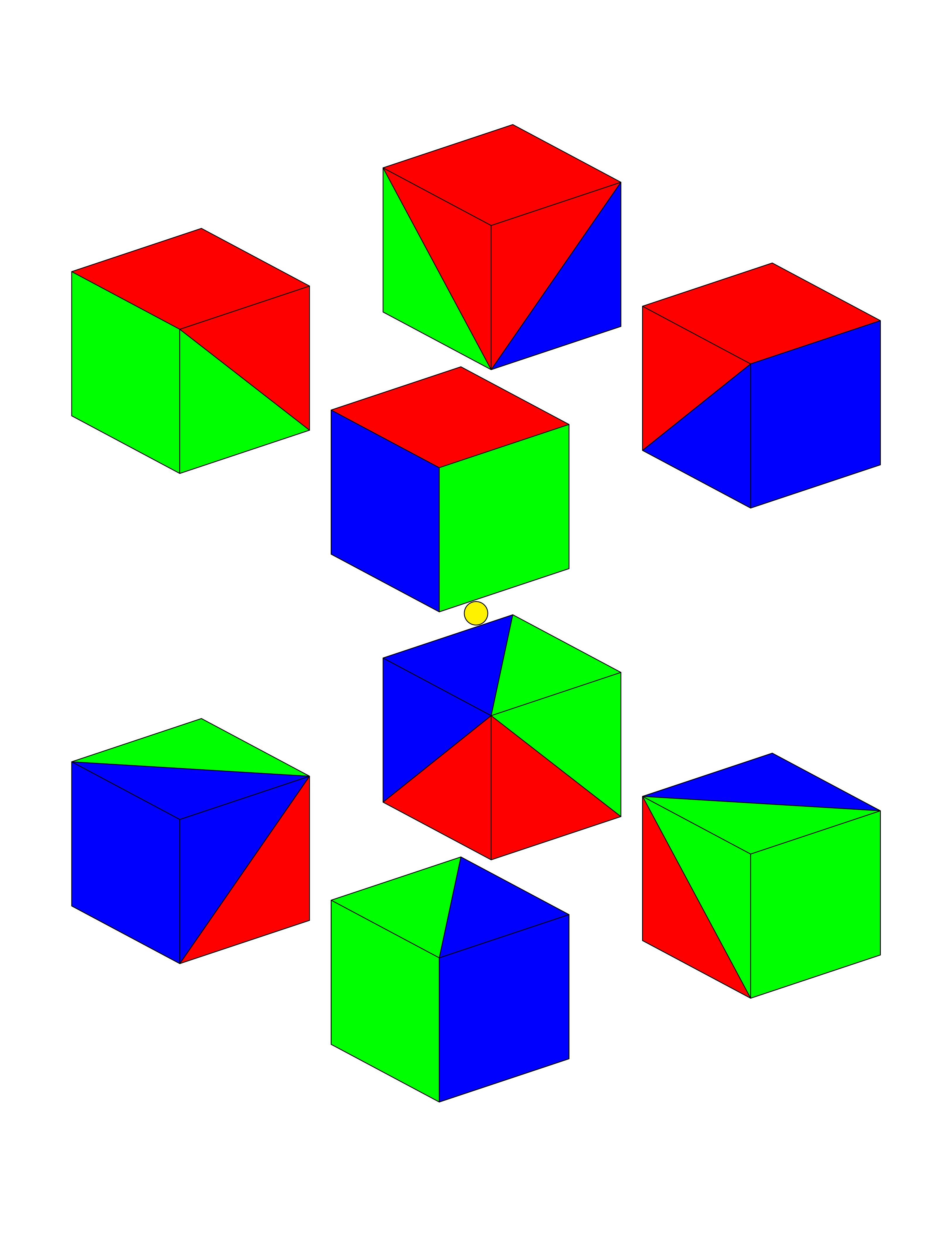}
\caption{Each octant is further decomposed into 3 pyramidal
  components. Here we use red, green and blue for the different
  pyramidal components to easily distinguish them. Note that the
  source is located at the vertex of each of the 24 pyramidal
  components.}
\label{fig:decomposition_24_separated}
\end{figure}

\begin{figure}
\centering
\includegraphics[width=0.5\textwidth]{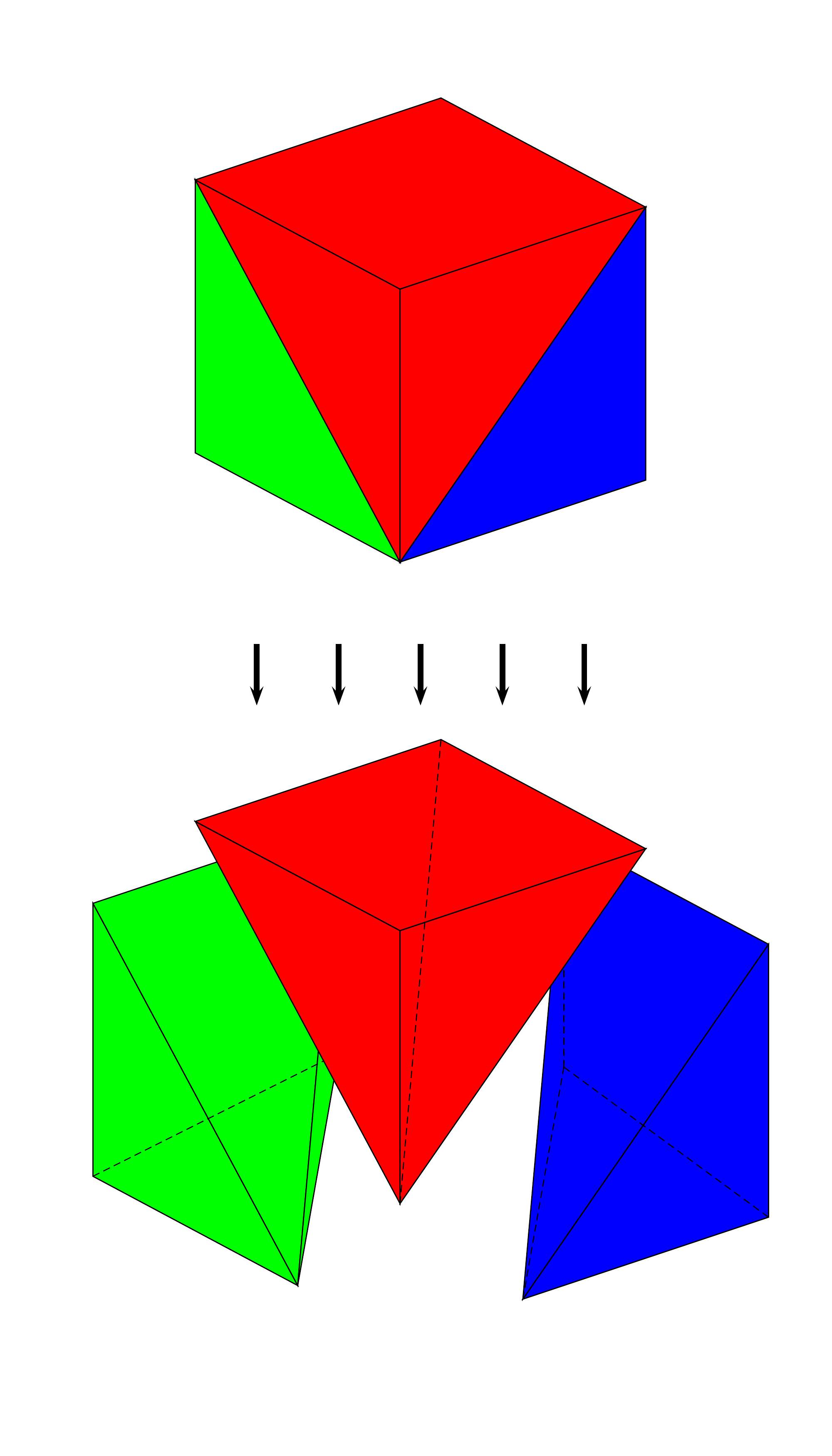}
\caption{Detailed decomposition into pyramidal components.}
\label{fig:decomposition_3_separated}
\end{figure}

Using this coordinate system we then construct 24 pyramidal components
by first dividing the simulation volume into 8 octants (see Figure
\ref{fig:decomposition_8_intact} for the case of a source in the centre of the simulation volume) and next dividing each octant into 3
pyramidal components (see Figure \ref{fig:decomposition_24_separated} and
\ref{fig:decomposition_3_separated}, also for a centrally located source).  A pyramidal component consists
of a rectangular base and a vertex where the source lies.  All the
pyramidal components have the source at their vertices.

There are two reasons for setting up such a pyramidal
decomposition of the grid. First, a pyramidal component is the
smallest repeated sub-domain unit in our ray-tracing scheme. The
problem is thus reduced to identifying NICs in one pyramidal
component. Simply stated, if the NICs all lie at the tip of the
pyramid, there is no need to trace the rest of the pyramid. Second,
the pyramidal ray-tracing process is highly parallelizable as the
ray-tracing in one pyramidal component is independent of the ray-tracing 
in the other ones.

To distinguish the different components we introduce a nomenclature
for the pyramidal components. This takes the form $[A,B,C]$, where $A$
indicates the direction of the primary axis, $B$ indicates the
direction of the secondary axis and $C$ indicates the direction of the
tertiary axis. For example for the octant of positive $x$, $y$
and $z$ coordinates, there are three pyramidal components, one with
the primary axis along the $x$ axis, $[+x,+y,+z]$, one with the
primary axis along the $y$ axis, $[+y,+x,+z]$ and one with the
primary axis along the $z$ axis, $[+z,+y,+x]$. If in
Fig.~\ref{fig:decomposition_3_separated} $x$, $y$ and $z$ axes run
along the blue, green and red sides of the cube, these three
pyramidal components correspond to the blue, green and red pyramids,
respectively. This nomenclature not only defines the pyramidal components
but is also useful to describe the ray-tracing within each
component in a coordinate independent way.

Table \ref{table:pyramid table} defines the full set of pyramidal
components. The first column shows the names of the 24 pyramidal
components. Each component consists of a set of cells $(i,j,k)$'s.
The parameters $i$, $j$ and $k$ of a pyramidal component satisfy the
following bounds, $l_i\leq i \leq u_i$, $l_j\leq j \leq u_j$ and
$l_k\leq k \leq u_k$.  The functional values of
$l_i,u_i,l_j,u_j,l_k,u_k$ need not to be constants.  They are defined
according to the second to the seventh columns of Table
\ref{table:pyramid table}. We note that with this definition of the
pyramidal components, they do not necessarily form a complete
pyramid. That is, some part of the volume may be cropped so that the
base is not necessarily a square.

\begin{table*}
\begin{center}
\begin{tabular}{ccccccc}
\hline\hline
component & $l_i$ & $u_i$ & $l_j$ & $u_j$ & $l_k$ & $u_k$ \\ \hline\hline
$[+x,+z,+y]$ & 0 & $M_x^+$ & 0 & $\min\{i,M_y^+\}$ & 0 & $\min\{i,M_z^+\}$ \\ \hline
$[+y,+x,+z]$ & 0 & $\min\{j,M_x^+\}$ & 0 & $M_y^+$ & 0 & $\min\{j,M_z^+\}$ \\ \hline
$[+z,+y,+x]$ & 0 & $\min\{k,M_x^+\}$ & 0 & $\min\{k,M_y^+\}$ & 0 & $M_z^+$ \\ \hline
$[-x,+z,+y]$ & $-M_x^-$ & 0 & 0 & $\min\{-i,M_y^+\}$ & 0 & $\min\{-i,M_z^+\}$ \\ \hline
$[+y,-x,+z]$ & $\max\{-j,-M_x^-\}$ & 0 & 0 & $M_y^+$ & 0 & $\min\{j,M_z^+\}$ \\ \hline
$[+z,+y,-x]$ & $\max\{-k,-M_x^-\}$ & 0 & 0 & $\min\{k,M_y^+\}$ & 0 & $M_z^+$ \\ \hline
$[+x,+z,-y]$ & 0 & $M_x^+$ & $\max\{-i,-M_y^-\}$ & 0 & 0 & $\min\{i,M_z^+\}$ \\ \hline
$[-y,+x,+z]$ & 0 & $\min\{-j,M_x^+\}$ & $-M_y^-$ & 0 & 0 & $\min\{-j,M_z^+\}$ \\ \hline
$[+z,-y,+x]$ & 0 & $\min\{k,M_x^+\}$ & $\max\{-k,-M_y^-\}$ & 0 & 0 & $M_z^+$ \\ \hline
$[-x,+z,-y]$ & $-M_x^-$ & 0 & $\max\{i,-M_y^-\}$ & 0 & 0 & $\min\{-i,M_z^+\}$ \\ \hline
$[-y,-x,+z]$ & $\max\{j,-M_x^-\}$ & 0 & $-M_y^-$ & 0 & 0 & $\min\{-j,M_z^+\}$ \\ \hline
$[+z,-y,-x]$ & $\max\{-k,-M_x^-\}$ & 0 & $\max\{-k,-M_y^-\}$ & 0 & 0 & $M_z^+$ \\ \hline
$[+x,-z,+y]$ & 0 & $M_x^+$ & 0 & $\min\{i,M_y^+\}$ & $\max\{-i,-M_z^-\}$ & 0 \\ \hline
$[+y,+x,-z]$ & 0 & $\min\{j,M_x^+\}$ & 0 & $M_y^+$ & $\max\{-j,-M_z^-\}$ & 0 \\ \hline
$[-z,+y,+x]$ & 0 & $\min\{-k,M_x^+\}$ & 0 & $\min\{-k,M_y^+\}$ & $-M_z^-$ & 0 \\ \hline
$[-x,-z,+y]$ & $-M_x^-$ & 0 & 0 & $\min\{-i,M_y^+\}$ & $\max\{i,-M_z^-\}$ & 0 \\ \hline
$[+y,-x,-z]$ & $\max\{-j,-M_x^-\}$ & 0 & 0 & $M_y^+$ & $\max\{-j,-M_z^-\}$ & 0 \\ \hline
$[-z,+y,-x]$ & $\max\{k,-M_x^-\}$ & 0 & 0 & $\min\{-k,M_y^+\}$ & $-M_z^-$ & 0 \\ \hline
$[+x,-z,-y]$ & 0 & $M_x^+$ & $\max\{-i,-M_y^-\}$ & 0 & $\max\{-i,-M_z^-\}$ & 0 \\ \hline
$[-y,+x,-z]$ & 0 & $\min\{-j,M_x^+\}$ & $-M_y^-$ & 0 & $\max\{j,-M_z^-\}$ & 0 \\ \hline
$[-z,-y,+x]$ & 0 & $\min\{-k,M_x^+\}$ & $\max\{k,-M_y^-\}$ & 0 & $-M_z^-$ & 0 \\ \hline
$[-x,-z,-y]$ & $-M_x^-$ & 0 & $\max\{i,-M_y^-\}$ & 0& $\max\{i,-M_z^-\}$ & 0 \\ \hline
$[-y,-x,-z]$ & $\max\{j,-M_x^-\}$ & 0 & $-M_y^-$ & 0 & $\max\{j,-M_z^-\}$ & 0 \\ \hline
$[-z,-y,-x]$ & $\max\{k,-M_x^-\}$ & 0 & $\max\{k,-M_y^-\}$ & 0 & $-M_z^-$ & 0 \\
\hline
\end{tabular}
\caption{Definitions of the 24 pyramidal components.
The first column contains the names of the pyramidal components. Each set of three constitutes one of the eight octants, see Fig.~\ref{fig:decomposition_8_intact}. The three pyramidal components for one octant are illustrated in Fig.~\ref{fig:decomposition_24_separated}.
Columns 2 through 7 list the coordinates defining each pyramidal component: lower bound of $\mathbin{x}$, upper bound of $\mathbin{x}$, lower bound of $\mathbin{y}$, upper bound of $\mathbin{y}$, lower bound of $\mathbin{z}$, upper bound of $\mathbin{z}$, respectively. 
To see if a point $(i,j,k)$ is part of a given pyramidal component, 
check whether the values of $i$, $j$ and $k$ fall within the listed bounds.}
\label{table:pyramid table}
\end{center}
\end{table*}

Using the pyramidal decomposition of the volume, we construct the list
of ICs as follows. Let the pyramidal component $[A,B,C] = [\pm a,\pm
b,\pm c]$ have $\mathbin{N_t}$ ICs. The IC list has therefore size $N_t$.  '
We scan through the whole component cell by cell in a specific order.
Starting from the source, we proceed to the next cell along the
primary axis ($A$ direction). If an IC is encountered, we store its
coordinates in the list of ICs using an ascending order of
indices. After we have scanned the whole row, we move to the
neighbouring row, which is parallel to the previous row but displaced
in the direction of the secondary axis ($B$ direction). We perform the
scanning of this row from the cell closest to the source, and then
proceed to the next cell in the $A$ direction until we finish the
whole row. We repeat this procedure until the first plane of cells has
been scanned, after which we proceed to the next plane which is
parallel to the first plane but displaced along the tertiary axis
($C$) direction. This second plane is scanned in the same way as the
first one. We scan subsequent planes until all the cells of the
pyramidal component have been scanned. After this the IC list stores
the coordinates of the $N_t$ ICs in an order beneficial for the NIC
identification method we describe in Section~\ref{sec:NIC identification}.


\subsection{NIC identification}\label{sec:NIC identification}

To find the NICs from the list of ICs for a given pyramidal component,
we use the following algorithm. Let $M_b =
\min\{M^{\pm}_a,M^{\pm}_b\}$ and $M_c = \min\{M^{\pm}_a,M^{\pm}_c\}$,
where the $M^{\pm}_a$, $M^{\pm}_b$ and $M^{\pm}_c$ follow the signs of
the pyramidal component $[A,B,C] = [\pm a,\pm b,\pm c]$.  The base of
the pyramidal component is a plane of $(M_b+1)\times (M_c+1)$ cells.
We define the coordinates of these base cells as
$(i_{\mbox{\scriptsize{secondary}}},i_{\mbox{\scriptsize{tertiary}}})$.
The base cell closest to the source has coordinate $(0,0)$ and the
base cell furthest from the source has coordinate $(\pm
M_{\mbox{\scriptsize{b}}},\pm M_{\mbox{\scriptsize{c}}})$. In this
$\pm M_b$ and $\pm M_c$ have the same signs as $B = \pm b$ and $C =
\pm c$, respectively. 

In our method, we send rays into a
pyramidal component following the algorithm given in Section~\ref{sec:Ray
  updating algorithm}.  The rays start from the source cell and pass
through the planes that lie parallel to the base. We choose to indicate
the direction of a ray with the coordinates of its crossing point with the base plane
$(i_{\mbox{\scriptsize{secondary}}},
i_{\mbox{\scriptsize{tertiary}}})$ which is not necessarily a 2-tuple of integers.  
In addition, the directions of rays are not constrained by the base plane
of $(M_b+1)\times (M_c+1)$ cells, but by a slightly larger plane of
$(M^{\pm}_a+1)\times (M^{\pm}_a+1)$ cells. One can imagine that this
larger plane and the position of the source form a complete pyramid
which includes the pyramidal component $[A,B,C]$. Both the complete
pyramid and the pyramidal component $[A,B,C]$ span the same solid
angle from their vertices.

We check for each ray if it passes through the ICs stored in the IC
list in an ascending order.  That is, we start from the first element
of the IC array, and then the second, and so on. Due to the order
in which the ICs are stored, the first IC to pass this test is the NIC.

For a ray to pass through a cell, it has to pass through a cell face 
parallel to the base plane, either the plane facing the source or the plane
facing away from the source, or both. For an IC with centre coordinates 
$(i_x,i_y,i_z)$ located in the pyramidal component $[A,B,C]$
these two faces are defined by $j_a=i_a+\mbox{sgn}(A)/2$ (i.e.\ $a^+$
plane) and $j_a=i_a-\mbox{sgn}(A)/2$ (i.e.\ $a^-$ plane). We remind the reader
that the variables $a\in\{ x,y,z\}$ and $A=\pm a$ follow from the primary axis
and its sign of the pyramidal component, so for example in pyramidal component 
$[-y,x,z]$, $a=y$ and therefore the $a^+$ plane is defined by $j_y=i_y-1/2$.

\begin{figure}
\centering
\includegraphics[width=0.5\textwidth]{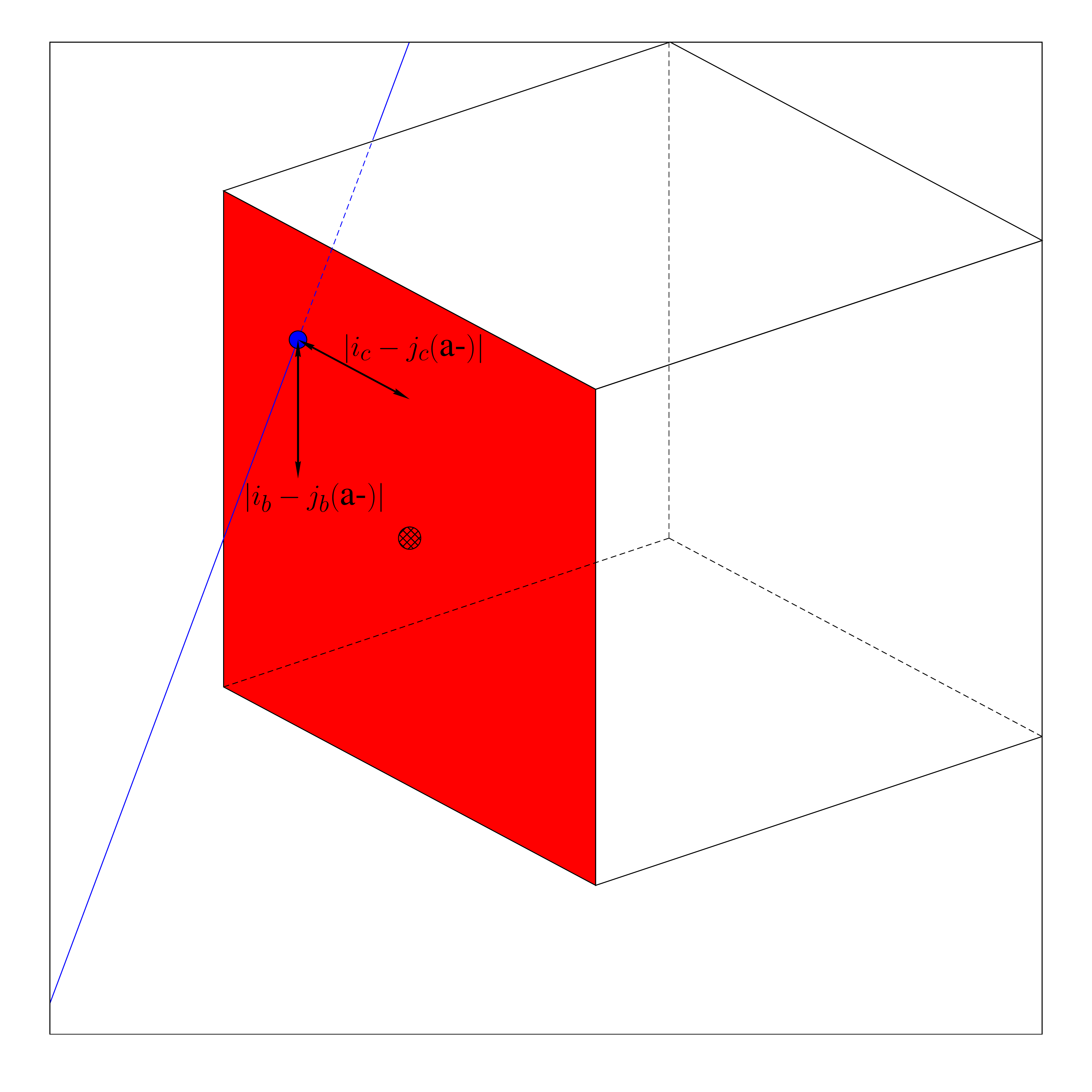}
\caption{A ray traverses an IC with centre coordinates $(i_x,i_y,i_z)$
at the $a^-$ plane (the red area). The entry point $(j_a,j_b,j_c)$
is indicated by a dot. The center of the IC's $a^-$ face $(j_a,i_b,i_c)$ is
indicated by a crosshatched circle. The distances between the entry point
and the cell centre, projected along the $b$ and $c$ directions, are
shown by the black arrows and the quantities next to the black
arrows.}
\label{fig:a_minus_plane}
\end{figure}

\begin{figure}
\centering
\includegraphics[width=0.5\textwidth]{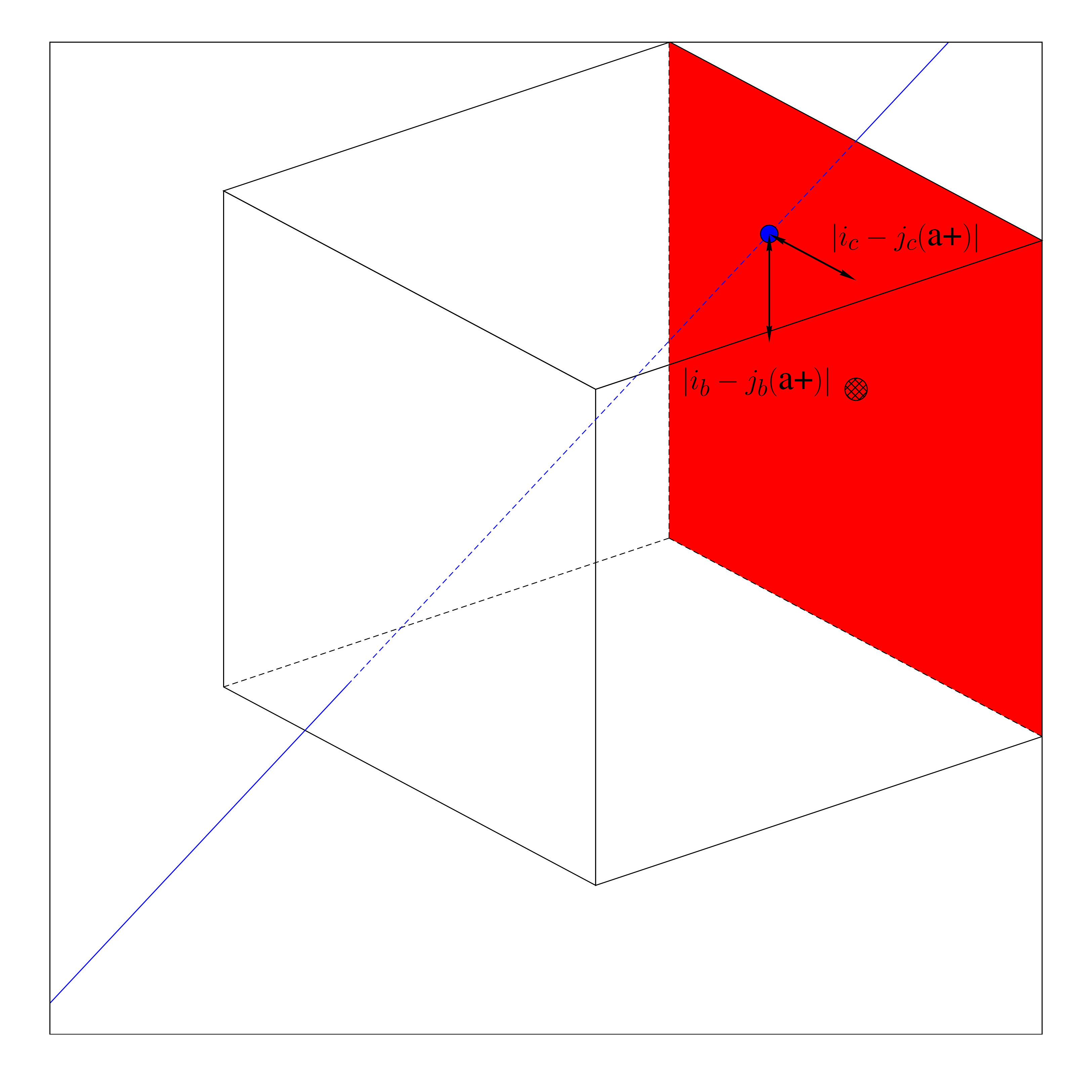}
\caption{Same as Figure~\ref{fig:a_minus_plane} but for the $a^+$ plane.}
\label{fig:a_plus_plane}
\end{figure}

Given the location of these two planes, the ray equation for a ray
$(i_{\mbox{\scriptsize{secondary}}},i_{\mbox{\scriptsize{tertiary}}})$,
provides us with the other two coordinates of the location where
the ray crosses these planes, denoted as $j_b$'s and $j_c$'s (in the
example given above these would be $j_x$ and
$j_z$). Figure~\ref{fig:a_minus_plane} and \ref{fig:a_plus_plane}
illustrate how a ray crosses an IC at the $a^-$ and $a^+$ planes and
also show the geometrical meaning of the different coordinates. To
establish whether the ray crosses the IC we just need to compare
these to the coordinates of the IC.
If the ray passes
through this IC, at least one of the following two conditions
has to be satisfied.
\begin{enumerate}[leftmargin=*]
\item The ray traverses into the IC at the $a^-$ plane if and only if
$\vert i_b-j_b(\mbox{$a^-$})\vert\leq 1/2$ and
$\vert i_c-j_c(\mbox{$a^-$})\vert\leq 1/2$.
\item The ray traverses into the IC at the $a^+$ plane if and only if
$\vert i_b-j_b(\mbox{$a^+$})\vert\leq 1/2$ and
$\vert i_c-j_c(\mbox{$a^+$})\vert\leq 1/2$.
\end{enumerate}

Since the first IC to pass the test is the NIC, we then stop the
NIC identification procedure for this ray and proceed to the next
ray. We explain the selection of this next ray in
Section~\ref{sec:Ray updating algorithm}.

\subsection{Ray updating algorithm}\label{sec:Ray updating algorithm}

\begin{figure}
\centering
\includegraphics[width=0.5\textwidth]{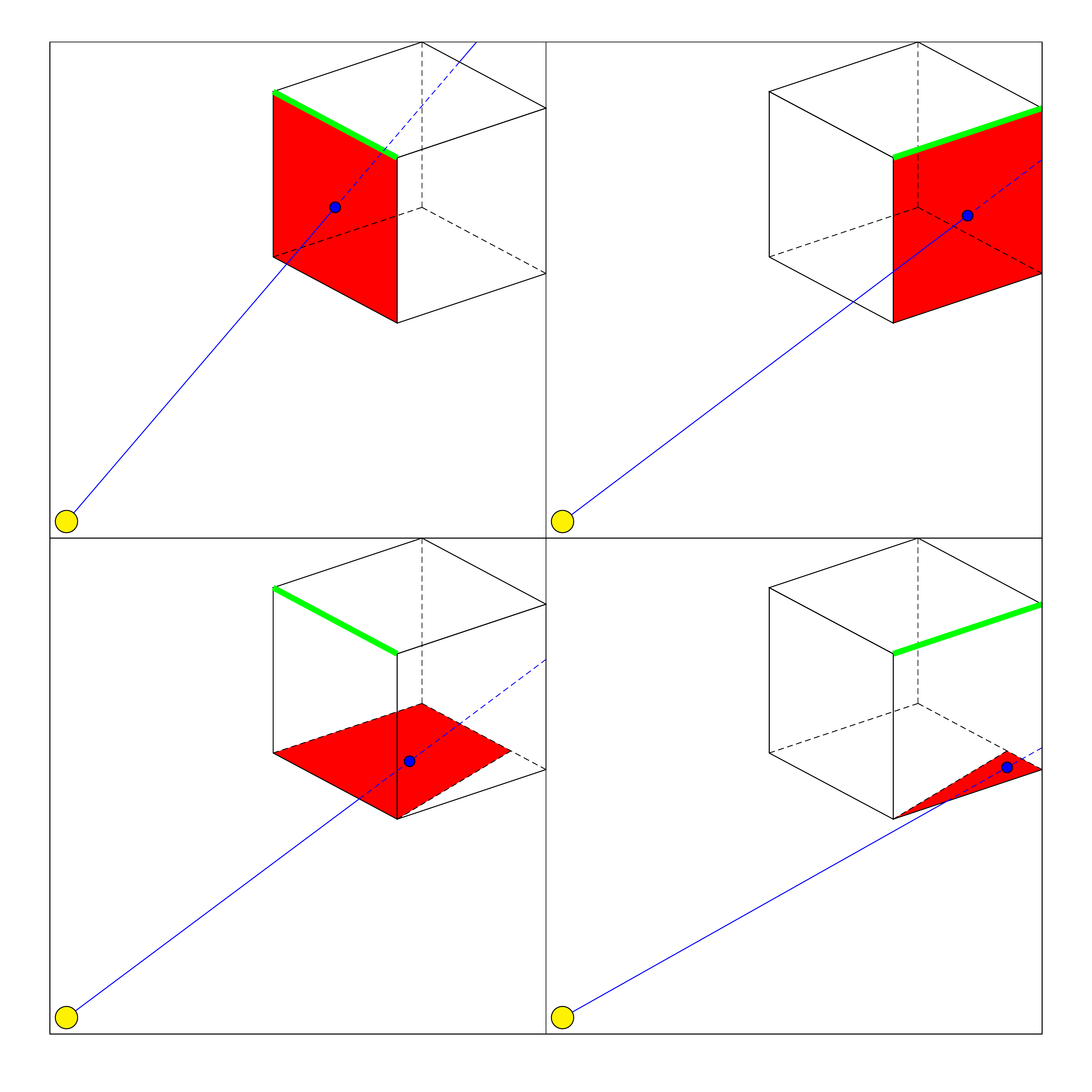}
\caption{A ray enters into a cell through one of the four faces.
  Top-left: the ray passes through front face. Top-right: the ray
  passes through side face.  Bottom-left: the ray passes through
  bottom face 1. Bottom-right: the ray passes through bottom
  face 2.  The thick green line in each subplot indicates the verge of
  the cell through which the next ray will pass.}
\label{surface_front_side_bottom}
\end{figure}

The first ray considered is $(i_{\mbox{\scriptsize{secondary}}},
i_{\mbox{\scriptsize{tertiary}}})=(0,0)$. However, in order to
minimize the number of rays to check we do not step through all
possible values $(i_{\mbox{\scriptsize{secondary}}},
i_{\mbox{\scriptsize{tertiary}}})$ but instead select the next ray
depending on whether the previous one traversed a NIC or not. If the
previous ray $(i_{\mbox{\scriptsize{secondary}}},
i_{\mbox{\scriptsize{tertiary}}})$ did not cross an IC, the next ray
$(i'_{\mbox{\scriptsize{secondary}}},
i_{\mbox{\scriptsize{tertiary}}})$ will simply have
$i'_{\mbox{\scriptsize{secondary}}} =
i_{\mbox{\scriptsize{secondary}}} + \mbox{sgn}(B)$. If it did, the
selection of the next ray depends on the location of the entry point
into the IC. This entry point lies on one of the faces of the NIC
facing the source, either the front, the side, or the bottom
face, as shown in Figure~\ref{surface_front_side_bottom}. The
bottom face is divided into two parts (labelled 1 and 2) to aid
the ray-updating algorithm. From the ray equation of the
ray $(i_{\mbox{\scriptsize{secondary}}},
i_{\mbox{\scriptsize{tertiary}}})$ we can check which face the previous ray
passed through.
\begin{enumerate}[leftmargin=*]
\item If the previous ray entered through the front face, then the
intersection $(j_x,j_y,j_z)$ of the ray to the plane
$j_a=i_a-\mbox{sgn}(A)/2$ satisfied $\vert i_b-j_b\vert\leq 1/2$ and
$\vert i_c-j_c\vert\leq 1/2$.
\item If the previous ray entered through the side face, then the
intersection $(j_x,j_y,j_z)$ of the ray to the plane
$c=i_c-\mbox{sgn}(C)/2$ satisfied $\vert i_a-j_a\vert\leq 1/2$ and
$\vert i_b-j_b\vert\leq 1/2$.
\item If the previous ray entered through the bottom face, then the
intersection $(j_x,j_y,j_z)$ of the ray to the plane
$b=i_b-\mbox{sgn}(B)/2$ satisfied $\vert i_a-j_a\vert\leq 1/2$ and
$\vert i_c-j_c\vert\leq 1/2$.
\begin{enumerate}[leftmargin=*]
\item For the same ray, if its intersection $(k_x,k_y,k_z)$ to the
plane $a=i_a-\mbox{sgn}(A)/2$ satisfied $\vert i_c-k_c\vert\leq
1/2$, it entered through bottom face 1.
\item Otherwise, it entered through bottom face 2.
\end{enumerate}
\end{enumerate}

We then choose the next ray 
such that it will touch on the verge (green line in
Figure~\ref{surface_front_side_bottom}) of the traversed IC
$(i_x,i_y,i_z)$. This is achieved by setting the value of
$i'_{\mbox{\scriptsize{secondary}}}$ according to {the following rules:}
\begin{enumerate}[leftmargin=*]
\item If the previous ray passed through the front face or the
bottom face 1, the new ray is selected to touch the line
$\frac{a}{i_a-\mbox{\scriptsize{sgn}}(A)/2}=
\frac{b}{i_b+\mbox{\scriptsize{sgn}}(B)/2}$. This implies
$i'_{\mbox{\scriptsize{secondary}}}=\pm
M^{\pm}_a\frac{i_b+\mbox{\scriptsize{sgn}}(B)/2}
{i_a-\mbox{\scriptsize{sgn}}(A)/2}$.  Both $\pm$ follow the sign of
$A = \pm a$.
\item If the previous ray passed through the side face or the
bottom face 2, the new ray is selected to touch the line
$\frac{b}{i_b+\mbox{\scriptsize{sgn}}(B)/2}=
\frac{c}{i_c-\mbox{\scriptsize{sgn}}(C)/2}$. This implies
$i'_{\mbox{\scriptsize{secondary}}}=i_{\mbox{\scriptsize{tertiary}}}
\frac{i_b+\mbox{\scriptsize{sgn}}(B)/2}{i_c-\mbox{\scriptsize{sgn}}(C)/2}$.
\end{enumerate}

If the new ray $(i'_{\mbox{\scriptsize{secondary}}},
i_{\mbox{\scriptsize{tertiary}}})$ falls outside the angular range of
the pyramidal component, i.e.\ if $\vert
i'_{\mbox{\scriptsize{secondary}}} \vert > M^{\pm}_a+1$, we skip
NIC identification for this ray and proceed to the ray
$(0,i_{\mbox{\scriptsize{tertiary}}}+\mbox{sgn}(C))$, the first ray of
the next slice. The ray updating terminates once the ray coordinates
reach $(i''_{\mbox{\scriptsize{secondary}}},i''_
{\mbox{\scriptsize{tertiary}}})$ where $\vert
i''_{\mbox{\scriptsize{tertiary}}}\vert > M^{\pm}_{a}+1$.  This ray
updating algorithm ensures that we use the minimum number of rays
required to trace all the NICs.

\subsection{IC removal in the array}\label{sec:IC removal in the array}
The ray-tracing proceeds slice by slice in a pyramidal component,
i.e.\ the unity increment of $\vert
i_{\mbox{\scriptsize{tertiary}}}\vert$ described in the last
paragraph of the previous section implies proceeding to the
neighbouring slice.  After a slice has been scanned, the list of
ICs can contain ICs from this slice which are not NICs. Since these
ICs will never become NICs for other slices, it is best to
remove them from the list of ICs so that they no longer need to
be considered.
  
\begin{figure*}
\centering
\includegraphics[width=0.8\textwidth]{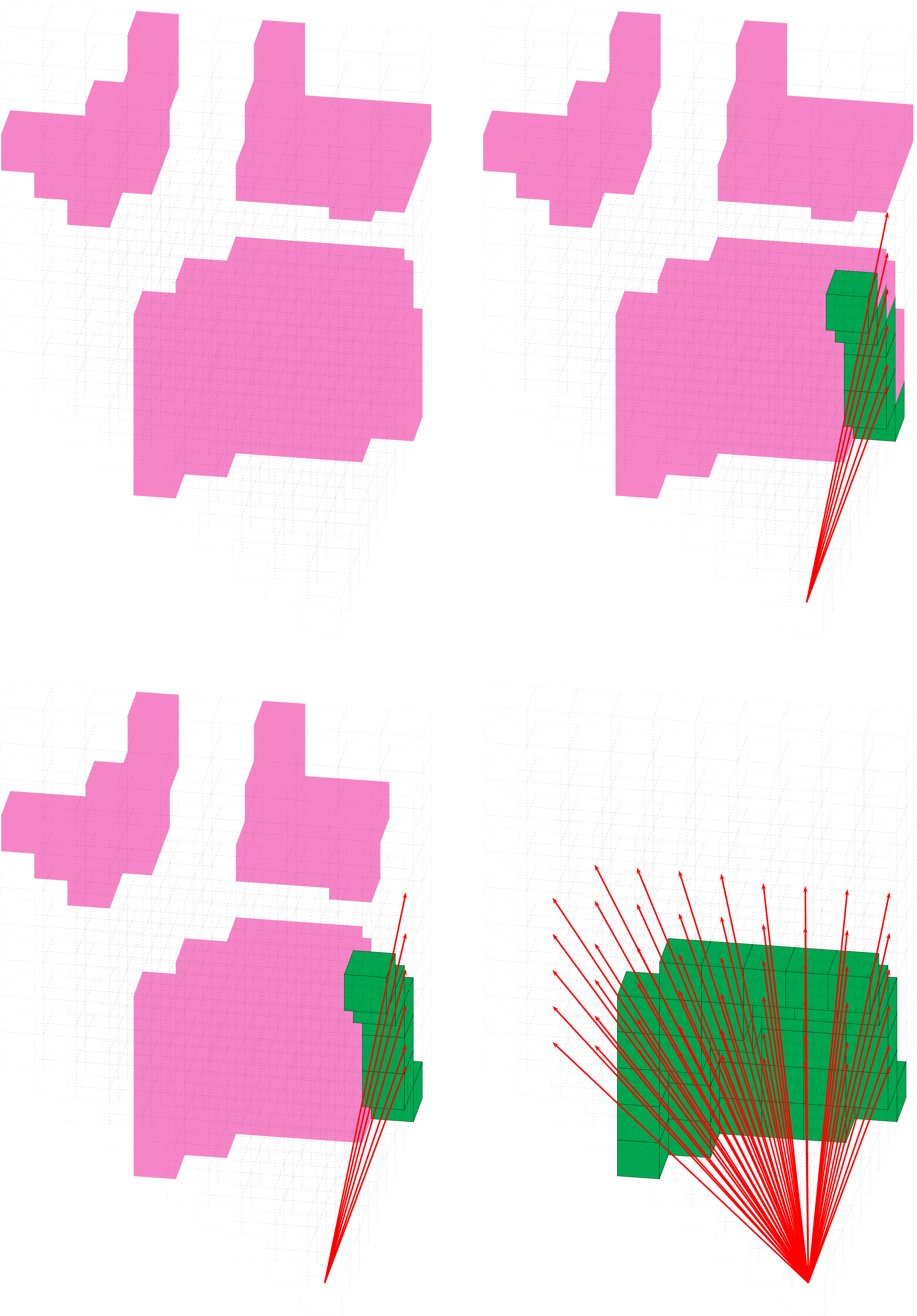}
\caption{Top left panel: A graphical example of a pyramidal component
  which contains a number of ICs (pink cells).  The source is located
  at the vertex position.  Top right panel: a series of rays (red
  arrows) are sent through the first slice following the procedure
  specified in Section \ref{sec:Ray updating algorithm}.  The NICs are
  shown in green and the other ICs in the first slice remain pink.
  Bottom left panel: The other ICs (pink cells) of the first slice are
  removed as described in Section \ref{sec:IC removal in the array}.
  Bottom right panel: At the end of the procedure, all the slices in
  the pyramidal components have been scanned by rays. All NICs (green
  cells) have been identified and all other ICs (pink cells) have
  been removed.}
\label{fig:pyramid4to1}
\end{figure*}

Figure~\ref{fig:pyramid4to1} shows an illustration of NIC
identification and the removal of the other ICs in one pyramidal
component. In the top left panel of Figure~\ref{fig:pyramid4to1}, a
pyramidal component contains a number of ICs (pink cells). We
consider the case of multiple sources of ionization so that there
are ionized regions away from the source being considered. The
source is located at the vertex position.  The ICs are stored in the
IC list in the order explained in Section \ref{sec:Fast ray-tracing
method}. The top right panel of Figure~\ref{fig:pyramid4to1} shows
how the first few rays are sent through the first slice of cells.  The
NIC identification procedure (Section \ref{sec:NIC identification})
identifies the NICs, indicated by the green cells. The bottom left
panel of Figure~\ref{fig:pyramid4to1} shows how the ICs which are not
NICs from the first slice are removed from the list of ICs to avoid
their use in the NIC identification procedure for the subsequent
slices. The bottom right panel of Figure~\ref{fig:pyramid4to1}
shows the end stage of the ray-tracing method: the IC list only stores
NICs which are then used to estimate the adaptive time-step for the
next evolution cycle. The total number of rays used
in this example is 53 and the number of base cells is 100,
illustrating the efficiency of our ray selection algorithm.

\subsection{Multiple-source treatment}

The previous sections (\ref{sec:Fast ray-tracing method} --
\ref{sec:IC removal in the array}) have explained in detail the adaptive
time-step algorithm for the single-source case.  In the
multiple-source case, we repeat the same procedure to
obtain photo-ionization rates of NICs for each
source. The sum of the photo-ionization rates
contributed by all sources is then used to calculate the time-step
$\Delta t_k$ for each NIC $k$. The adaptive
time-step $\Delta t$ is the minimum of all $\Delta t_k$'s.

\section{Asynchronized evolution}\label{sec:Asynchronized evolution}


\subsection{Introduction}
\label{sec:Asynchronized evolution introduction}
In most types of simulations all computational cells use the same
time-step, be it a constant time-step or an adaptive time-step.
Section \ref{sec:Adaptive time-step} outlined how the
photo-heating rates $\mathcal{H}$'s of all cells can be calculated
accurately using an adaptive time-step. The main problem with
this approach is that the adaptive time-step is almost always much smaller than
the recombination timescale, thus making the simulation more
expensive.  However, the photo-heating events occur only in particular
cells in the simulation volume, so not all the cells necessarily need
to use this relatively small time-step for an accurate photo-heating
calculation.  Consider for example an ionized cell which is close to
thermal and ionization equilibrium. In this cell the evolution of the
ionization and temperature will be slow and
the optical depths across this cell
will essentially be constant.
Using a relatively short time-step on these ionized
cells is a waste of computational resources.

Asynchronous evolution is the application of different time-steps to
different cells. The general idea is that cells which are close to
equilibrium use relatively long time-steps while the rest use
relatively short time-steps. This approach will dedicate most of the
computational resources to the cells that require it thus saving a
substantial amount of computational time.

For an expanding H{\sc II} region, there are three types of
cells. Inside the H{\sc II} region there are the ionized cells which
are close to equilibrium. Next there are the ionization front cells
which are far from equilibrium. Beyond the ionization front lie the
neutral cells which may or may not be in equilibrium and thus require
careful consideration. For a long time-step, the ionization front may
advance substantially into the neutral region and thus the cells that
were neutral and in equilibrium at the start of the time-step may not 
remain so during the time-step. Furthermore, if the source contains
sufficiently energetic photons these will penetrate far beyond the
ionization front and start to heat and ionize the neutral material, 
in some cases establishing an ionization front which is as broad as
the entire computational domain.

For these reasons we decide to only consider the ionized cells to be
in equilibrium and to group the neutral cells together with the ionization
front cells as non-equilibrium cells. This means that when the
ionized regions cover only a small fraction of the computational domain,
the gain from using asynchronous evolution will be small. Large gains
are only achieved once the ionized regions have grown substantially.

Both reasons for including the neutral cells in the set of non-equilibrium
cells can in principle be dealt with, for example by forecasting the
extent by which the ionized regions will grow during a time-step and
measuring the photo-rates in the neutral cells. However, we postpone
an exploration of this to future work.


\subsection{Asynchronous evolution algorithm}

Let us assume that we start our simulation at $t=0$ and want to
know the state of our computational domain at a time $\Delta t$. We
categorize the cells into three sets.
\begin{itemize}
\item $\mathcal{L}$ is the set of ionized cells which are in thermal equilibrium 
and their clocks are $t<\Delta t$ (i.e.\ ionized cells).
\item $\mathcal{S}$ is the set of non-$\mathcal{L}$ cells 
and their clocks clocks are $t<\Delta t$ (i.e.\ ionization
front and neutral cells).
\item $\mathcal{E}$ is the set of cells whose clocks are $t=\Delta t$ (i.e.\ all cells at the end of the simulation step).
\end{itemize}
The criterion for thermal equilibrium of the ionized cells is discussed in
Section~\ref{sec:Criteria of thermal equilibrium}.

At the start of the simulation, all the cells' clocks are synchronized
at $t=0$. We identify which cells belong to sets $\mathcal{L}$ or
$\mathcal{S}$. The $\mathcal{L}$ cells are assigned a time-step
$\Delta t$ while the $\mathcal{S}$ cells are assigned a time-step
$\Delta t_1$ where $\Delta t_1$ is the adaptive time-step of the first
evolution stage. In the first evolution stage, only $\mathcal{S}$
cells are updated and their clocks are set to $t=\Delta t_1$. The
$\mathcal{L}$ cells are not updated and so their clocks remain $t=0$.

Next, we identify the new sets of $\mathcal{L}$ cells and
$\mathcal{S}$ cells. The new $\mathcal{L}$ cells are assigned a
time-step $\Delta t-\Delta t_1$ while $\mathcal{S}$ cells are assigned
a time-step $\Delta t_2$ where $\Delta t_2$ is the adaptive time-step
of the second evolution stage. 
Also in the second evolution stage, only $\mathcal{S}$
cells are updated and their clocks reach $t=\Delta t_1+\Delta t_2$.
The $\mathcal{L}$ cells keep their clocks at $t=0$.
Note that the $\mathcal{L}$ cells from the first stage are assigned a time-step $\Delta t$ while the those identified in the second stage are assigned a time-step $\Delta t-\Delta t_1$.

This procedure is repeated until all the clocks of the
$\mathcal{S}$ cells from the previous stage reach $t=\Delta t$.  In
the other words, the procedure is repeated until the first
$\mathcal{E}$ cells appear. Only after this we apply the different
assigned time-steps to the ${\mathcal{L}}$ cells. This final stage
is performed simultaneously for all ${\mathcal{L}}$ cells. With this
all cells have become $\mathcal{E}$ cells with their clocks
synchronized at $t=\Delta t$.

For the evolution of $\mathcal{S}$ cells, we calculate the
photo-rates only for the $\mathcal{S}$ cells contributed by each
source. The standard iterative method is used to get converged
results for the multiple sources \citep[see][]{C2-Ray2}. For the
evolution of the $\mathcal{L}$ cells, we calculate the photo-rates
only for the $\mathcal{L}$ cells and perform just one iteration on
them and ignore the convergence issue. This is acceptable since for
cells in both ionization and thermal equilibrium, further evolution
does not produce fluctuating evolution results during several
iterations.

\begin{figure*}
  \centering
    \includegraphics[width=0.7\textwidth]{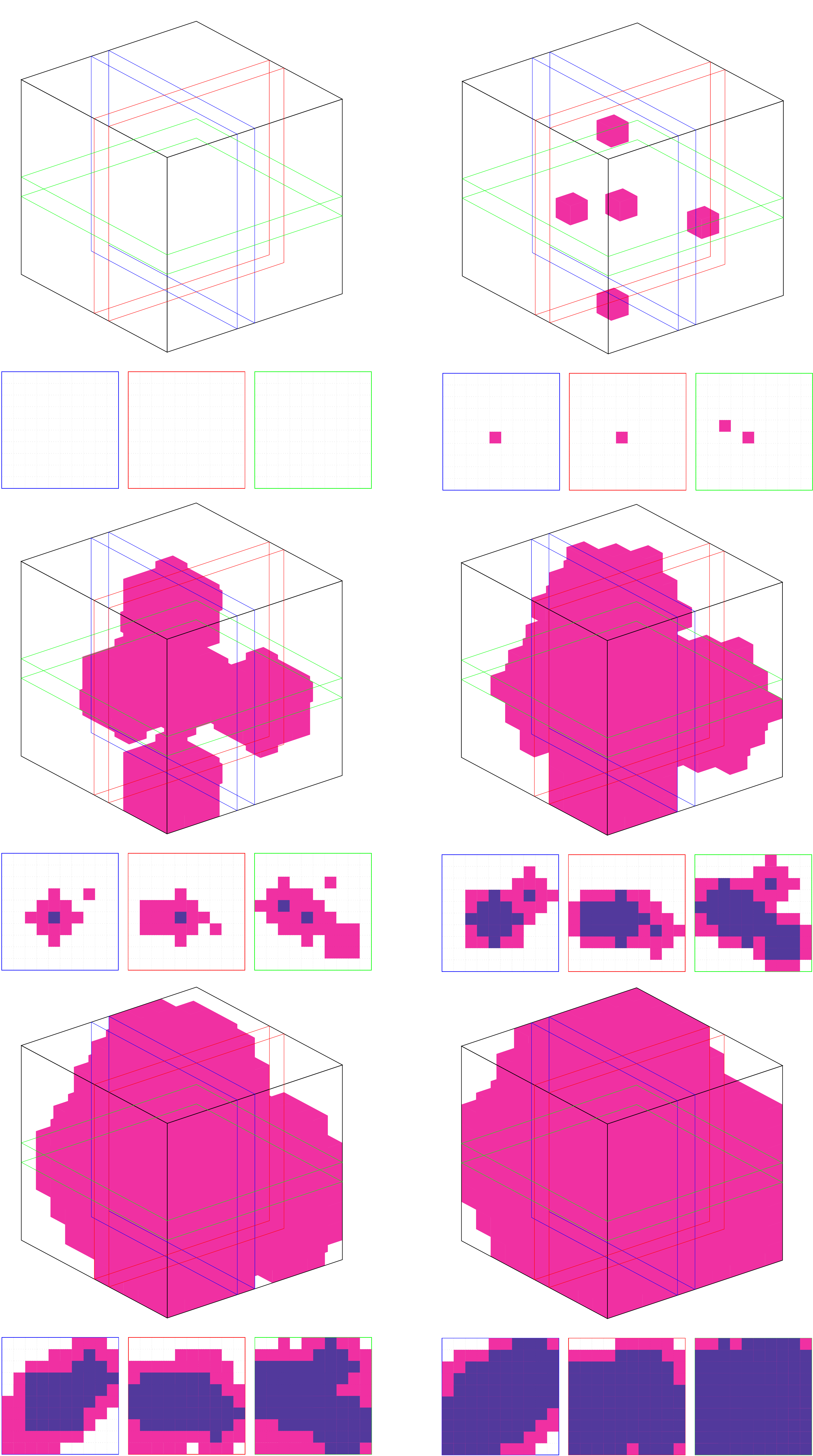}
  \caption{An example of asynchronous evolution. 
White: neutral $\mathcal{S}$ cells. 
Pink: newly identified $\mathcal{L}$ cells. 
Purple: previously identified $\mathcal{L}$ cells. Three slices through the source are shown separately.
From top to bottom, from left to right shows the asynchronous evolution during five time-steps.}
  \label{fig:asynchronous6to1}
\end{figure*}

Figure \ref{fig:asynchronous6to1} shows an example of applying
asynchronous evolution on a simulation volume.  All the neutral cells
are shown as white and they are all classified as $\mathcal{S}$ cells.
Ionized cells are shown as pink or purple, depending on if they are
newly identified $\mathcal{L}$ cells or previously identified
$\mathcal{L}$ cells. As the ionized bubbles grow, more and more cells
become $\mathcal{L}$ cells, which consume less computational
resources.  When the simulation volume is close to fully ionized, the
computational costs to evolve it further are substantially reduced.

\subsection{Criteria for thermal equilibrium}
\label{sec:Criteria of thermal equilibrium}

We need to define a good criterion of thermal equilibrium to assign
cells to set $\mathcal{L}$. The best criterion would be one that
is based on the values of the rates, as equilibrium is a state in
which all rates are balanced. However, since it is the rates which we
are trying to determine, we do not have access to them. Furthermore,
their values may change over the period of a time-step. 

As explained in Section~\ref{sec:Asynchronized evolution introduction}
we only consider the ionized cells to be in equilibrium and hence
a criterion based on the neutral fraction in a cell may be useful.
Experiments show that this is indeed the case
although the precise criterion is found to depend on the nature of the
sources.  Hard photons, for example, are able to change the thermal
state of cells containing only trace amounts of photon absorbers. To
define appropriate criteria, we have experimented with a range of different
sources. For example, we have found that if the sources have black body
spectra with a temperature of $T=50000K$ (an approximation of
star-forming galaxies), letting cells which have
$x_{\mbox{\scriptsize{HI}}}\leq 1\times 10^{-4}$ belong to set
$\mathcal{L}$ and otherwise belong to set $\mathcal{S}$, gives
satisfactory results.  If the sources have power-law spectra with a
spectral index of $1.5$ (an approximation of quasars where the
emission is dominated by an accreting black-hole), letting cells which
have $x_{\mbox{\scriptsize{HI}}}\leq 1\times 10^{-4}$ and
$x_{\mbox{\scriptsize{HeII}}}\leq 1\times 10^{-4}$ belong to set
$\mathcal{L}$ and otherwise belong to set $\mathcal{S}$, gives
reasonable results.

These criteria based on the ionization fractions are only approximate 
and other options, such as using rates from previous time-steps, may
be worth exploring in future work.

We would like to point out that our asynchronous evolution
strategy works even when some of the ionized cells start
recombining, for example in a radiation hydrodynamic simulation or
one where sources can turn off. The reason is that the recombining
cells evolve on a timescale set by the recombination time and can
therefore be evolved with a long time-step as recombination times
are typically much longer than ionization times.

\section{Parallelization strategy}\label{sec:Parallelization strategy}

The performance of \textsc{C}$^2$\textsc{-Ray} benefits from a high
degree of parallelization.  Its computational efficiency scales
linearly with the number of cores used as long as the number of
sources is larger than the number of cores.  This is particularly
important for large scale reionization simulations as they require
substantial amounts of computing time and a large amount of memory.

\textsc{C}$^2$\textsc{-Ray} uses a hybrid parallelization scheme that
makes use of both distributed memory parallelization (employing the
Message Passing Interface (MPI) library) and shared memory
parallelization (based on the application programming interface
OpenMP). An example of the scalability of the hydrogen-only version
can be found in figure~1 of \citet{2014MNRAS.439..725I}. Also the new
version presented here has several procedures in which the
contribution from one source is calculated independently of the
existence of other sources. These procedures are the optical depth
calculation by short characteristics ray-tracing, the photo-rates
calculation, the NIC identification and the adaptive time-step
calculation. Therefore it is advantageous to parallelize these
procedures over the sources where one MPI process deals with one
source at a time.

After having looped through all the sources, the MPI processes communicate
with each other to exchange information. Examples of this are the
summation of the photo-rates from all the sources
and obtaining the adaptive time-step $\Delta t$ by extracting the
minimum one among the $\Delta t_s$'s where $s$ is the index of the
sources. 

Since most of the procedures in \textsc{C}$^2$\textsc{-Ray}
can be parallelized over the sources, it possesses a high scaling
efficiency when there are many sources.  Furthermore, because sources
emit radially, the radiative properties of the cells depend only on
their neighbouring cells which are closer to the source in the radial
directions.  Examples of this are the optical depth calculation and
the NIC identification.  For this reason, we can further parallelize
the procedures by domain decomposition and assign the decomposed
domains to different OpenMP threads.

Of the procedures which are suitable for OpenMP parallelization the
domain decomposition of short-characteristics ray-tracing is eight-fold
(the eight octants), the photo-rates calculation can be
$\mathbin{P^3}$-fold where
$\mathbin{P^3}$ is the number of cells in the
simulation box, the NIC identification is twenty four-fold (the 24
pyramidal components) and the adaptive time-step calculation is
$m$-fold where $m$ is the number of NICs.  The asynchronous evolution
of $\mathcal{L}$ and $\mathcal{S}$ cells are independent of other
cells, which makes the evolution procedure parallelizable.  If
$\mathbin{r}$ cells evolve with the same time-step,
we distribute these cells into $\mathbin{r\approx
p\times q}$ equal parts where $p$ is the number of MPI processes and
$q$ is the number of OpenMP threads of each process. The evolution
procedure of each part is assigned to each thread and the results from
the different MPI processes are
exchanged afterwards.

We end this section by concluding that \textsc{C}$^2$\textsc{-Ray} is a
highly parallelized radiative transfer code.  It can run on many different types
of parallel machines with different distributed memory (any number
of distributed memory cores) and shared memory (2, 4, 8, 16, 24
threads per core) configurations.  For example, a configuration which
contains any number of distributed memory processors less than the
number of sources and $2^i$ or $24\times 2^j$ threads per processor
will fully utilize the computational resources. This high scaling
efficiency is one of the important advantages of
\textsc{C}$^2$\textsc{-Ray}.

\section{\textsc{C}$^2$\textsc{-Ray} algorithm}\label{sec:C2-Ray algorithm}

\begin{figure}
   \centering
   \includegraphics[width=0.5\textwidth]{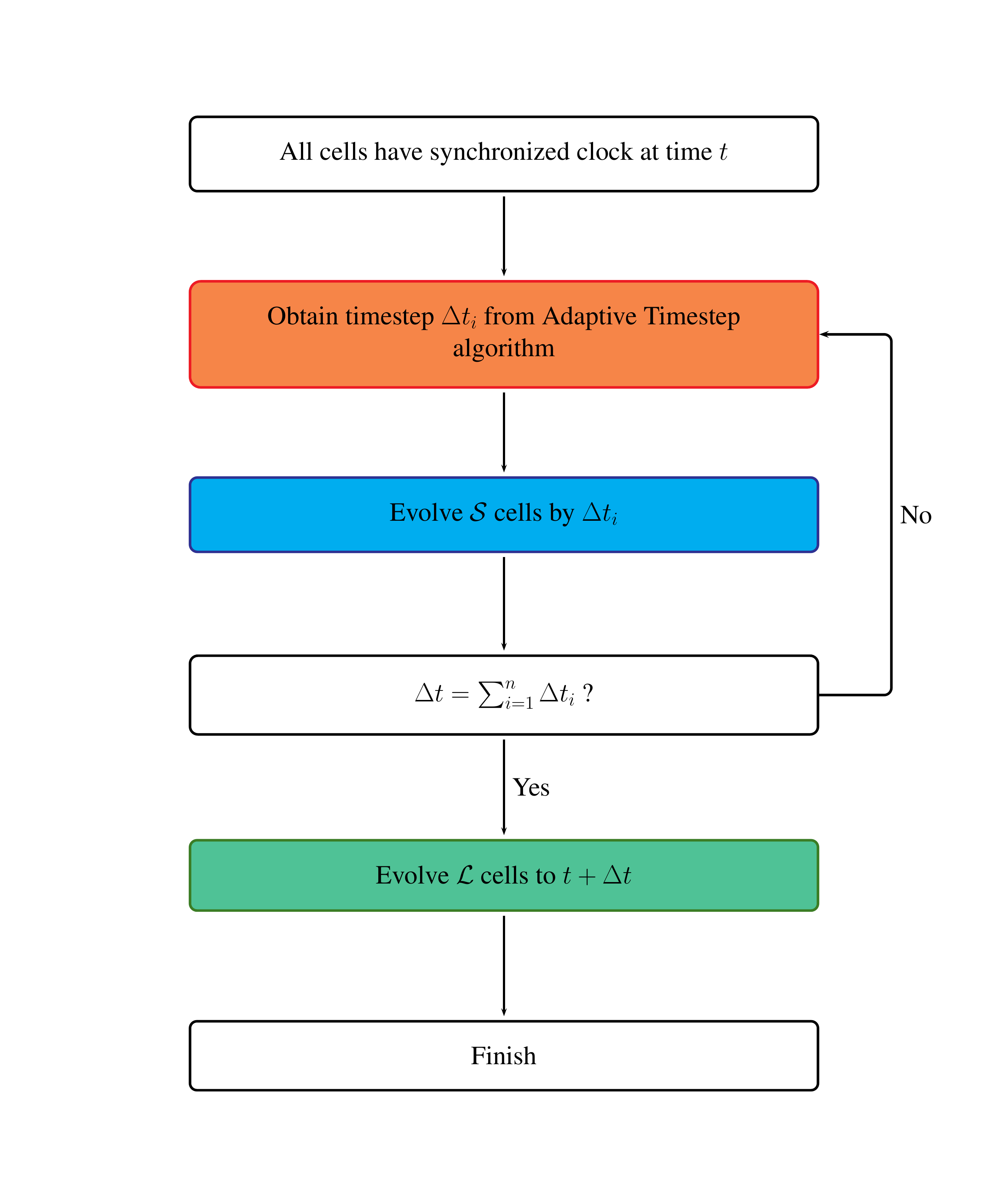}
   \caption{Overall algorithm of \textsc{C}$^2$\textsc{-Ray}. }
   \label{fig:flow_chart_general}
\end{figure}

\begin{figure}
   \centering
   \includegraphics[width=0.5\textwidth]{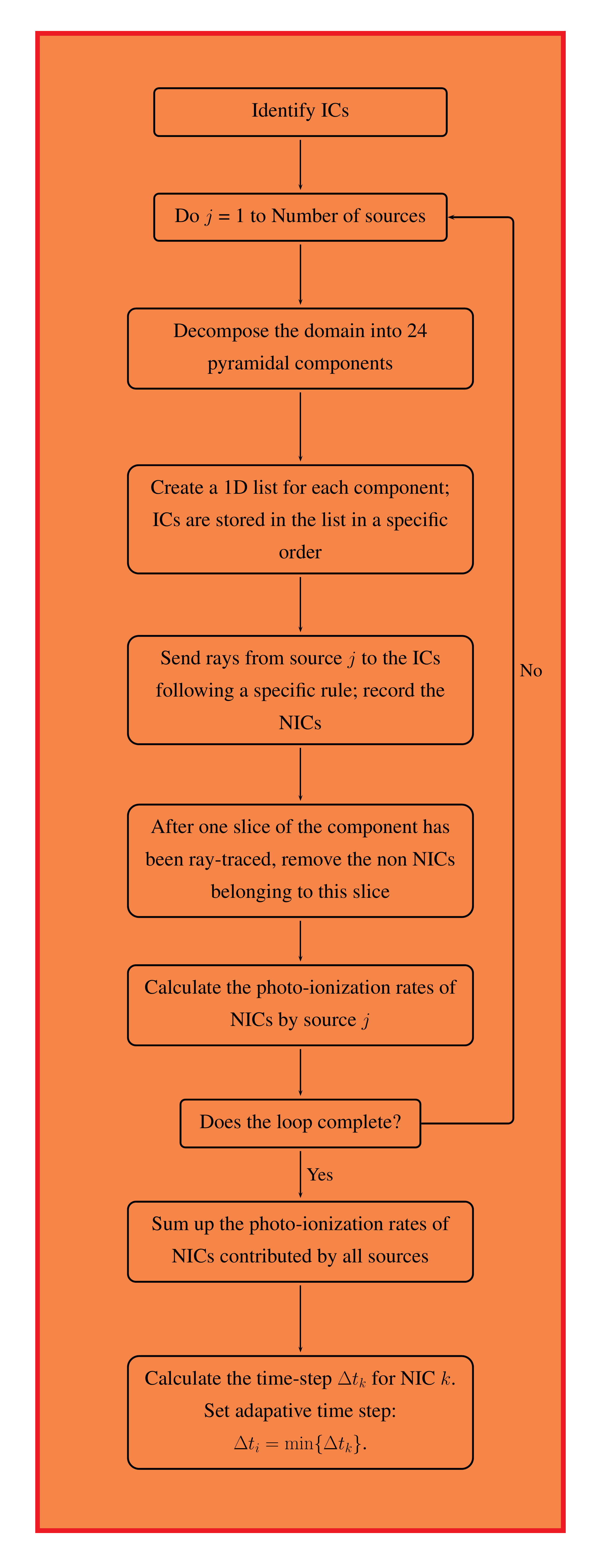}
   \caption{Adaptive time-step algorithm.}
   \label{fig:flow_chart_AT}
\end{figure}

\begin{figure}
   \centering
   \includegraphics[width=0.5\textwidth]{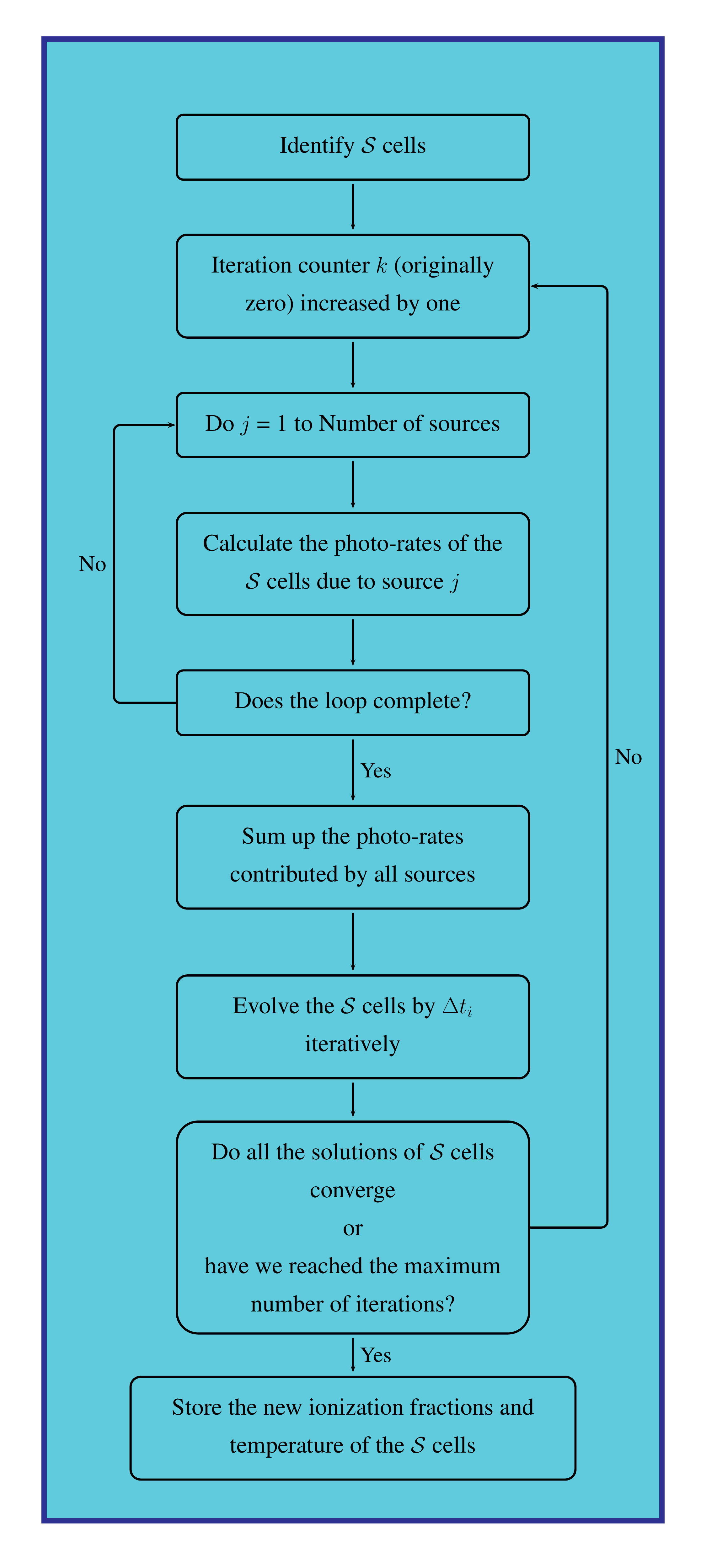}
   \caption{Asynchronous evolution algorithm of of $\mathcal{S}$ cells.}
   \label{fig:flow_chart_ADP}
\end{figure}

\begin{figure}
   \centering
   \includegraphics[width=0.5\textwidth]{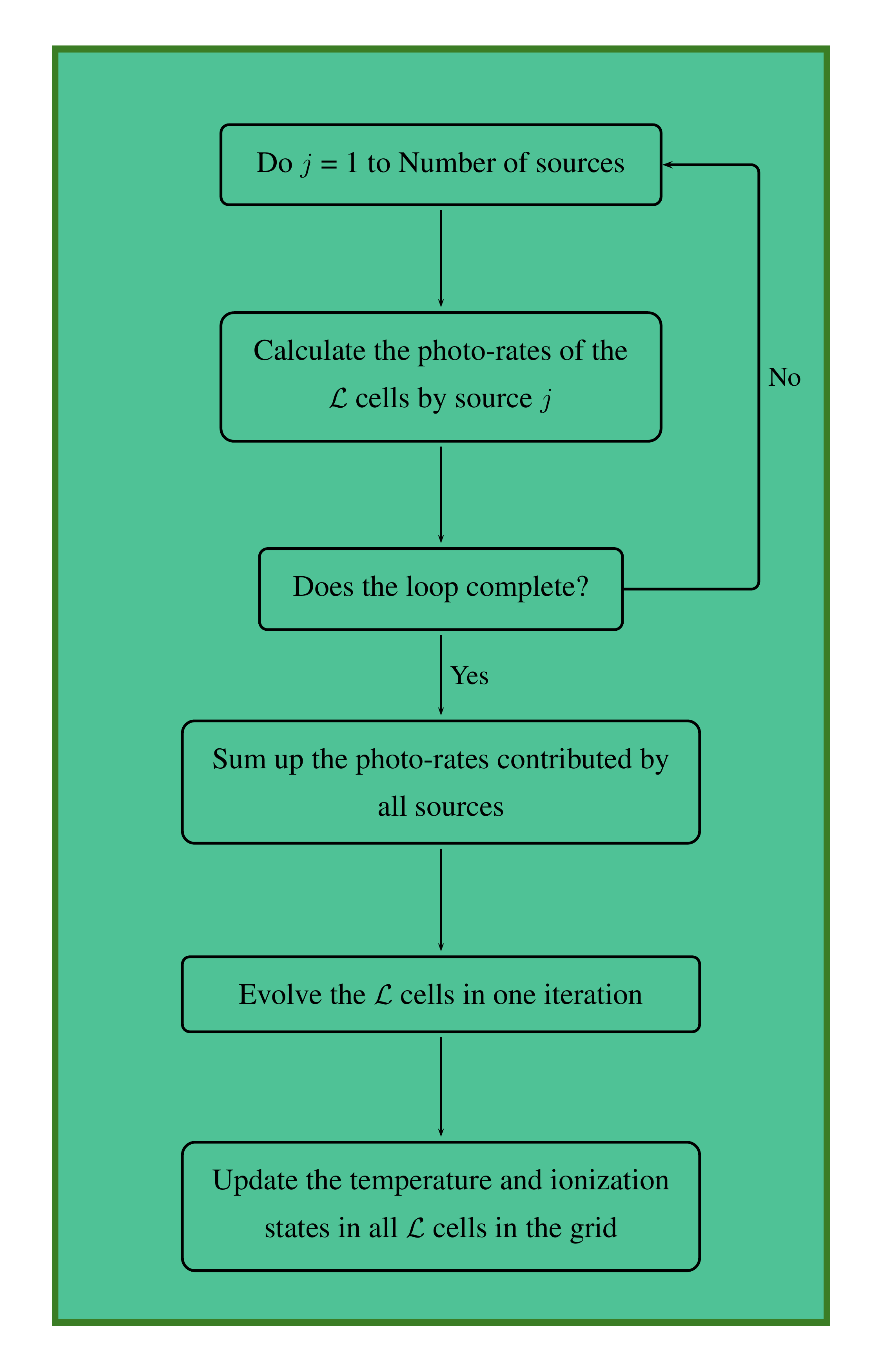}
   \caption{Asynchronous evolution algorithm of of $\mathcal{L}$ cells.}
   \label{fig:flow_chart_LTE}
\end{figure}

We summarize the updated \textsc{C}$^2$\textsc{-Ray} algorithm in
Figures \ref{fig:flow_chart_general} -- \ref{fig:flow_chart_LTE}.
These show how \textsc{C}$^2$\textsc{-Ray} deals with a given time
step $\Delta t$ which is too large for a correct photo-heating rate calculation.
Figure~\ref{fig:flow_chart_general} shows the overall algorithm and
Figures \ref{fig:flow_chart_AT} to \ref{fig:flow_chart_LTE} show in
more detail the stages of the adaptive time-step calculation,
evolution of the $\mathcal{S}$ cells and evolution of the
$\mathcal{L}$ cells.
 
In the entire algorithm, the clocks of the cells are important. 
At time $t$ the clocks in all cells are synchronized. At the
end they are synchronized again, at time $t + \Delta t$. The adaptive
time-step algorithm goes through a series of $n$ stages, each with
its own optimal time-step $\Delta t_i$, $\Delta t=\sum_{i=1}^{n}
\Delta t_i$ and the evolution during these stages is performed
asynchronously. At stage $i$, the set of $\mathcal{S}$ cells is
evolved with a time-step $\Delta t_i$. Accurate photo-heating rates are
obtained using this time-step and the clocks of $\mathcal{S}$ cells
are increased by $\Delta t_i$. 
The convergence criteria stated in Figure~\ref{fig:flow_chart_ADP} is that
the relative change of $x_{\mbox{\scriptsize{HI}}}$,
$x_{\mbox{\scriptsize{HeI}}}$ and $x_{\mbox{\scriptsize{HeIII}}}$
compared to the previous evolution cycle have to be smaller than a specified
tolerance. For all the work in this paper the value used is 0.01.

This process is repeated until the
clocks of the last set of $\mathcal{S}$ cells reaches $t+\Delta t$. 
After this the set of $\mathcal{L}$ cells are evolved
with longer time-steps $\Delta t-\sum_{i=1}^{m-1}\Delta t_i$, where
$m$ is the number of the stage after which they were identified as
$\mathcal{L}$ cells. At the end of the simulation step the clocks of
all cells are synchronized again at $t+\Delta t$.

\section{Tests}\label{sec:Tests}

We perform four tests to evaluate and validate our new efficient
photo-heating algorithm. In Tests 1 and 2, we consider a single source in a
homogeneous density environment. In Test 3, we simulate multiple sources
in a cosmological density field, using the standard set-up introduced
by \cite{comparison}. For these three tests we use both a constant
time-step and our adaptive and asynchronous time-stepping approach.
In Test 4, we compare \textsc{C}$^2$\textsc{-Ray} results to
those of another time-dependent photo-ionization code developed for
supernova applications. The codes use different ionization and
thermal solvers and this comparison thus serves to validate the results
of \textsc{C}$^2$\textsc{-Ray}.


\subsection{Test 1 - Single source in a homogeneous environment}\label{test 1}

We start with a simple test to show the difference between the
constant time-step and the adaptive time-step approach. The set-up is
as follows. We assume a homogeneous density field of $n_\mathrm{H}=1.92\times 10^{-4}$~cm$^{-3}$ (the mean density at $z=9$ in a  
$\Omega_\mathrm{b}=0.043$, $H_0=70$~km~s$^{-1}$ universe),  temperature
$T=100$~K and neutral fractions $x_{\mbox{\scriptsize{HI}}}=1$ and $x_{\mbox{\scriptsize{HeI}}}=1$. We put
a single source at the center of a three-dimensional volume of
size 4.2~Mpc. The
resolution used is $30^3$. One model (1A) uses a black body source
($50000$~K) representing a Pop~II galaxy and one model (1B) uses a
power-law source $L_{\nu}\propto \nu^{-1.5}$ representing quasar
emission.  Both models assume an ionizing photon emission rate of
$5\times 10^{55}$ s$^{-1}$.  The constant time-step approach uses a
time-step of $\Delta t=10$~Myr and the adaptive time-step approach uses $f=4$,
meaning that the hydrogen ionized fraction can change by $\sim$25\% in a
time-step. The choice of $f=4$ is discussed below. This test does not
use asynchronous evolution.

\begin{figure}
   \centering
   \includegraphics[width=0.5\textwidth]{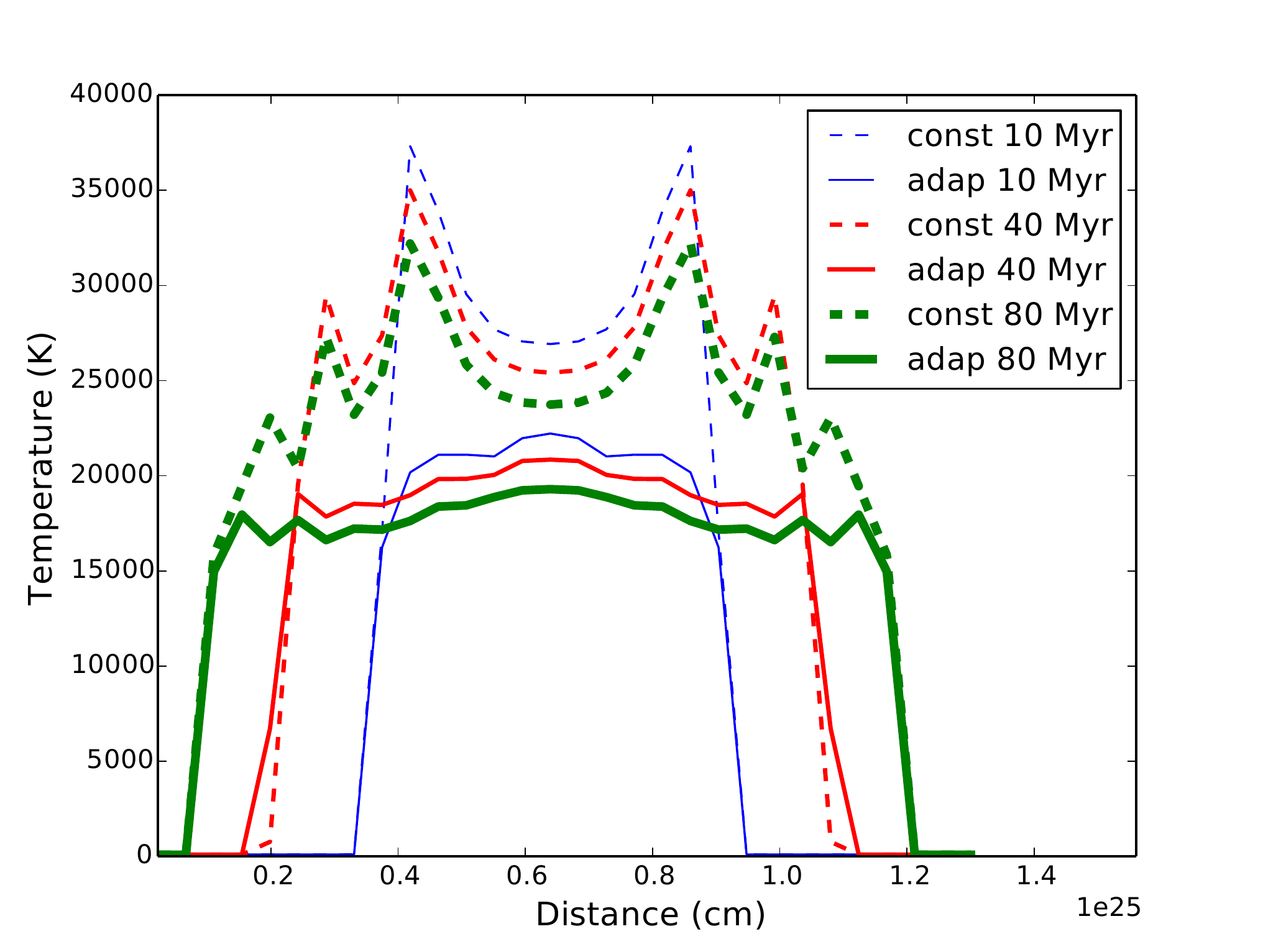}
   \caption{Test 1A: Thermal evolution of constant density with a
     black body source of 50000~K.  Blue, red and green lines (of
     increasing thickness) are the results for $t=10$~Myr, $t=40$~Myr,
     $t=80$~Myr, respectively.  Solid lines show the results from a
     simulation using the adaptive time-step approach and dashed lines
     from a simulation using a constant time-step of $\Delta
     t=10$~Myr. Note the wavy temperature profiles for the constant
     time-step model.}
   \label{fig:Test1_bb}
\end{figure}

\begin{figure}
   \centering
   \includegraphics[width=0.5\textwidth]{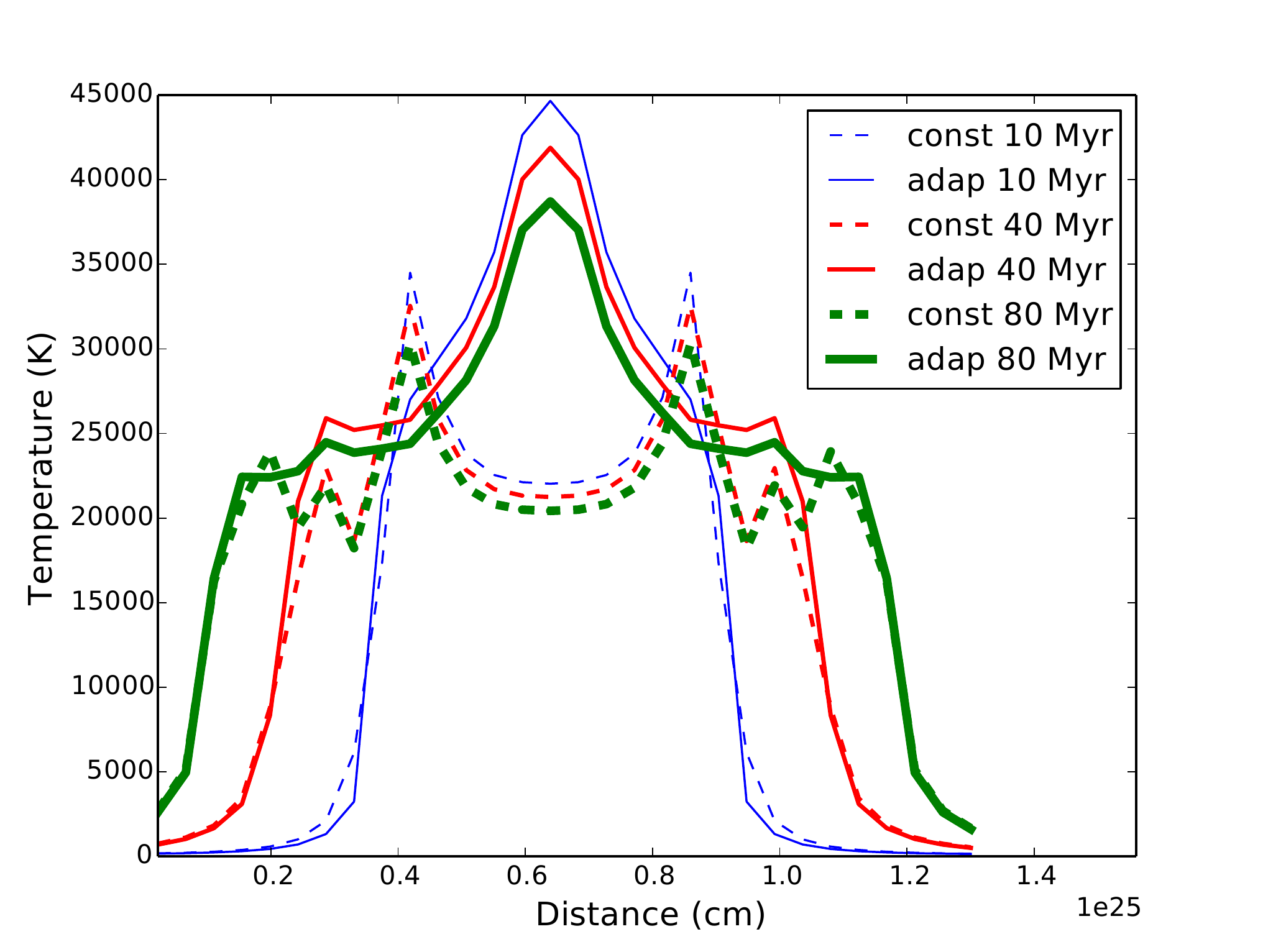}
   \caption{Same as Figure~\ref{fig:Test1_bb}, but for test 1B: a power-law source with index 1.5.}
   \label{fig:Test1_pl}
\end{figure}

Figures~\ref{fig:Test1_bb} and \ref{fig:Test1_pl} illustrate the
evolution of the temperature along a line going through the source
location. The different curves show the temperature profiles at 10~Myr
(blue thin line), 40~Myr (red thicker line) and 80~Myr (green thickest line). The constant
time-step approach (dashed lines) shows temperature spikes at several
positions.  These positions correspond to the positions of the
ionization front after each time-step (10~Myr).  These temperature
spikes are the result of incorrect photo-heating rates, as
explained in Section~\ref{sec:effect_of_time_step}.  The adaptive
time-step approach (solid lines) shows smoother temperature
profiles. These profiles peak at the source and decrease outward.
The small wave-like structures are the result of slight
deviations of photo-heating rate estimates from the correct
photo-heating rate.

\begin{figure}
   \centering
   \includegraphics[width=0.5\textwidth]{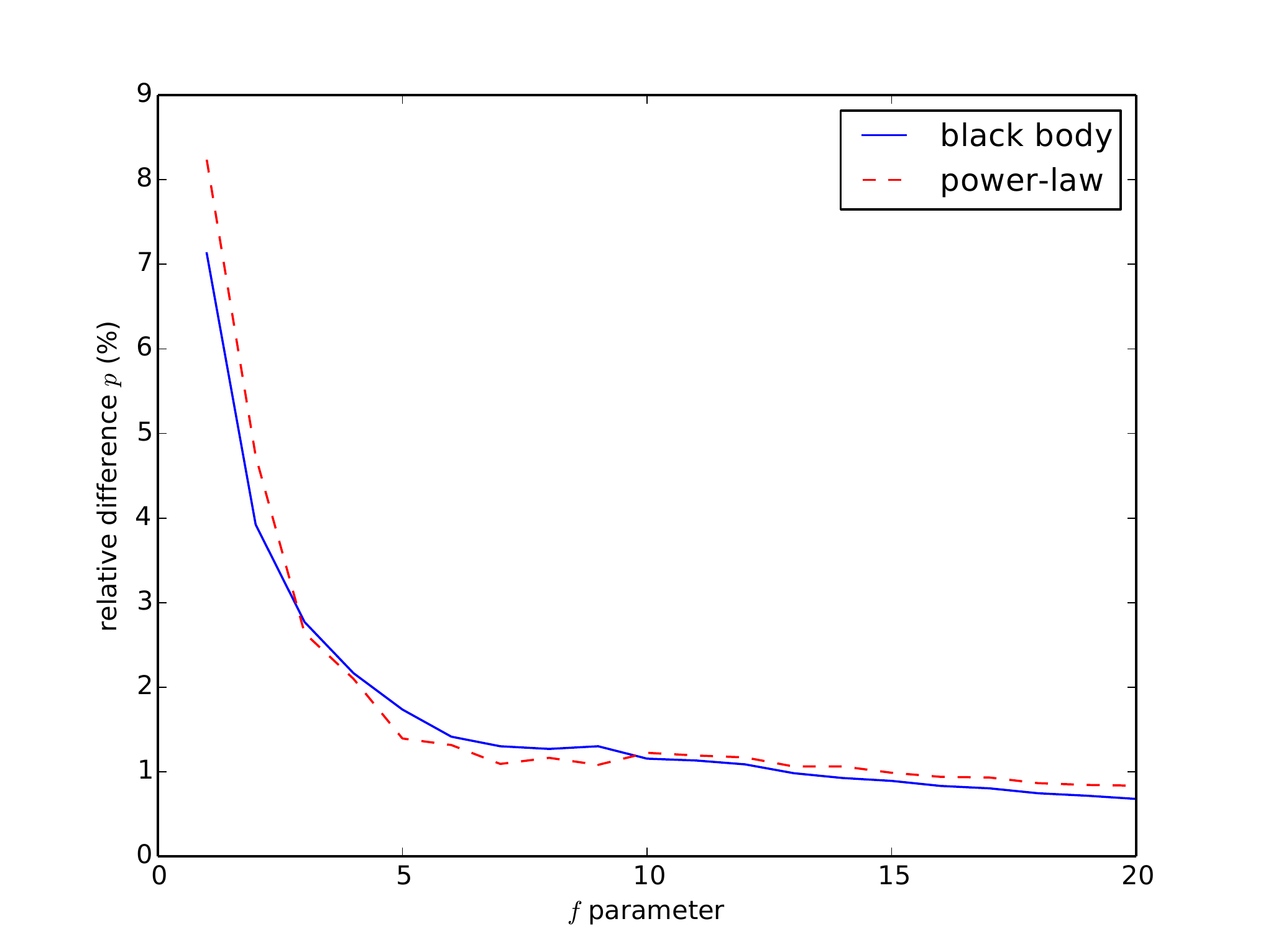}
   \caption{Absolute value of the relative difference $p$ between the
     results of Test 1 at time 80~Myr for the constant time-step
     approach (with time-step $5\times 10^{-4}$~Myr) and the adaptive
     time-step approach ($1\leq f\leq 20$). The absolute value of the
     relative difference is calculated relative to the constant
     time-step simulation result. The $p$ values of model (1A) (blue solid line)
     and model (1B) (red dashed line) are shown.  }
   \label{fig:Test1_f}
\end{figure}

To determine which $f$ values produce accurate results, we explore a
range of $f$ values (1 - 20) as well as a constant time-step simulation with a
time-step of $5\times 10^{-4}$~Myr. 
We use
the absolute value of the relative difference $p$ between the
temperature fields of the $\Delta t=5\times 10^{-4}$~Myr and the
adaptive time-step simulations as a measure of the accuracy of the
latter. This quantity $p$ is calculated from
\begin{equation}
p=\frac{\sum_{x,y,z} \vert T_{a}(x,y,z)-T_{c}(x,y,z)\vert}{\sum_{x,y,z} T_{c}(x,y,z)},
\end{equation}
where $T_a$ and $T_c$ are the temperature fields of adaptive and
constant time-step approaches respectively.  The results for $p$ of model (1A)
and (1B) for output time 80~Myr and $1\leq f\leq 20$ are displayed in
Figure~\ref{fig:Test1_f}.  For $f=1$, the $p$ value is the highest for
both models (7.12\% for model (1A) and 8.24\% for model (1B)). The $p$ value
declines when $f$ increases. $p$ becomes less than $\sim$2\% when
$f>4$ and after that only slowly declines for larger $f$.  We
choose $f=4$ which corresponds to $p\sim 2\%$ to be the modest choice
and the results for $f=4$ are shown in Figure~\ref{fig:Test1_bb}
and \ref{fig:Test1_pl}.

\subsection{Test 2 - Efficiency test}

The next test serves to illustrate the efficiency of our new method.
In a homogeneous density field with the same parameters, size and
resolution as in Test 1, we place 27 power-law sources. The
  sources are placed in a regular grid of $3\times 3\times 3$ with
  one source in the centre and the others located on the faces and
  corners of the volume.  This configuration ensures that after
  $80$~Myr the grid has just achieved a fully ionized state, mimicking
  the end of reionization.
  We use the same thermal equilibrium criteria as stated in
  Section~\ref{sec:Criteria of thermal equilibrium} for the power-law
  source. We compare the computational time required to run the test
with and without asynchronous evolution.  A range of $f$ parameters
are used in the test to show its influence.

\begin{figure}
   \centering
   \includegraphics[width=0.5\textwidth]{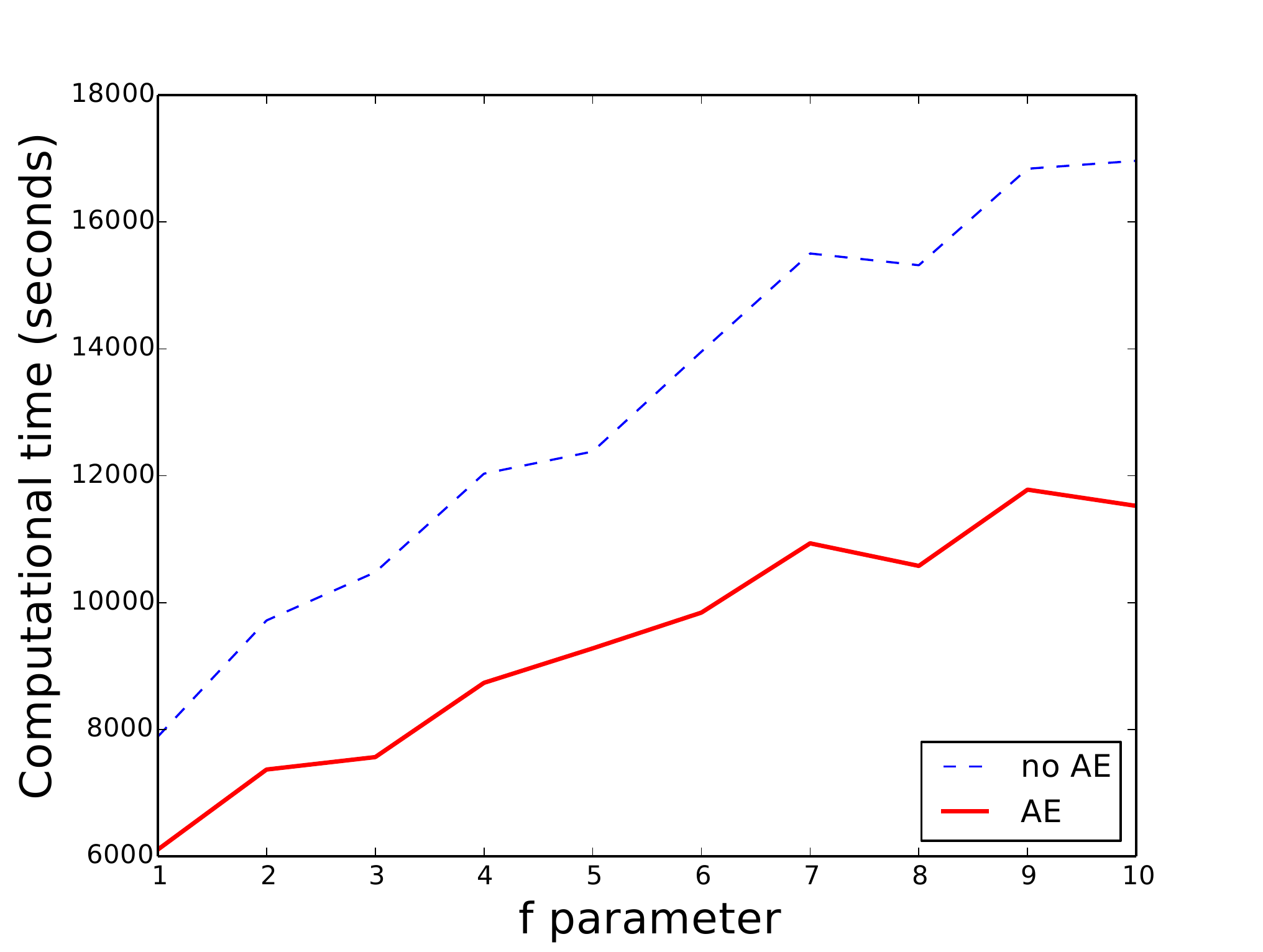}
   \caption{Test 2: Comparison of the computational time between a
     simulation with and without asynchronous evolution. The blue dashed line
     shows the results for the case without asynchronous evolution and
     the red solid line the same for the case with asynchronous
     evolution. For this test, the gain is between 33\% and 40\%.}
   \label{fig:Test2_efficiency}
\end{figure}

Figure~\ref{fig:Test2_efficiency} shows the computational time as
a function of $f$ for both cases. The mostly positive slope of both
curves is expected because higher $f$ values imply shorter
time-steps and therefore more computational time.  
The model not using
asynchronous evolution clearly requires more computational
time than the model using using asynchronous evolution, as expected.
The gain is between 33\% and 40\%, with slightly higher values corresponding
to larger values of $f$.

\subsection{Test 3 - Multiple sources in a cosmological field}

This test is the standard cosmological density field test from the
Cosmological Radiative Transfer Code Comparison Project
\citep{comparison}. The simulation volume contains a cosmological
density field of size €$0.5\,h^{-1}$ comoving Mpc.  The resolution is
$128^3$ cells. The initial temperature of the gas is $T=100$~K.
Sixteen ionizing sources are placed at the location of the 16 most
massive haloes in the simulation volume.  Each source produces
$f_{\gamma}=250$ ionizing photons per atom over a time interval
$t_s=3$ Myr. The ionizing photon production rate of each source is 
calculated from
\begin{equation}
\dot{N}_{\gamma}=f_{\gamma}\frac{M\Omega_\mathrm{b}}{\Omega_\mathrm{m}m_\mathrm{p}t_\mathrm{s}}\,,
\end{equation}
with $M$ the halo mass and $m_\mathrm{p}$ the mass of a proton. All sources
have a black body spectrum with $T_\mathrm{eff}=10^5$~K. The
cosmological parameters used are $\Omega_{\Lambda}=0.73$,
$\Omega_\mathrm{m}=0.27$, $\Omega_\mathrm{b}=0.043$, $h=0.7$.  The total evolution time is
0.40 Myr. 
For our asynchronous evolution approach we use
$x_{\mbox{\scriptsize{HI}}}\leq 1\times 10^{-4}$ and
$x_{\mbox{\scriptsize{HeII}}}\leq 3\times 10^{-4}$ as the criteria for
equilibrium of the ionized cells.

\begin{figure*}
   \centering
   \includegraphics[width=\textwidth]{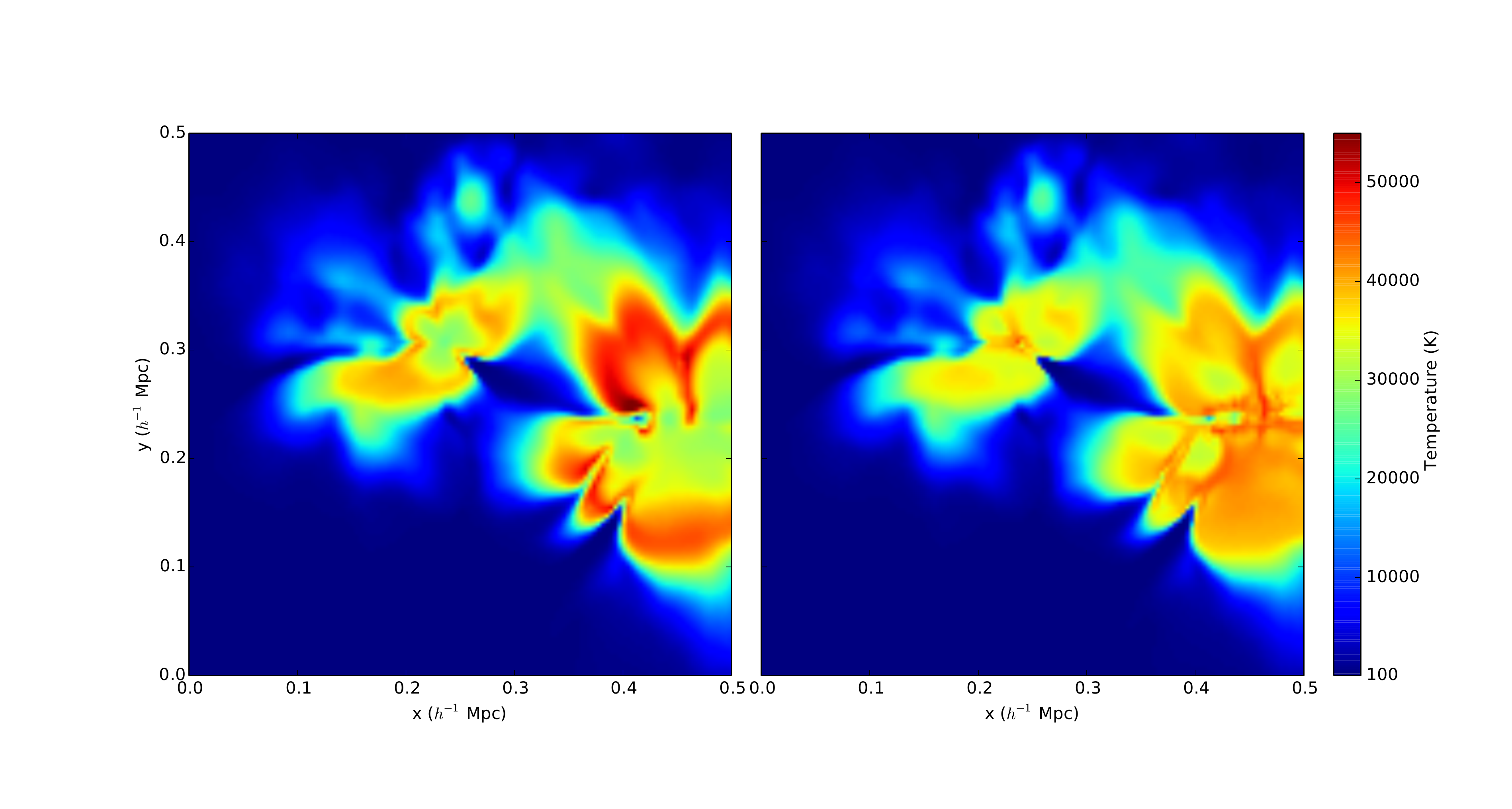}
   \caption{Test 3: Temperature structure cross section at
     $t=0.05$~Myr.  Left panel: constant time-step ($\Delta t=
     5.0\times 10^{-2}$~Myr). Right panel: adaptive time-step ($f=2$).}
   \label{fig:Test4_3Dcomparison_005Myr}
\end{figure*}

\begin{figure*}
   \centering
   \includegraphics[width=\textwidth]{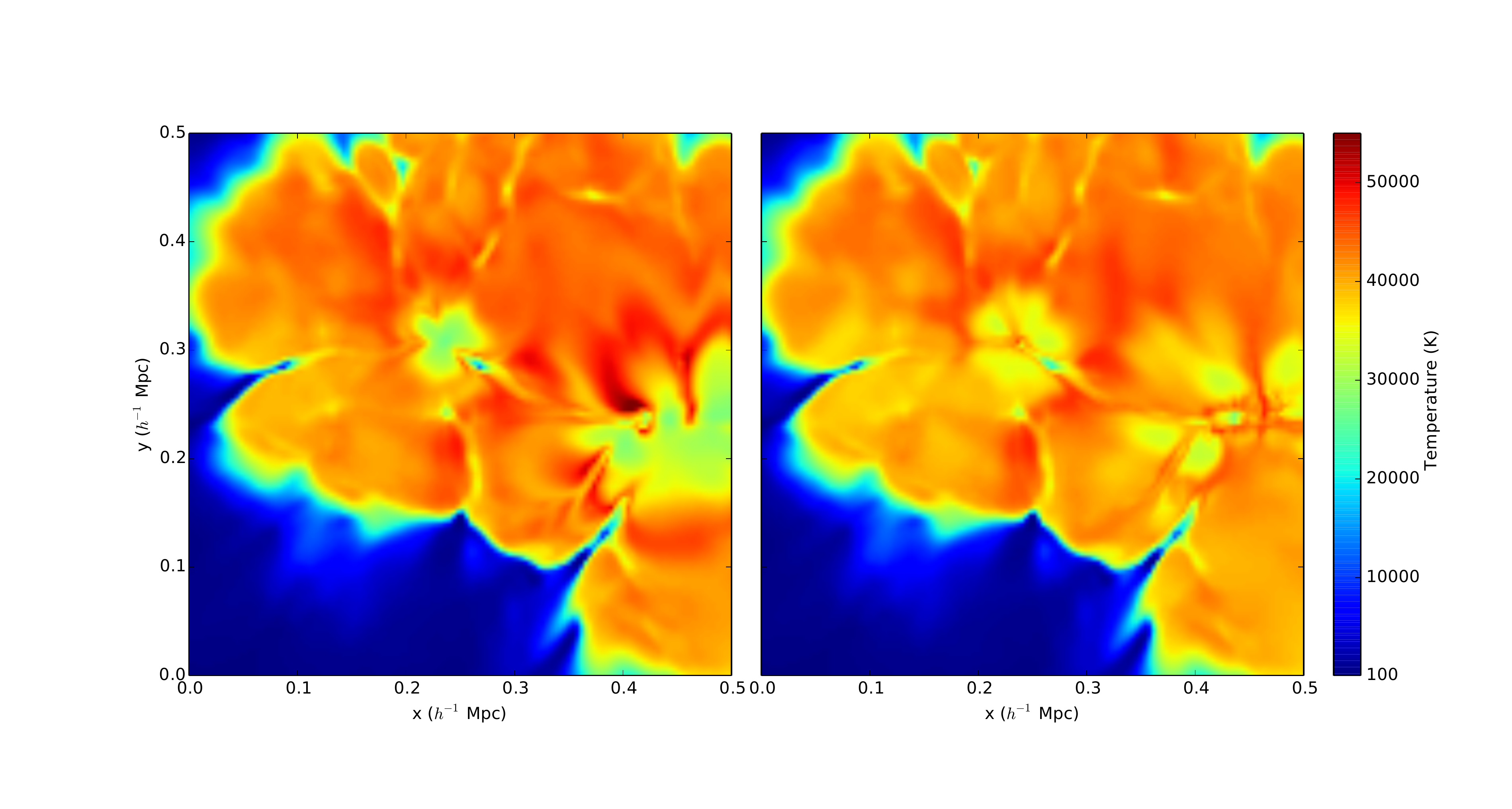}
   \caption{Test 3: Temperature structure cross section at
     $t=0.20$~Myr.  Left panel: constant time-step ($\Delta t=
     5.0\times 10^{-2}$~Myr). Right panel: adaptive time-step ($f=2$).}
   \label{fig:Test4_3Dcomparison_020Myr}
\end{figure*}

First, we compare the thermal results between a constant time-step
approach and the adaptive time-step approach. For the constant
time-step case we use a time-step of $0.05$ Myr. For the adaptive
time-step approach we use $f=2$.  Figures
\ref{fig:Test4_3Dcomparison_005Myr} and
\ref{fig:Test4_3Dcomparison_020Myr} show our temperature results of a
central slice at $t=0.05$ Myr and $0.2$ Myr, respectively.

Figure \ref{fig:Test4_3Dcomparison_005Myr} shows the typical butterfly
shape of the temperature distribution associated with this test. The
left plot displays the result of a single time-step evolution
(constant time-step approach). We see that the temperature has its
local maximum at the ionization front similar to what we saw in the
results for Test 1 above.  The right plot shows the result of adaptive
time-step approach.  Here there is no pronounced thermal maximum at
the ionization front and no abrupt change in the thermal structure in
the ionized regions.

Figure \ref{fig:Test4_3Dcomparison_020Myr} shows the late time results
($t=0.2$ Myr). The left plot shows the temperature distribution after
4 constant time-steps. We can discern multiple sharp layers
corresponding to the different ionization front positions at those
four time-steps.  The right plot shows the result for the adaptive
time-step approach.  We see some thermal structures in the large
ionized bubble.  However, these structures do not trace previous
ionization front positions. Instead, they correlate with the gas
distribution in the sense that higher density regions attain higher
temperatures than lower density regions.

\begin{figure}
   \centering
   \includegraphics[width=0.5\textwidth]{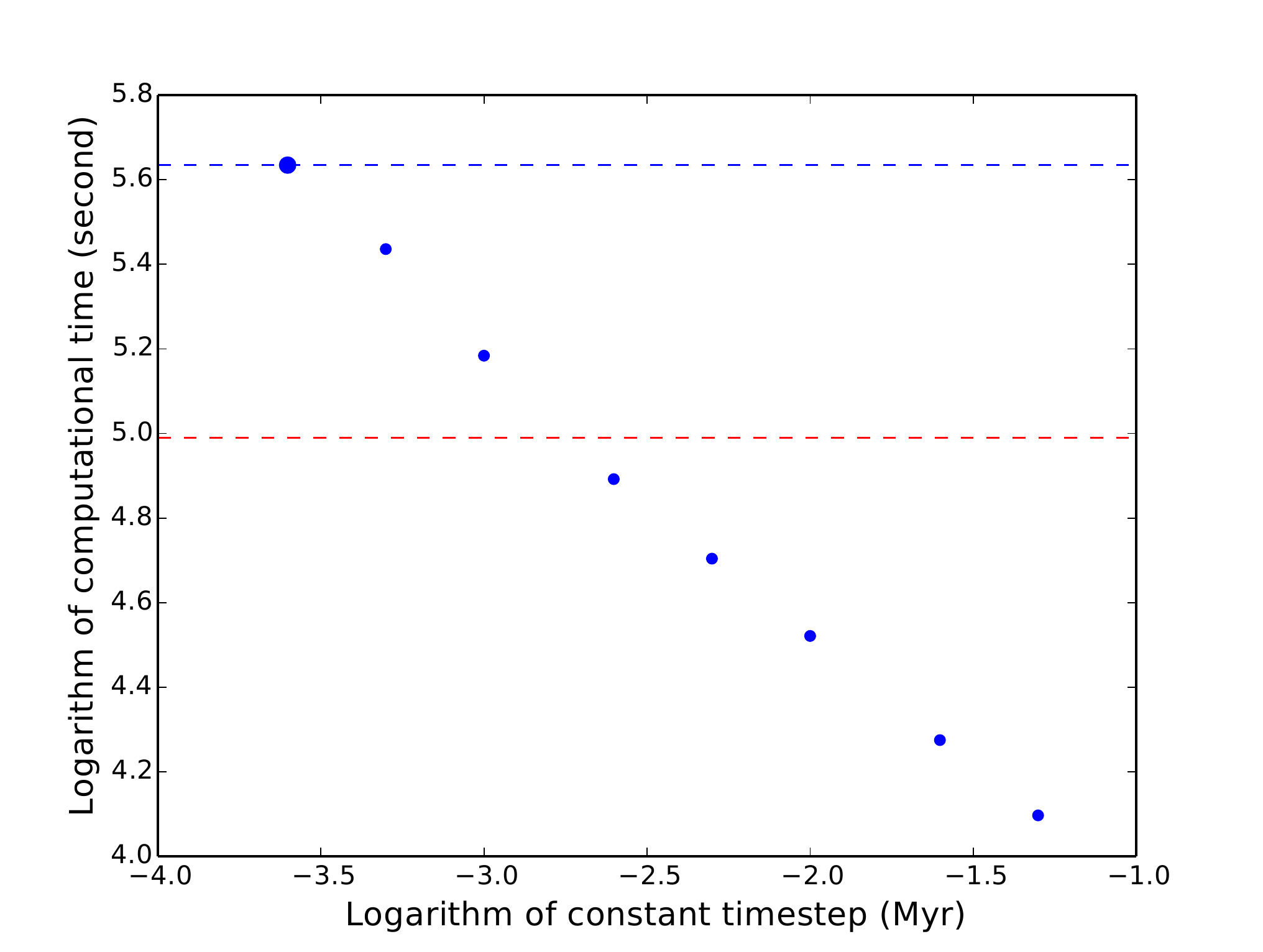}
   \caption{Test 3: A plot of computational time against different
     choices of constant time-steps. The 8 blue dots correspond to the
     $\Delta t$ values (from left to right) \{$2.5\times 10^{-4}$,
     $5.0\times 10^{-4}$, $1.0\times 10^{-3}$, $2.5\times 10^{-3}$,
     $5.0\times 10^{-3}$, $1.0\times 10^{-2}$, $2.5\times 10^{-2}$,
     $5.0\times 10^{-2}$\} Myr.  The red dashed and blue dashed lines
     (crossing the big blue dot) show the computational
     time required to run the adaptive time-step model and a
     converged constant time-step model, respectively.}
   \label{fig:Test4_computational_time}
\end{figure}

We have also used this test problem to evaluate the performance of the
adaptive time-step approach compared to the constant time-step
approach. We find that converged results require a value of
$\Delta t=2.5\times 10^{-4}$ Myr in case of the constant time-step
approach and a value of $f=2$ in case of the adaptive time-step
approach. The dots in Figure~\ref{fig:Test4_computational_time} show
the computational time consumed by adopting different constant
time-steps. The converged simulation which uses a constant time-step
of $2.5\times 10^{-4}$ Myr consumed 431056 seconds (CPU time) and is
indicated by a larger dot and a dashed line. The value for the
simulation with an adaptive time-step ($f=2$), 97636 seconds, is
indicated by the lower horizontal dashed line. Therefore we conclude
that for this problem the adaptive time-step approach is a factor 4.4
more efficient than the constant time-step approach.  From the figure
it can also be seen that the computational time consumed by the
adaptive time-step approach is equivalent to the case of a constant
time-step $2.1\times 10^{-3}$ Myr.

\begin{figure}
   \centering
   \includegraphics[width=0.5\textwidth]{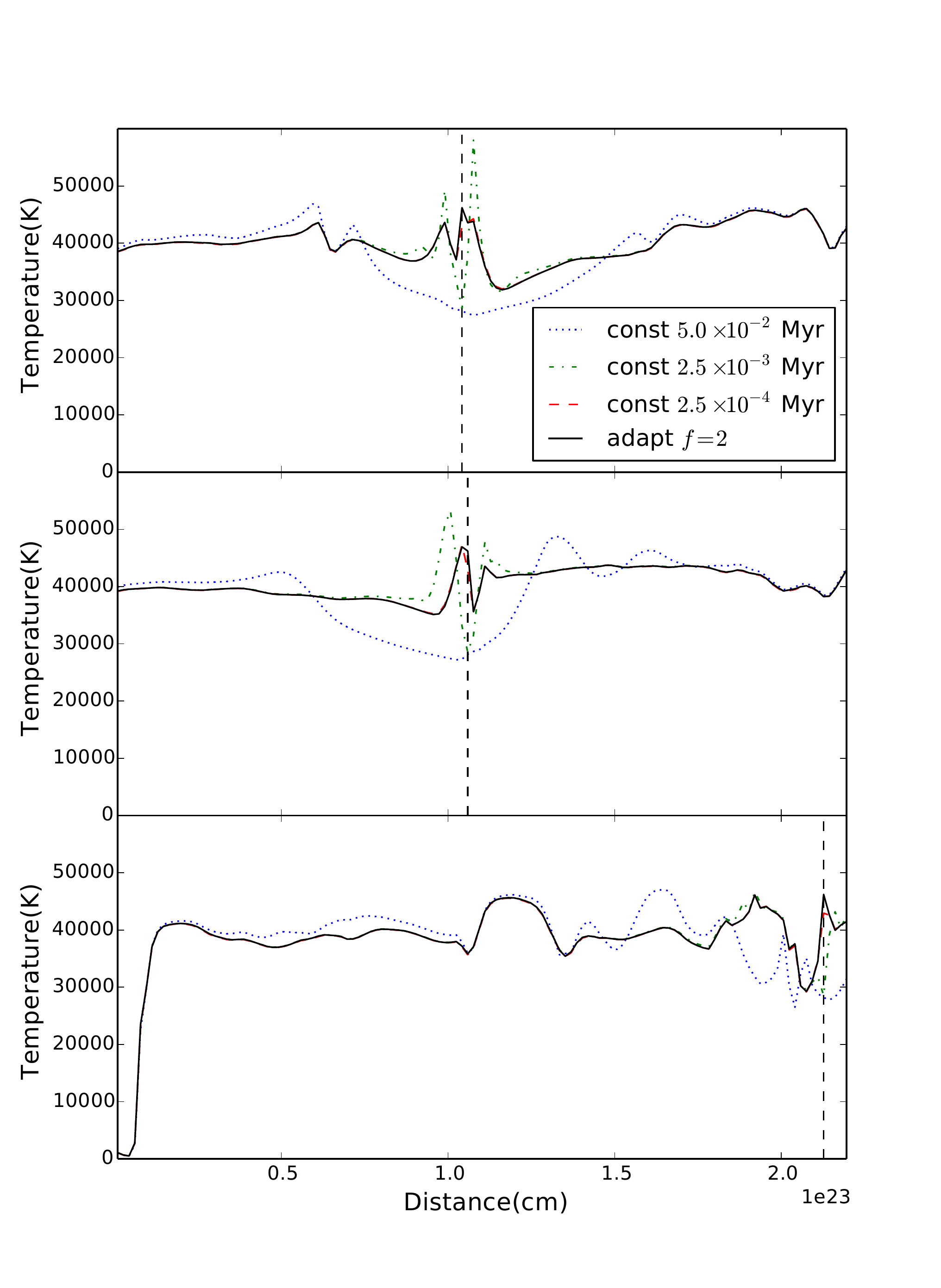}
   \caption{Test 3: Temperature structures along three perpendicular
     directions through one of the sixteen sources at $t=0.40$~Myr.
     Blue dotted line: $\Delta t=0.05$~Myr, green dot-dashed line: $\Delta t=2.5\times
     10^{-3}$~Myr, red dashed line: $\Delta t=2.5\times 10^{-4}$~Myr, black
     solid line: adaptive time-step with $f=2$.}
   \label{fig:Test4_1Dcomparison_040Myr}
\end{figure}

Lastly, we compare the temperature results of three constant time-step
simulations ($\Delta t=0.05, 2.5\times 10^{-3}$ and $2.5\times
10^{-4}$~Myr) with those from the adaptive time-step simulation.
Figure~\ref{fig:Test4_1Dcomparison_040Myr} shows the temperature
profiles along the three different coordinate directions traversing
one of the sources. The long constant time-step case ($\Delta
t=0.05$~Myr, blue dotted line) produces a large scale uneven thermal
distribution unrelated to the gas density. The converged case 
($\Delta t=2.5\times 10^{-4}$~Myr, red dashed line) and the adaptive time-step
simulation (black solid line) produce essentially the same thermal
profiles. The constant time-step case with an equivalent computational
time to the adaptive time-step simulation ($\Delta t=2.5\times
10^{-3}$~Myr, green dot-dashed line) produces better results than the long
time-step case but still does not produce a correct thermal profile in
a region approximately one-fifth of the box size centered around the
source. This shows the advantage of the adaptive time-step approach
over the constant time-step approach.

\subsection{Test 4 - \textsc{C}$^2$\textsc{-Ray} versus \textsc{SNC}}

Lastly, we compare the \textsc{C}$^2$\textsc{-Ray} results with those
obtained with a different photo-ionization code, namely \textsc{SNC}.
The base of this code is briefly described in \citet{lun88}, while
updates were given in \citet{lun91,lun96} and
\citet{matt10}. \textsc{SNC} is a time-dependent photo-ionization code
which includes elements from hydrogen to iron, with several of the ions
being treated with multi-level atoms. It has been mainly used to
calculate the time-dependent ionization and temperature of gas around
supernovae and in supernova remnants, but has also been recently used in a
cosmological application to study time-dependent effects of ionization
by Population III stars \citep{ryd13}. The code generally deals with
spherically symmetric nebulae.

Time-steps in \textsc{SNC} are calculated for each radial shell, based
on the accepted fastest changes in degrees of ionization/recombination
and temperature. Normally, an accepted change is a few per cent per
time-step for the ionization and about a factor of ten smaller change
in temperature. A time-step also cannot be longer than the
light-crossing time of a shell. Just as the new version of
\textsc{C}$^2$\textsc{-Ray} the conditions in each shell are not
calculated every time-step.  However, even if the selected time-step
for a particular shell could formally be substantially longer than at
the ionization front, the largest time step allowed for all shells 
is $k$ times the shortest time step; here we use $k=7$.

In terms of the atomic physics, both \textsc{C}$^2$\textsc{-Ray} and
\textsc{SNC} calculate all ions time-dependently. However, \textsc{SNC} also
treats all included levels of H{\sc I}, He{\sc I} and He{\sc II}
time-dependently. This is however, of minor importance for cosmological
applications.  Of greater importance is the treatment of the diffuse
emission. For this \textsc{SNC} uses a modified on-the-spot approximation
in which recombination photons which can escape from the local environment are
allowed to do so. If they can not, they are absorbed locally. However, escaping
recombination photons are assumed to escape from the entire system and
cannot be absorbed elsewhere.

The \textsc{SNC} code treats photon escape in spectral lines using
escape probabilities. For spectral lines capable of ionizing H{\sc I}
or He{\sc I}, the probability of the photon being trapped by the
continuum is also calculated.  The fractions absorbed by the continuum
are included in the ionization balance, and the ones scattered by
lines are fed into the calculations of the level populations of the
multi-level atoms of H and He. For the results discussed below, no
velocity field is assumed, so the only broadening mechanisms are
thermal broadening and natural line broadening. 

To calculate the
collisional excitation rates of H and He, \textsc{SNC} has since
\citet{lun96} been updated with the effective collision strengths of
\citet{and00} and \citet{kis95} for H{\sc I} and He{\sc II},
respectively. \textsc{SNC} also includes heating, excitation and
further ionization due to fast photo-electrons, just as
\textsc{C}$^2$\textsc{-Ray} does. \textsc{SNC} does this by using
interpolation between the calculated Spencer-Fano results of
\citet{koz96}, where the inputs are the local number densities of
electrons, H{\sc I} and He{\sc I}.

To compare the results of \textsc{C}$^2$\textsc{-Ray} and \textsc{SNC},
we have performed a 1D simulation of a spherical nebula with a particle density of
$7\times 10^{-5}$~cm$^{-3}$ and consisting of He and H with a number
density ratio of 0.08. The domain size is 3~Mpc,
divided into 1000 cells. The value of $f$ used is 2 and for SNC
the time steps are calculated according to the criteria outlined above.
The initial temperature is 100~K. For
the ionizing spectrum we use a power-law source with energy index
1.5. The total rate of ionizing photons is $10^{56}$~s$^{-1}$. We
evolve the nebula for $10$~Myrs.

\begin{figure}
   \centering
   \vskip -1cm
   \includegraphics[width=0.5\textwidth]{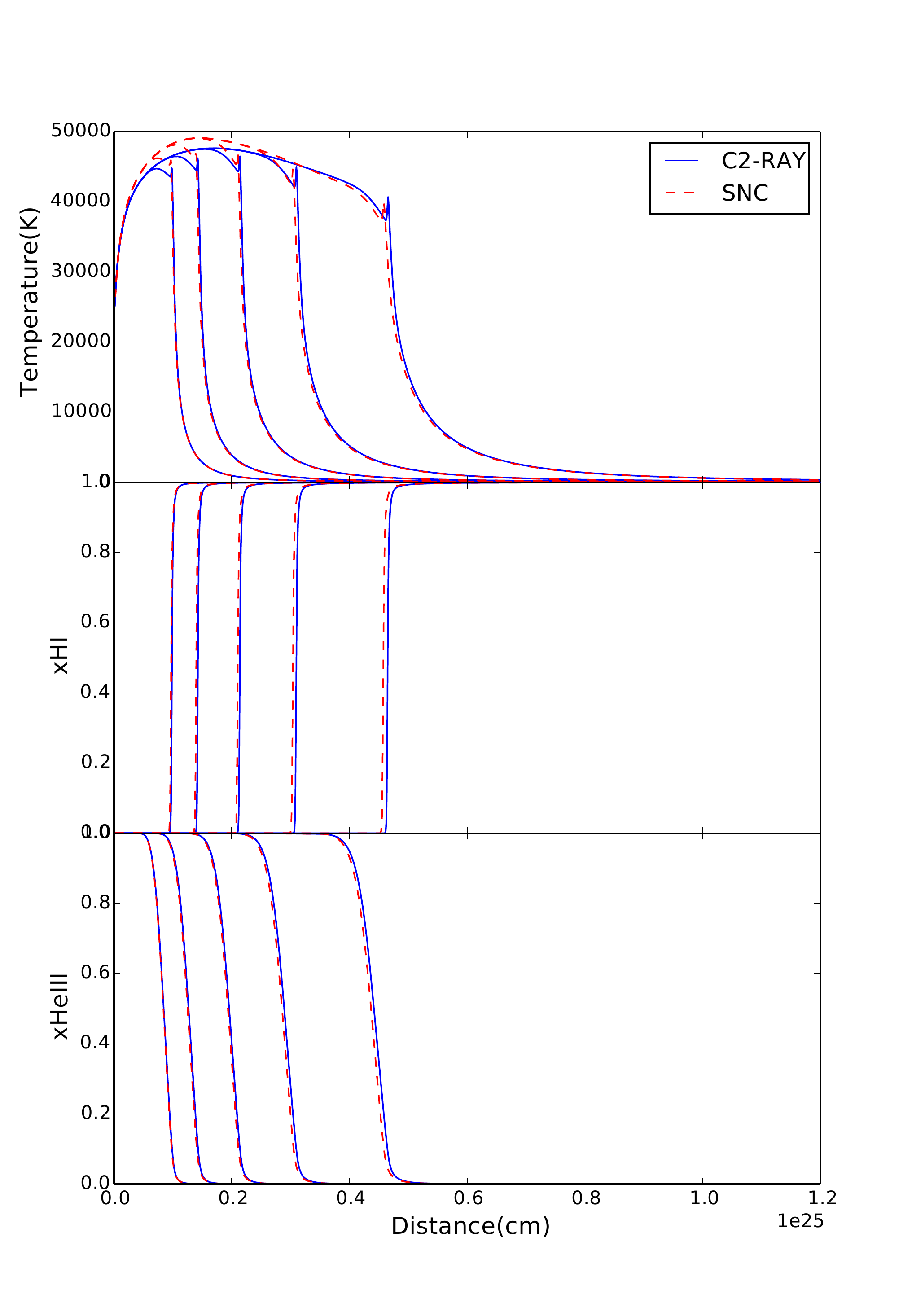}
   \caption{Test 4: Comparison of the \textsc{C}$^2$\textsc{-Ray} and the \textsc{SNC} results for the temperature, $x_{\mbox{\scriptsize{HI}}}$ and  $x_{\mbox{\scriptsize{HeIII}}}$ evolution. 
Five time instances are shown in the plots. They correspond to quasar ages of $0.1, 0.3, 1.0, 3.0, 10.0$ Myr.}
   \label{fig:Test3_T}
\end{figure}

Figure~\ref{fig:Test3_T} shows
our results for the thermal and ionization
evolution at $0.1, 0.3, 1.0, 3.0, 10.0$ Myr. Concentrating first on
the temperature results, we see that 
the general shapes of the thermal profiles
are similar. The temperature climbs to its maximum in the region between
the source and the He{\sc III} ionization front.
The value of the peak temperature in the \textsc{SNC}
results is some 4\% higher than in \textsc{C}$^2$\textsc{-Ray}, which
is the largest difference seen between the two results.

The lower temperatures close to the source are caused by the
lower column densities there. Due to these the gas close to the
source encounters a larger number of low energy photons. As these
are preferentially absorbed due to their higher ionization cross
section, this leads to lower temperatures in the innermost parts of
the ionized region. This is seen in both simulations.  Outward from
the peak, the temperature decreases because of the presence of
substantial amounts of helium gas that have not been fully ionized and
heated yet. Both sets of results show this feature. The
temperature decreases rapidly at the H{\sc II} region boundary and
exhibits a smoothly decreasing tail. This gas outside the H{\sc II}
region is not ionized but heated by the high energy photons which have
a high mean free path in the neutral gas. Both experiments
capture the transfer of high energy photons in the neutral gas and the
thermal response when these photons are gradually absorbed by the gas.

The lower two panels of Figure \ref{fig:Test3_T} show the
evolution of $x_{\mbox{\scriptsize{HI}}}$ and
$x_{\mbox{\scriptsize{HeIII}}}$ respectively.  The
$x_{\mbox{\scriptsize{HI}}}$ evolution of both experiments
agree.  Although \textsc{SNC} gives a slightly smaller H{\sc II}
region than \textsc{C}$^2$\textsc{-Ray}, their profiles at
$x_{\mbox{\scriptsize{HI}}}\ge 0.99$ coincide with each other. 
The agreement on the $x_{\mbox{\scriptsize{HeIII}}}$ evolution is
excellent and the profiles are almost indistinguishable. 

Overall, both sets of results show a high level of consistency
with each other. We therefore conclude that the results of both codes
agree with each other.  Given the different algorithms used in the two
codes, as well as their different origins, this increases the
likelihood that the results are correct.

\section{Conclusions}
\label{sec:Conclusions}

Especially for photo-ionization calculations of the low density IGM, an
accurate determination of the photo-heating rate is important. Since the
photo-heating rate depends sensitively on the energy of the photons
responsible for the photo-ionization, an accurate calculation of the
photo-heating rate is more challenging than an accurate calculation of the
photo-ionization rate. Specifically, the large time-step approach used
by previous versions of \textsc{C}$^2$\textsc{-Ray} provides an
accurate determination of the photo-ionization rates but does not yield
accurate photo-heating rates. The latter requires time-steps which
temporally resolve the ionization of each cell.

We present a new version of \textsc{C}$^2$\textsc{-Ray} which
calculates the required time-step on the fly and at the same time
implements two methods to reduce the computational costs
associated with this. Our adaptive time-step algorithm calculates an
optimal time-step by only considering the relevant cells for this
calculation. These cells are identified by a fast and parallelizable
ray-tracing method based on a pyramidal decomposition of the
volume. To further reduce the computational cost, the code also uses
asynchronous evolution which means that different cells are evolved
with different time-steps. Specifically, cells which have already been
ionized are evolved with much longer time-steps than the cells which
still need to be ionized. The clocks of the different cells are
synchronized at the end of each simulation step.

These techniques ensure that most of the computational resources are
allocated to cells that potentially undergo substantial thermal evolution. The
updated version of the code retains the efficient parallelization of
the previous versions, using a combination of distributed and shared
memory parallelization.

We have tested the new version of \textsc{C}$^2$\textsc{-Ray} against
the older version and showed that it achieves the same accuracy at a
much lower computational cost. We also compare the results to another
time-dependent photo-ionization code (\textsc{SNC}) and show that the two
codes show good agreement both on the ionization fractions and the
temperature.

The new version constitutes a major improvement over the previous
versions. However, it is important to point out that this improvement
only affects calculations of the temperature evolution. To date, all
\textsc{C}$^2$\textsc{-Ray} simulations of reionization did not
consider the thermal evolution and therefore those results remain
valid \citep[e.g.][]{21cm2,2012MNRAS.424..762D,
  2014MNRAS.439..725I}. Since the recombination rates only weakly
depend on the temperature, the thermal state of the gas does not
influence the reionization process much. Also, under the assumption of
a spin temperature well above the CMB temperature, the redshifted 21cm
signal does not actually depend on the gas temperature
\citep{2006PhR...433..181F}. Thermal calculations are therefore not
necessary for the later stages of reionization.  The new version of
\textsc{C}$^2$\textsc{-Ray} is useful to explore the earlier phases of
reionization when hard photons from x-ray sources had not yet fully
raised the temperature of the IGM above the CMB temperature as well as
other problems in which the temperature of the IGM is important, for
example the structure of near zones around high redshift quasars.

We note that the pyramidal ray tracing scheme may have applications
beyond the one presented here. It is a very efficient method of finding
cells nearest to a given position which have a certain property. For
example, it could be employed to find the probability distribution function of
distances to edges of ionized regions, one of the methods to characterize
the size distribution of ionized regions in reionization simulations
\citep{2007ApJ...669..663M}

This paper contains a comparison of the new version of 
\textsc{C}$^2$\textsc{-Ray} with the time-dependent photo-ionization
code \textsc{SNC}. However, it would be
very useful to compare a wider range of codes developed for
photo-ionization calculations of the IGM in the context of a code
comparison. As the hydrogen-only results of the Cosmological Radiative
Transfer Comparison Project show \citep{comparison}, there can be
considerable spread in the temperature results. A more detailed
evaluation, including not only the temperature, but for example also
the photo-heating rates, would be useful, both for the existing codes
as well as for the development of new codes.


\section*{Acknowledgments} 
We thank the referee, Joakim Rosdahl, for his thorough review of the
paper which has led to substantial improvements. This study was
supported in part by the Swedish Research Council grants 2012-4144 (PI
Mellema) and 2011-5251 (PI Lundqvist).

\bibliographystyle{mn2e}
\bibliography{paper,extra}

\begin{thebibliography}{47}
\expandafter\ifx\csname natexlab\endcsname\relax\def\natexlab#1{#1}\fi

\bibitem[{Anderson {et~al}\mbox{.}(2000)Anderson, Ballance, Badnell, \&
  Summers}]{and00}
Anderson H., Ballance C.~P., Badnell N.~R., Summers H.~P., 2000, \jpb, 33, 1255

\bibitem[{{Bolton} {et~al}\mbox{.}(2012){Bolton}, {Becker}, {Raskutti},
  {Wyithe}, {Haehnelt}, \& {Sargent}}]{Bolton_temperature2}
{Bolton} J.~S., {Becker} G.~D., {Raskutti} S., {Wyithe} J.~S.~B., {Haehnelt}
  M.~G., {Sargent} W.~L.~W., 2012, \mnras, 419, 2880

\bibitem[{{Bolton} {et~al}\mbox{.}(2010){Bolton}, {Becker}, {Wyithe},
  {Haehnelt}, \& {Sargent}}]{Bolton_temperature}
{Bolton} J.~S., {Becker} G.~D., {Wyithe} J.~S.~B., {Haehnelt} M.~G., {Sargent}
  W.~L.~W., 2010, \mnras, 406, 612

\bibitem[{{Bolton} \& {Haehnelt}(2007)}]{Bolton_z6}
{Bolton} J.~S., {Haehnelt} M.~G., 2007, \mnras, 374, 493

\bibitem[{{Bolton} {et~al}\mbox{.}(2011){Bolton}, {Haehnelt}, {Warren},
  {Hewett}, {Mortlock}, {Venemans}, {McMahon}, \& {Simpson}}]{Bolton_z7085}
{Bolton} J.~S., {Haehnelt} M.~G., {Warren} S.~J., {Hewett} P.~C., {Mortlock}
  D.~J., {Venemans} B.~P., {McMahon} R.~G., {Simpson} C., 2011, \mnras, 416,
  L70

\bibitem[{{Cantalupo} \& {Porciani}(2011)}]{RADAMESH}
{Cantalupo} S., {Porciani} C., 2011, \mnras, 411, 1678

\bibitem[{{Cen} {et~al}\mbox{.}(2009){Cen}, {McDonald}, {Trac}, \&
  {Loeb}}]{Cen2009}
{Cen} R., {McDonald} P., {Trac} H., {Loeb} A., 2009, \apjl, 706, L164

\bibitem[{{Datta} {et~al}\mbox{.}(2012){Datta}, {Friedrich}, {Mellema},
  {Iliev}, \& {Shapiro}}]{2012MNRAS.424..762D}
{Datta} K.~K., {Friedrich} M.~M., {Mellema} G., {Iliev} I.~T., {Shapiro} P.~R.,
  2012, \mnras, 424, 762

\bibitem[{{Fan} {et~al}\mbox{.}(2006){Fan}, {Strauss}, {Becker}, {White},
  {Gunn}, {Knapp}, {Richards}, {Schneider}, {Brinkmann}, \& {Fukugita}}]{Fan}
{Fan} X. {et~al.}, 2006, \aj, 132, 117

\bibitem[{{Ferland} {et~al}\mbox{.}(2013){Ferland}, {Porter}, {van Hoof},
  {Williams}, {Abel}, {Lykins}, {Shaw}, {Henney}, \&
  {Stancil}}]{2013RMxAA..49..137F}
{Ferland} G.~J. {et~al.}, 2013, \rmxaa, 49, 137

\bibitem[{{Friedrich} {et~al}\mbox{.}(2012){Friedrich}, {Mellema}, {Iliev}, \&
  {Shapiro}}]{C2-Ray2}
{Friedrich} M.~M., {Mellema} G., {Iliev} I.~T., {Shapiro} P.~R., 2012, \mnras,
  421, 2232

\bibitem[{{Furlanetto} {et~al}\mbox{.}(2006){Furlanetto}, {Oh}, \&
  {Briggs}}]{2006PhR...433..181F}
{Furlanetto} S.~R., {Oh} S.~P., {Briggs} F.~H., 2006, \physrep, 433, 181

\bibitem[{{Graziani} {et~al}\mbox{.}(2013){Graziani}, {Maselli}, \&
  {Ciardi}}]{CRASH3}
{Graziani} L., {Maselli} A., {Ciardi} B., 2013, \mnras, 431, 722

\bibitem[{{Hui} \& {Haiman}(2003)}]{Hui2003}
{Hui} L., {Haiman} Z., 2003, \apj, 596, 9

\bibitem[{{Iliev} {et~al}\mbox{.}(2006{\natexlab{a}}){Iliev}, {Ciardi},
  {Alvarez}, {Maselli}, {Ferrara}, {Gnedin}, {Mellema}, {Nakamoto}, {Norman},
  {Razoumov}, {Rijkhorst}, {Ritzerveld}, {Shapiro}, {Susa}, {Umemura}, \&
  {Whalen}}]{2006MNRAS.371.1057I}
{Iliev} I.~T. {et~al.}, 2006{\natexlab{a}}, \mnras, 371, 1057

\bibitem[{{Iliev} {et~al}\mbox{.}(2006{\natexlab{b}}){Iliev}, {Ciardi},
  {Alvarez}, {Maselli}, {Ferrara}, {Gnedin}, {Mellema}, {Nakamoto}, {Norman},
  {Razoumov}, {Rijkhorst}, {Ritzerveld}, {Shapiro}, {Susa}, {Umemura}, \&
  {Whalen}}]{comparison}
{Iliev} I.~T. {et~al.}, 2006{\natexlab{b}}, \mnras, 371, 1057

\bibitem[{{Iliev} {et~al}\mbox{.}(2014){Iliev}, {Mellema}, {Ahn}, {Shapiro},
  {Mao}, \& {Pen}}]{2014MNRAS.439..725I}
{Iliev} I.~T., {Mellema} G., {Ahn} K., {Shapiro} P.~R., {Mao} Y., {Pen} U.-L.,
  2014, \mnras, 439, 725

\bibitem[{{Iliev} {et~al}\mbox{.}(2006{\natexlab{c}}){Iliev}, {Mellema}, {Pen},
  {Merz}, {Shapiro}, \& {Alvarez}}]{2006MNRAS.369.1625I}
{Iliev} I.~T., {Mellema} G., {Pen} U.-L., {Merz} H., {Shapiro} P.~R., {Alvarez}
  M.~A., 2006{\natexlab{c}}, \mnras, 369, 1625

\bibitem[{{Iliev} {et~al}\mbox{.}(2012){Iliev}, {Mellema}, {Shapiro}, {Pen},
  {Mao}, {Koda}, \& {Ahn}}]{21cm2}
{Iliev} I.~T., {Mellema} G., {Shapiro} P.~R., {Pen} U.-L., {Mao} Y., {Koda} J.,
  {Ahn} K., 2012, \mnras, 423, 2222

\bibitem[{{Iliev} {et~al}\mbox{.}(2011){Iliev}, {Moore}, {Gottl{\"o}ber},
  {Yepes}, {Hoffman}, \& {Mellema}}]{2011MNRAS.413.2093I}
{Iliev} I.~T., {Moore} B., {Gottl{\"o}ber} S., {Yepes} G., {Hoffman} Y.,
  {Mellema} G., 2011, \mnras, 413, 2093

\bibitem[{{Iliev} {et~al}\mbox{.}(2009){Iliev}, {Whalen}, {Mellema}, {Ahn},
  {Baek}, {Gnedin}, {Kravtsov}, {Norman}, {Raicevic}, {Reynolds}, {Sato},
  {Shapiro}, {Semelin}, {Smidt}, {Susa}, {Theuns}, \&
  {Umemura}}]{2009MNRAS.400.1283I}
{Iliev} I.~T. {et~al.}, 2009, \mnras, 400, 1283

\bibitem[{Kisielius {et~al}\mbox{.}(1995)Kisielius, Berrington, \&
  Norrington}]{kis95}
Kisielius R., Berrington K.~A., Norrington P.~H., 1995, \jpb, 28, 2459

\bibitem[{{Kozma} \& {Fransson}(1998)}]{koz96}
{Kozma} C., {Fransson} C., 1998, \apj, 496, 946

\bibitem[{{Kunasz} \& {Auer}(1988)}]{short_characteristics}
{Kunasz} P., {Auer} L.~H., 1988, \jqsrt, 39, 67

\bibitem[{{Lundqvist} \& {Fransson}(1988)}]{lun88}
{Lundqvist} P., {Fransson} C., 1988, \aap, 192, 221

\bibitem[{{Lundqvist} \& {Fransson}(1991)}]{lun91}
{Lundqvist} P., {Fransson} C., 1991, \apj, 380, 575

\bibitem[{{Lundqvist} \& {Fransson}(1996)}]{lun96}
{Lundqvist} P., {Fransson} C., 1996, \apj, 464, 924

\bibitem[{{Mackey}(2012)}]{2012A&A...539A.147M}
{Mackey} J., 2012, \aap, 539, A147

\bibitem[{{Mattila} {et~al}\mbox{.}(2010){Mattila}, {Lundqvist},
  {Gr{\"o}ningsson}, {Meikle}, {Stathakis}, {Fransson}, \& {Cannon}}]{matt10}
{Mattila} S., {Lundqvist} P., {Gr{\"o}ningsson} P., {Meikle} P., {Stathakis}
  R., {Fransson} C., {Cannon} R., 2010, \apj, 717, 1140

\bibitem[{{Mellema} {et~al}\mbox{.}(2006){Mellema}, {Iliev}, {Alvarez}, \&
  {Shapiro}}]{C2-Ray}
{Mellema} G., {Iliev} I.~T., {Alvarez} M.~A., {Shapiro} P.~R., 2006, \na, 11,
  374

\bibitem[{{Mesinger} \& {Furlanetto}(2007)}]{2007ApJ...669..663M}
{Mesinger} A., {Furlanetto} S., 2007, \apj, 669, 663

\bibitem[{{Mesinger} \& {Haiman}(2007)}]{Mesinger07}
{Mesinger} A., {Haiman} Z., 2007, \apj, 660, 923

\bibitem[{{Norman} {et~al}\mbox{.}(1998){Norman}, {Paschos}, \&
  {Abel}}]{cosmological_radiative_transfer}
{Norman} M.~L., {Paschos} P., {Abel} T., 1998, \memsai, 69, 455

\bibitem[{{Paciga} {et~al}\mbox{.}(2013){Paciga}, {Albert}, {Bandura}, {Chang},
  {Gupta}, {Hirata}, {Odegova}, {Pen}, {Peterson}, {Roy}, {Shaw}, {Sigurdson},
  \& {Voytek}}]{2013MNRAS.433..639P}
{Paciga} G. {et~al.}, 2013, \mnras, 433, 639

\bibitem[{{Parsons} {et~al}\mbox{.}(2010){Parsons}, {Backer}, {Foster},
  {Wright}, {Bradley}, {Gugliucci}, {Parashare}, {Benoit}, {Aguirre}, {Jacobs},
  {Carilli}, {Herne}, {Lynch}, {Manley}, \& {Werthimer}}]{2010AJ....139.1468P}
{Parsons} A.~R. {et~al.}, 2010, \aj, 139, 1468

\bibitem[{{Pawlik} \& {Schaye}(2011)}]{TRAPHIC2}
{Pawlik} A.~H., {Schaye} J., 2011, \mnras, 412, 1943

\bibitem[{{Pritchard} \& {Furlanetto}(2007)}]{Pritchard2007}
{Pritchard} J.~R., {Furlanetto} S.~R., 2007, \mnras, 376, 1680

\bibitem[{{Rai{\v c}evi{\'c}} {et~al}\mbox{.}(2014){Rai{\v c}evi{\'c}},
  {Pawlik}, {Schaye}, \& {Rahmati}}]{recombination_radiation}
{Rai{\v c}evi{\'c}} M., {Pawlik} A.~H., {Schaye} J., {Rahmati} A., 2014,
  \mnras, 437, 2816

\bibitem[{{Raskutti} {et~al}\mbox{.}(2012){Raskutti}, {Bolton}, {Wyithe}, \&
  {Becker}}]{Raskutti_temperature}
{Raskutti} S., {Bolton} J.~S., {Wyithe} J.~S.~B., {Becker} G.~D., 2012, \mnras,
  421, 1969

\bibitem[{{Ricotti} {et~al}\mbox{.}(2002){Ricotti}, {Gnedin}, \&
  {Shull}}]{secondary_ionization}
{Ricotti} M., {Gnedin} N.~Y., {Shull} J.~M., 2002, \apj, 575, 33

\bibitem[{{Rydberg} {et~al}\mbox{.}(2013){Rydberg}, {Zackrisson}, {Lundqvist},
  \& {Scott}}]{ryd13}
{Rydberg} C.-E., {Zackrisson} E., {Lundqvist} P., {Scott} P., 2013, \mnras,
  429, 3658

\bibitem[{{Schroeder} {et~al}\mbox{.}(2013){Schroeder}, {Mesinger}, \&
  {Haiman}}]{Schroeder13}
{Schroeder} J., {Mesinger} A., {Haiman} Z., 2013, \mnras, 428, 3058

\bibitem[{{Theuns} {et~al}\mbox{.}(2002){Theuns}, {Schaye}, {Zaroubi}, {Kim},
  {Tzanavaris}, \& {Carswell}}]{Theuns2002}
{Theuns} T., {Schaye} J., {Zaroubi} S., {Kim} T.-S., {Tzanavaris} P.,
  {Carswell} B., 2002, \apjl, 567, L103

\bibitem[{{Tingay} {et~al}\mbox{.}(2013){Tingay}, {Goeke}, {Bowman}, {Emrich},
  {Ord}, {Mitchell}, {Morales}, {Booler}, {Crosse}, {Wayth}, {Lonsdale},
  {Tremblay}, {Pallot}, {Colegate}, {Wicenec}, {Kudryavtseva}, {Arcus},
  {Barnes}, {Bernardi}, {Briggs}, {Burns}, {Bunton}, {Cappallo}, {Corey},
  {Deshpande}, {Desouza}, {Gaensler}, {Greenhill}, {Hall}, {Hazelton}, {Herne},
  {Hewitt}, {Johnston-Hollitt}, {Kaplan}, {Kasper}, {Kincaid}, {Koenig},
  {Kratzenberg}, {Lynch}, {Mckinley}, {Mcwhirter}, {Morgan}, {Oberoi},
  {Pathikulangara}, {Prabu}, {Remillard}, {Rogers}, {Roshi}, {Salah}, {Sault},
  {Udaya-Shankar}, {Schlagenhaufer}, {Srivani}, {Stevens}, {Subrahmanyan},
  {Waterson}, {Webster}, {Whitney}, {Williams}, {Williams}, \&
  {Wyithe}}]{2013PASA...30....7T}
{Tingay} S.~J. {et~al.}, 2013, \pasa, 30, 7

\bibitem[{{Trac} \& {Gnedin}(2011)}]{2011ASL.....4..228T}
{Trac} H.~Y., {Gnedin} N.~Y., 2011, Advanced Science Letters, 4, 228

\bibitem[{{van Haarlem} {et~al}\mbox{.}(2013){van Haarlem}, {Wise}, {Gunst},
  {Heald}, {McKean}, {Hessels}, {de Bruyn}, {Nijboer}, {Swinbank}, {Fallows},
  {Brentjens}, {Nelles}, {Beck}, {Falcke}, {Fender}, {H{\"o}randel},
  {Koopmans}, {Mann}, {Miley}, {R{\"o}ttgering}, {Stappers}, {Wijers},
  {Zaroubi}, {van den Akker}, {Alexov}, {Anderson}, {Anderson}, {van Ardenne},
  {Arts}, {Asgekar}, {Avruch}, {Batejat}, {B{\"a}hren}, {Bell}, {Bell}, {van
  Bemmel}, {Bennema}, {Bentum}, {Bernardi}, {Best}, {B{\^i}rzan}, {Bonafede},
  {Boonstra}, {Braun}, {Bregman}, {Breitling}, {van de Brink}, {Broderick},
  {Broekema}, {Brouw}, {Br{\"u}ggen}, {Butcher}, {van Cappellen}, {Ciardi},
  {Coenen}, {Conway}, {Coolen}, {Corstanje}, {Damstra}, {Davies}, {Deller},
  {Dettmar}, {van Diepen}, {Dijkstra}, {Donker}, {Doorduin}, {Dromer}, {Drost},
  {van Duin}, {Eisl{\"o}ffel}, {van Enst}, {Ferrari}, {Frieswijk}, {Gankema},
  {Garrett}, {de Gasperin}, {Gerbers}, {de Geus}, {Grie{\ss}meier}, {Grit},
  {Gruppen}, {Hamaker}, {Hassall}, {Hoeft}, {Holties}, {Horneffer}, {van der
  Horst}, {van Houwelingen}, {Huijgen}, {Iacobelli}, {Intema}, {Jackson},
  {Jelic}, {de Jong}, {Juette}, {Kant}, {Karastergiou}, {Koers}, {Kollen},
  {Kondratiev}, {Kooistra}, {Koopman}, {Koster}, {Kuniyoshi}, {Kramer},
  {Kuper}, {Lambropoulos}, {Law}, {van Leeuwen}, {Lemaitre}, {Loose}, {Maat},
  {Macario}, {Markoff}, {Masters}, {McFadden}, {McKay-Bukowski}, {Meijering},
  {Meulman}, {Mevius}, {Middelberg}, {Millenaar}, {Miller-Jones}, {Mohan},
  {Mol}, {Morawietz}, {Morganti}, {Mulcahy}, {Mulder}, {Munk}, {Nieuwenhuis},
  {van Nieuwpoort}, {Noordam}, {Norden}, {Noutsos}, {Offringa}, {Olofsson},
  {Omar}, {Orr{\'u}}, {Overeem}, {Paas}, {Pandey-Pommier}, {Pandey}, {Pizzo},
  {Polatidis}, {Rafferty}, {Rawlings}, {Reich}, {de Reijer}, {Reitsma},
  {Renting}, {Riemers}, {Rol}, {Romein}, {Roosjen}, {Ruiter}, {Scaife}, {van
  der Schaaf}, {Scheers}, {Schellart}, {Schoenmakers}, {Schoonderbeek},
  {Serylak}, {Shulevski}, {Sluman}, {Smirnov}, {Sobey}, {Spreeuw}, {Steinmetz},
  {Sterks}, {Stiepel}, {Stuurwold}, {Tagger}, {Tang}, {Tasse}, {Thomas},
  {Thoudam}, {Toribio}, {van der Tol}, {Usov}, {van Veelen}, {van der Veen},
  {ter Veen}, {Verbiest}, {Vermeulen}, {Vermaas}, {Vocks}, {Vogt}, {de Vos},
  {van der Wal}, {van Weeren}, {Weggemans}, {Weltevrede}, {White}, {Wijnholds},
  {Wilhelmsson}, {Wucknitz}, {Yatawatta}, {Zarka}, {Zensus}, \& {van
  Zwieten}}]{2013A&A...556A...2V}
{van Haarlem} M.~P. {et~al.}, 2013, \aap, 556, A2

\bibitem[{{Willott} {et~al}\mbox{.}(2007){Willott}, {Delorme}, {Omont},
  {Bergeron}, {Delfosse}, {Forveille}, {Albert}, {Reyl{\'e}}, {Hill},
  {Gully-Santiago}, {Vinten}, {Crampton}, {Hutchings}, {Schade}, {Simard},
  {Sawicki}, {Beelen}, \& {Cox}}]{Willott}
{Willott} C.~J. {et~al.}, 2007, \aj, 134, 2435

\end{thebibliography}

\end{document}